\documentclass[12pt]{article}
\usepackage{amsfonts}
\usepackage{amsmath}
\usepackage{amssymb}

\newtheorem{example}{Example}

\newtheorem{lemma}{Lemma}

\newtheorem{theorem}{Theorem}
\newtheorem{remark}{Remark}

\numberwithin{equation}{section}
\numberwithin{lemma}{section}
\numberwithin{theorem}{section}
\numberwithin{remark}{section}
\numberwithin{corollary}{section}
\numberwithin{proposition}{section}
\numberwithin{definition}{section}
\numberwithin{example}{section}
\numberwithin{table}{section}
\input{tcilatex}
\begin{document}

\title{Numerical integration of Heath-Jarrow-Morton model of interest rates}
\author{M. Krivko and M.V. Tretyakov \\
%EndAName
{\small Department of Mathematics, University of Leicester, Leicester
LE1~7RH, UK}\\
{\small \ E-mail: mk211@le.ac.uk and M.Tretyakov@le.ac.uk}}
\maketitle

\begin{abstract}
We propose and analyze numerical methods for the Heath-Jarrow-Morton (HJM)
model. To construct the methods, we first discretize the infinite
dimensional HJM equation in maturity time variable using quadrature rules
for approximating the arbitrage-free drift. This results in a finite
dimensional system of stochastic differential equations (SDEs) which we
approximate in the weak and mean-square sense using the general theory of
numerical integration of SDEs. The proposed numerical algorithms are highly
computationally efficient due to the use of high-order quadrature rules
which allow us to take relatively large discretization steps in the maturity
time without affecting overall accuracy of the algorithms. Convergence
theorems for the methods are proved. Results of some numerical experiments
with European-type interest rate derivatives are presented.


\end{abstract}

\section{Introduction}

The framework proposed by Heath, Jarrow and Morton \cite{HJM1992} -- HJM
henceforth -- models the evolution of the term structure of interest rates
through the dynamics of the forward rate curve. These dynamics are described
by a multifactor infinite dimensional stochastic equation with the entire
forward rate curve as state variable. Under no-arbitrage conditions, the HJM
model is fully characterized by specifying forward rate volatility functions
and the initial forward curve. The original HJM\ framework is used for
modelling fixed income markets (see \cite%
{HJM1992,BrigoMercurio,CarmonaTehranchi,TalayIR} and also references
therein). Recently, the HJM philosophy has been extended to credit and
equity markets (see, e.g. the recent review \cite{Carmona}).

The HJM model has closed-form solutions only for some special cases of
volatility, and valuations under the HJM framework usually require a
numerical approximation. As far as we know, the literature on numerics for
the HJM model is rather sparse. The common approach (see, e.g. \cite%
{HJM1990,Jarrow2002,Glasserman,BjorkSzepessyTempone} and the references
therein) is to take coinciding grids in the running time $t$ and in the
maturity time $T.$ The known methods differ in the way they approximate the
integral in the arbitrage-free drift of the HJM model while they all use
Euler-type schemes for discretization in time. In \cite%
{HJM1990,Jarrow2002,Glasserman} approximations of the arbitrage-free drift
are chosen so that the overall discrete approximations of the HJM equation
preserve the martingale property for the discretized discounted bond
process. A different numerical approach, based on a functional backward
Kolmogorov equation, is considered in \cite{Marcozzi}. In comparison with
other works, the papers \cite{BjorkSzepessyTempone,Marcozzi} rigorously
prove weak convergence of the proposed numerical methods.

In this paper we propose and analyze a new class of effective numerical
methods for the HJM equation exploiting the idea of the method of lines.
These methods can be used for simulating HJM models of various
specifications. Our main focus is on the weak-sense numerical methods which
can be used for valuing a broad range of interest rate products. To
construct the numerical methods, we first discretize the infinite
dimensional HJM equation in maturity time variable $T$ using quadrature
rules for approximating the arbitrage-free drift. This results in a finite
dimensional system of stochastic differential equations (SDEs). As we show
in the paper, if we take a quadrature rule of order $p,$ the solution of
this finite dimensional system of SDEs converges to the HJM\ solution with
mean-square order $p$ in the maturity time discretization step $\Delta $.
From the method of lines point of view, we interpret the maturity time $T$
as a \textquotedblleft space\textquotedblright\ variable and the running
time $t$ as a \textquotedblleft time\textquotedblright\ variable. To get
fully discrete methods (discrete in both $T$ and $t),$ we approximate the
obtained finite dimensional system of SDEs in the weak and mean-square
senses using the general theory of numerical integration of SDEs (see, e.g. 
\cite{MT1}). The proposed numerical algorithms are computationally highly
efficient due to the use of high-order quadrature rules which allow us to
take relatively large discretization steps in the maturity time without
affecting overall accuracy of the algorithms, i.e., the number of forward
rates we need to approximate at each time moment $t$ is significantly less
than it is usually required in the case when the time grids for $t$ and $T$
coincide. Further, since we exploit the method of lines, we have flexibility
in choosing appropriate approximations in \textquotedblleft
space\textquotedblright\ and \textquotedblleft time\textquotedblright\
separately. As we will see, in practice it is beneficial to use higher order
rules for integration with respect to maturity time $T$ and lower order
numerical schemes for integration with respect to time $t$.

The rest of the paper is organized as follows. In Section~\ref{sec:hjm} we
recall the HJM framework. Section~\ref{sec:met} deals with construction of a
class of numerical methods of a general form for the HJM equation. We start
with $T$-discretization of the HJM equation (Section~\ref{sec:Tdis}), then
consider discretization in time $t$ (Section~\ref{sec:tdis}), and finally,
based on the results of the previous subsections, we obtain approximations
suitable for evaluating prices of European-type interest rate contracts
under the HJM framework (Section~\ref{sec:appyz}). In Section~\ref{sec:err}
we prove convergence theorems for the methods constructed in Section~\ref%
{sec:met}. We first prove convergence theorems for the HJM\ approximation
discrete in the maturity time $T$ only (Section~\ref{sec:errMd}). Then we
analyze weak convergence of fully discrete methods to the approximations
discrete in the maturity time (Section~\ref{sec:errtd}). We show that this
convergence is uniform in the maturity time discretization step $\Delta $ in
order to obtain weak convergence of the fully-discrete numerical methods to
the solution of the HJM equation. We note that both the considered class of
numerical methods and proof of their convergence include the known numerical
schemes for HJM as, e.g. those from \cite{Glasserman}. In Section~\ref%
{sec:alg} we illustrate the introduced class of numerical methods from
Section~\ref{sec:met} by presenting some particular algorithms of various
accuracy orders, which are ready for implementation. In Section~\ref{sec:msq}
we propose a fully-discrete mean-square approximation of the HJM equation
and prove the corresponding mean-square convergence theorem. In Section~\ref%
{sec:exp} we test the proposed numerical algorithms on the Vasicek and
proportional volatility models.

\section{HJM framework of the instantaneous forward rate dynamics\label%
{sec:hjm}}

Throughout this paper we assume that there exists an arbitrage-free market
with a frictionlessly traded continuum of default-free zero-coupon bonds $%
\{P(t,T),$ $t\leq T,$ $T\in \left[ t_{0},T^{\ast }\right] ,$ $t\in \left[
t_{0},t^{\ast }\right] \},$ where $P(t,T)$ denotes the price at time $t$ of
a bond with maturity $T$. We require that $P(T,T)=1$ and $P(t,T)$ is
sufficiently smooth in the maturity variable $T.$

A convenient, albeit a theoretical concept, the forward rate $f(t,T)$, $%
t\leq T,$ $T\in \left[ t_{0},T^{\ast }\right] ,$ $t\in \left[ t_{0},t^{\ast }%
\right] $, represents the instantaneous continuously compounded interest
rate prevailing at time $t$ for riskless borrowing or lending over the
infinitesimal time interval $[T,T+dT].$ The relation between zero-coupon
bonds and instantaneous forward rates is given by%
\begin{equation}
P(t,T)=\exp \left( -\tint\limits_{t}^{T}f(t,u)du\right) .  \label{1}
\end{equation}%
The current instantaneous rate, or so-called short rate, is%
\begin{equation}
r(t):=f(t,t).  \label{2}
\end{equation}%
To represent the accumulating factor one can define the savings account 
\begin{equation}
B(t)=\exp \left( \tint\limits_{t_{0}}^{t}r(s)ds\right) .  \label{2a}
\end{equation}

The HJM framework \cite{HJM1992} models the dynamics of the forward curve%
\begin{equation*}
\left\{ f(t,T),\text{\ \ }t\leq T,\text{\ \ }T\in \left[ t_{0},T^{\ast }%
\right] ,\text{\ \ }t\in \left[ t_{0},t^{\ast }\right] \right\} .
\end{equation*}%
Given an integrable deterministic initial forward curve 
\begin{equation*}
f(t_{0},T)=f_{0}(T),
\end{equation*}%
the arbitrage-free dynamics of the forward curve under the risk-neutral
measure $Q$ associated with the numeraire $B(t)$ are modelled through an Ito
process of the form%
\begin{eqnarray}
f(t,T)-f_{0}(T) &=&\int_{t_{0}}^{t}\sigma ^{\top }(s,T)\left(
\int_{s}^{T}\sigma (s,u)du\right) ds  \label{3} \\
&&+\int_{t_{0}}^{t}\sigma ^{\top }(s,T)dW(s),\ \ t_{0}\leq t\leq t^{\ast
}\wedge T,\ \ t_{0}\leq T\leq T^{\ast },  \notag
\end{eqnarray}%
where $W(t)=\left( W_{1}(t),\ldots ,W_{d}(t)\right) ^{\top }$ is a $d$%
-dimensional standard Wiener process defined on a filtered probability space 
$\left( \Omega ,\mathcal{F},\left\{ \mathcal{F}_{t}\right\} _{t_{0}\leq
t\leq t^{\ast }},Q\right) $ satisfying the usual hypothesis; $\sigma (t,T)$
is an $\mathbb{R}^{d}$-valued $\mathcal{F}_{t}$-progressively measurable
stochastic process with $\int_{t_{0}}^{T}\left\vert \sigma (s,T)\right\vert
^{2}ds<\infty ;$ and $t^{\ast }\wedge T:=\min (t^{\ast },T).$

In general, the volatility $\sigma (t,T):=$ $\sigma (t,T,\omega )$ can
depend on the current and past values of forward rates. In this paper we
restrict ourselves to the case in which $\sigma $ depends on the current
forward rate only, i.e.,%
\begin{equation}
\sigma (t,T):=(\sigma _{1}(t,T,f(t,T)),\ldots ,\sigma
_{d}(t,T,f(t,T)))^{\top },  \label{3b}
\end{equation}%
where $\sigma _{i}(t,T,z),$\ $i=1,\ldots ,d,$ are deterministic functions
defined on $\left[ t_{0},t^{\ast }\right] \times \left[ t_{0},T^{\ast }%
\right] \times \mathbb{R}.$ Then the term $\int_{s}^{T}\sigma (s,u)du$ in (%
\ref{3}) can be written as $\int_{s}^{T}\sigma (s,u,f(s,u))du$, and,
consequently, (\ref{3})-(\ref{3b}) is an infinite-dimensional SDE. We impose
the following assumptions on the HJM model (\ref{3})-(\ref{3b}). \medskip

\noindent \textbf{Assumption 2.1 }\emph{The functions }$\sigma _{i}(t,T,z),$%
\emph{\ }$i=1,\ldots ,d,$\emph{\ are uniformly bounded, i.e., there is a
constant }$C>0$ \emph{such that}%
\begin{equation}
\left\vert \sigma _{i}(t,T,z)\right\vert \leq C,\text{ }(t,T,z)\in \left[
t_{0},t^{\ast }\right] \times \left[ t_{0},T^{\ast }\right] \times \mathbb{R}%
.  \label{4b}
\end{equation}

\noindent \textbf{Assumption 2.2 }\emph{For sufficiently large }$p_{1},$ $%
p_{2}\geq 1,$\emph{\ the partial derivatives}%
\begin{equation}
\frac{\partial ^{j+k+l}\sigma _{i}(t,T,z)}{\partial t^{j}\partial
T^{k}\partial z^{l}}\emph{,}\ 0\leq j\leq p_{1},\ 0\leq k+l\leq p_{2},\ \
i=1,\ldots ,d,  \label{4c}
\end{equation}%
\emph{are continuous and uniformly bounded in }$\left[ t_{0},t^{\ast }\right]
\times \left[ t_{0},T^{\ast }\right] \times \mathbb{R}$\emph{.}\medskip

\noindent \textbf{Assumption 2.3 }\emph{The initial forward curve} $%
f_{0}(T),\ T\in \left[ t_{0},T^{\ast }\right] ,$ \emph{is deterministic and
sufficiently smooth.}\medskip

The imposed conditions are sufficient to ensure that the SDE (\ref{3})-(\ref%
{3b}) has a unique strong solution $f(t,T),$ which is sufficiently smooth in
the last argument (see \cite{HJM1992,Wissel} and also \cite%
{Krilov,GihmanSkorohod} for differentiating SDE solutions with respect to a
parameter). Further, it is not difficult to show that they imply boundedness
of exponential moments of $f(t,T),$ i.e., for a $c\in \mathbb{R}$ there is a
constant $C>0$ such that%
\begin{equation}
E\exp (c|f(t,T)|)<C  \label{expmom}
\end{equation}%
for all $(t,T)\in \left[ t_{0},t^{\ast }\right] \times \left[ t_{0},T^{\ast }%
\right] .$ The constant $C$ in (\ref{expmom}) depends on the initial forward
curve $f_{0}(T),$ volatility $\sigma (t,T,z),$ and on $c.$

\begin{remark}
\label{Rem:as2}As it was shown in \cite{Wissel}, for the SDE $(\ref{3})$-$(%
\ref{3b})$ to have the unique strong solution it suffices to require a
weaker assumption than Assumption~2.1: 
\begin{equation*}
\left\vert \sigma _{i}(t,T,z)\right\vert \leq C\left( 1+\left\vert
z\right\vert ^{1/2}\right) .
\end{equation*}%
However, in the paper we restrict ourselves to the stronger set of
conditions which allow us to consider methods of higher order.
Assumptions~2.1-2.3 are sufficient for all the statements in this paper. The
choice of $p_{1}\ $and $p_{2}$ depends on a particular algorithm (as usual,
the more accurate an algorithm the more derivatives are needed). At the same
time, the imposed conditions are not necessary and the proposed numerical
methods themselves can be used under broader assumptions.
\end{remark}

We pay attention that Assumptions~2.1-2.3 do not guarantee positiveness of $%
f(t,T)$ which could be a desirable property taking into account the
financial context of the HJM model. One can notice that if we also require
that 
\begin{equation*}
f_{0}(T)\geq 0\text{ \ and \ }\sigma _{i}(t,T,0)=0,\ i=1,\ldots ,d,\
(t,T)\in \lbrack t_{0},t^{\ast }]\times \lbrack t_{0},T^{\ast }],
\end{equation*}%
then the forward rates are nonnegative $f(t,T)\geq 0$ for all $(t,T)\in
\lbrack t_{0},t^{\ast }]\times \lbrack t_{0},T^{\ast }].$\medskip\ 

Our main objective is to propose efficient numerical methods for pricing
interest rates derivatives. Among these instruments are interest rate caps,
floors, and swaptions \cite%
{BrigoMercurio,CarmonaTehranchi,Pelsser,Rebo,TalayIR}. A cap price is
obtained by summing up the prices of the underlying caplets. Consider a
caplet set at time $s_{k}$ with payment date at $s_{i}>s_{k},$ with strike $%
K $ and unit cap nominal value. Its price at time $t_{0}\leq s_{k}$ is given
by%
\begin{equation}
E\exp \left( -\int_{t_{0}}^{s_{k}}r(u)du\right) \left[ 1-(1+K(s_{i}-s_{k}))%
\exp \left( -\int_{s_{k}}^{s_{i}}f(s_{k},u)du\right) \right] _{+}\ .
\label{5}
\end{equation}%
Now consider a payer swaption of maturity $s_{k}$ and with underlying swap
maturity $s_{i}>s_{k}.$ Its price at time $t_{0}\leq s_{k}$ can be found as 
\begin{eqnarray}
&&E\exp \left( -\int_{t_{0\text{ }}}^{s_{k}}r(u)du\right) \left[ 1-\exp
\left( -\int_{s_{k}}^{s_{i}}f(s_{k},u)du\right) \right.  \label{6} \\
&&\left. -K\dsum\limits_{j=k+1}^{i}\left( s_{j}-s_{j-1}\right) \exp \left(
-\int_{s_{k}}^{s_{j}}f(s_{k},u)du\right) \right] _{+}\ .  \notag
\end{eqnarray}%
In (\ref{5}) and (\ref{6}) and in what follows expectation $E\left( \cdot
\right) $ without any index means expectation taken with respect to the
risk-neutral measure $Q$.

Let $G(z),$ $z\in \mathbb{R},$ be a payoff function satisfying the global
Lipschitz condition, i.e.,%
\begin{equation}
\left\vert G(z)-G(z^{\prime })\right\vert \leq K\left\vert z-z^{\prime
}\right\vert ,\ \ z,z^{\prime }\in \mathbb{R}.  \label{7a}
\end{equation}%
In this paper, motivated by the above examples, we consider the price of a
generic interest rate contract under risk-neutral measure of the form%
\begin{equation}
F(t_{0},f_{0}\left( \cdot \right) ;s_{k},s_{i})=E\exp (-Y(s_{k}))G\left(
P(s_{k},s_{i})\right) ,  \label{7}
\end{equation}%
where 
\begin{equation}
Y(s_{k})=\int_{t_{0}}^{s_{k}}r(u)du,  \label{8}
\end{equation}%
\begin{equation}
P(s_{k},s_{i})=\exp \left( -Z(s_{k},s_{i})\right) ,  \label{9a}
\end{equation}%
and%
\begin{equation}
Z(s_{k},s_{i})=\int_{s_{k}}^{s_{i}}f(s_{k},u)du.  \label{9}
\end{equation}

We note that (\ref{7}) does not cover the case of swaptions (\ref{6}). To
include swaptions, the payoff $G$ in (\ref{7}) should be of the form $%
G\left( P(s_{k},s_{k+1}),\ldots ,P(s_{k},s_{i})\right) $ and 
\begin{equation}
F(t_{0},f_{0}\left( \cdot \right) ;s_{k},s_{k+1},\ldots ,s_{i})=E\exp
(-Y(s_{k}))G\left( P(s_{k},s_{k+1}),\ldots ,P(s_{k},s_{i})\right) \ .
\label{7ext}
\end{equation}%
We limit ourselves in the paper to the payoff of the form (\ref{7}) for the
sake of transparent exposition. All the proposed numerical algorithms are
applicable to the more general form of the payoff (\ref{7ext}). Also, no
additional ideas are required to extend our theoretical analysis to the case
(\ref{7ext}).

\begin{remark}
\label{rem:forwmeas}(Forward measure pricing) The HJM dynamics can be
written under the $s_{k}$-forward measure (see details e.g., in \cite%
{BrigoMercurio,Pelsser,Glasserman}) instead of the risk-neutral measure. In
this case the corresponding SDE has the form $($cf. $(\ref{3}))$:%
\begin{eqnarray}
f(t,T)-f_{0}(T) &=&\int_{t_{0}}^{t}\sigma ^{\top }(s,T)\left(
\int_{s_{k}}^{T}\sigma (s,u)du\right) ds+\int_{t_{0}}^{t}\sigma ^{\top
}(s,T)dW^{s_{k}}(s),\ \ \ \ \   \label{3TF} \\
&&\ \ t_{0}\leq t\leq t^{\ast }\wedge T\wedge s_{k},\ \ \ \ t_{0}\leq T\leq
T^{\ast },  \notag
\end{eqnarray}%
with $W^{s_{k}}(s)$ being a $d$-dimensional standard Wiener process under
the $s_{k}$-forward measure $Q^{s_{k}}.$ The pricing formula for a generic
interest rate contract with payoff $G\left( P(s_{k},s_{i})\right) $ under $%
Q^{s_{k}}$ is $($cf. $(\ref{7}))$:%
\begin{equation}
F(t_{0},f_{0}\left( \cdot \right)
;s_{k},s_{i})=P(t_{0},s_{k})E^{s_{k}}\left( G\left( P(s_{k},s_{i})\right)
\right) .  \label{7TF}
\end{equation}%
This form is computationally simpler than $(\ref{7})$ since it does not
require evaluation of the short rate. At the same time we note that pricing
of some interest rate products (e.g., Eurodollar futures) require the use of
risk-neutral measure \cite{BrigoMercurio,NBS07}. In this paper we construct
numerical algorithms for approximating $(\ref{7}).$ Obviously, these
algorithms are readily (actually more easily) applicable to $(\ref{7TF}).$
\end{remark}

\section{Numerical method\label{sec:met}}

In this section we construct a numerical method for simulating (\ref{7})
with the forward rates $f(s_{k},u)$ satisfying the infinite-dimensional SDE (%
\ref{3})-(\ref{3b}). Examples of some particular algorithmic realizations of
this method are given in Section~\ref{sec:alg}.

This section is organized in the following way. We first introduce a
maturity time discretization ($T$-discretization) and arrive at a finite
dimensional approximation of (\ref{3})-(\ref{3b}), i.e., at a finite system
of SDEs (Section~\ref{sec:Tdis}). Then (Section~\ref{sec:tdis}) we
discretize time ($t$-discretization) and apply a weak-sense numerical
integrator to the obtained finite system of SDEs. Finally, Section~\ref%
{sec:appyz} deals with approximating the functionals $Y$ and $Z$ from (\ref%
{8})-(\ref{9}) and the option price (\ref{7}).

For the simplicity of presentation, we consider equally-spaced grids for
maturity time $T$ and time $t.$ A nonuniform discretization might be needed
in practical financial applications, and a generalization of the proposed
algorithms to nonuniform time grids is straightforward.

\subsection{$T$-discretization\label{sec:Tdis}}

Consider a uniform partition of the maturity time interval $[t_{0},T^{\ast
}] $ with a maturity time step ($T$-step) $\Delta =(T^{\ast }-t_{0})/N:$ 
\begin{equation}
t_{0}=T_{0}<\cdots <T_{N}=T^{\ast },\ \ T_{i}=i\Delta ,\ \ i=0,\ldots ,N.
\label{Tdis}
\end{equation}

Introduce the index notation which we will use throughout the paper. Denote
by $\ell (t)$ the auxiliary index dependent on time $t$ so that 
\begin{equation}
\ell (t)=\max \left\{ i=0,1,\ldots ,N:\text{ \ \ }t\geq T_{i}\right\} ,
\label{indl}
\end{equation}%
and by $\varrho (t)$ the auxiliary index dependent on time $t$ so that%
\begin{equation}
\varrho (t)=\min \left\{ i=0,1,\ldots ,N:\ \ \text{ }t<T_{i}\right\} ,
\label{indm}
\end{equation}%
i.e., $T_{\ell \left( t\right) }\leq t<T_{\varrho \left( t\right) }$ and $%
T_{\ell \left( t\right) }$ (or $T_{\varrho \left( t\right) }$) is the
closest node on the grid (\ref{Tdis}) to the time $t$ from the left (or from
the right). We also note that $\varrho (t)=\ell (t)+1.$

Further, we require for simplicity that $\Delta $ is sufficiently small so
that a number of nodes $T_{i}$ between $t^{\ast }$ and $T^{\ast }$ is
sufficient for realization of all the quadrature rules and
interpolation/extrapolation used in the method which we introduce in this
Section~\ref{sec:met}. We will pay attention to the required amount of nodes
between $t^{\ast }$ and $T^{\ast }$ in the method's description. At the same
time, if in practical realization the distance between $t^{\ast }$ and $%
T^{\ast }$ is relatively small in comparison with the chosen $T$-step $%
\Delta ,$ then one would need to run simulation for a slightly longer
maturity-time interval, extending it beyond $T^{\ast }$ by a few steps of $%
\Delta $ (see further explanation in Section~\ref{order4}).

For a node $T_{i}$, $i=0,\ldots ,N,$ on the maturity time grid (\ref{Tdis}),
we approximate the integrals in (\ref{3}): 
\begin{equation}
I_{j}(s,T_{i}):=\int_{s}^{T_{i}}\sigma _{j}(s,u)du,\text{ }j=1,\ldots ,d,%
\text{ }t_{0}\leq s\leq t^{\ast }\wedge T_{i},\ \ i=1,\ldots ,N,  \label{2.3}
\end{equation}%
by a composite quadrature rule $S_{I_{j}}(s,T_{i},\Delta ):$ 
\begin{equation}
I_{j}(s,T_{i})\approx S_{I_{j}}(s,T_{i},\Delta )=\Delta \dsum_{k=\varrho
(s)}^{\kappa (s,T_{i})}\gamma _{k}(s)\sigma _{j}(s,T_{k}),  \label{2.4}
\end{equation}%
where the quadrature rule's weights $\gamma _{k}(s)$ and the nodes $%
k=\varrho (s),\ldots ,\kappa (s,T_{i})$ are chosen so that under
Assumptions~2.1-2.3 the approximation is of order $O(\Delta ^{p})$ for a
given $p\geq 1,$ i.e., the numerical integration error is estimated as 
\begin{equation}
\left( E\left[ S_{_{I_{j}}}(s,T_{i},\Delta )-I_{j}(s,T_{i})\right]
^{2}\right) ^{1/2}\leq C\Delta ^{p}  \label{2.5}
\end{equation}%
with a constant $C>0$ independent of $\Delta ,$ $s,\ T_{i},$ $j.$ Some
examples of such quadratures are given in Section~\ref{sec:alg}. We note
(see details in Section~\ref{sec:alg}) that when $s$ and $T_{i}$ are close,
to approximate $I_{j}(s,T_{i})$ with a required accuracy the number $\kappa
(s,T_{i})$ in (\ref{2.4}) can be chosen larger than $i.$ We recall that
since we assumed that there is a sufficient number of nodes between $t^{\ast
}$ and $T^{\ast }$ the number $\kappa (s,T_{i})$ does not exceed $N.$ We
will also use the vector notation $I(s,T_{i}):=(I_{1}(s,T_{i}),\ldots
,I_{d}(s,T_{i}))^{\top }$ and $S_{I}(s,T_{i},\Delta
):=(S_{I_{1}}(s,T_{i},\Delta ),\ldots ,S_{I_{d}}(s,T_{i},\Delta ))^{\top }.$

For a fixed $T=T_{i},$ it is convenient for later purposes (namely, for
computing the short rate\textbf{\ }$r(t)=f(t,t)$ as it will become clear in
Section~\ref{sec:appyz}) to consider the SDE (\ref{3})-(\ref{3b}) on a
slightly larger time interval: $t_{0}\leq t\leq t^{\ast }\wedge
T_{(i+1)\wedge N},$ i.e., for $T_{i}<t^{\ast }$ (note that $t^{\ast }\leq $ $%
T_{N})$ we would like to extend the definition of $f(t,T_{i})$ from $t\in
\lbrack t_{0},T_{i}]$ to $t\in \lbrack t_{0},T_{i+1}].$ Though from the
point of view of financial applications the forward rate $f(t,T_{i})$ is not
defined on the interval $t\in (T_{i},T_{i+1}],$ Assumptions~2.1-2.3
guarantee that (\ref{3})-(\ref{3b}) has the strong solution on the extended
interval and, as it will be seen in future, this extension is beneficial
from the computational prospective (see also Remarks~\ref{rem:fict1} and~\ref%
{rem:fict2}). This extension requires from us to consider, in addition to (%
\ref{2.3}), the integrals 
\begin{equation}
I_{j}(s,T_{\ell (s)}):=\int_{s}^{T_{\ell (s)}}\sigma _{j}(s,u)du.
\label{2.5a}
\end{equation}%
We approximate these integrals by a quadrature rule analogous to the one in (%
\ref{2.4}) but with summation index $k$ starting\ from $\ell (s):$ 
\begin{equation}
I_{j}(s,T_{\ell (s)})\approx S_{I_{j}}(s,T_{\ell (s)},\Delta )=\Delta
\dsum_{k=\ell (s)}^{\kappa (s,T_{\ell (s)})}\gamma _{k}(s)\sigma
_{j}(s,T_{k}),  \label{2.44}
\end{equation}%
and we require that its error satisfies (\ref{2.5}). Combining (\ref{2.4})
and (\ref{2.44}), we will write in what follows that%
\begin{equation}
S_{I_{j}}(s,T_{i},\Delta )=\Delta \dsum_{k=\ell (s)}^{\kappa
(s,T_{i})}\gamma _{k}(s)\sigma _{j}(s,T_{k})  \label{2.444}
\end{equation}%
with the coefficient $\gamma _{\ell (s)}(s)=0$ if $i>\ell (s).$

Using (\ref{2.444}), we approximate the solution $f(t,T)$ of the
infinite-dimensional SDE (\ref{3})-(\ref{3b}) at the nodes $T=T_{0},\ldots
,T_{N},$ by the $N+1$-dimensional stochastic process $\tilde{f}%
^{i}(t)\approx f(t,T_{i}),$ $i=0,\ldots ,N,$ which satisfies the finite
system of coupled SDEs: 
\begin{eqnarray}
\tilde{f}^{i}(t)-f_{0}^{i} &=&\int_{t_{0}}^{t}\tilde{\sigma}_{i}^{\top }(s)%
\tilde{S}_{I}(s,T_{i},\Delta )ds+\int_{t_{0}}^{t}\tilde{\sigma}_{i}^{\top
}(s)dW(s),  \label{2.6} \\
\text{ }t_{0} &\leq &t\leq t^{\ast }\wedge T_{(i+1)\wedge N},\text{ }%
i=0,\ldots ,N,  \notag
\end{eqnarray}%
where 
\begin{eqnarray}
f_{0}^{i} &=&f_{0}(T_{i}),  \label{2.65} \\
\tilde{\sigma}_{i}(s) &=&(\tilde{\sigma}_{i,1}(s),\ldots ,\tilde{\sigma}%
_{i,d}(s))^{\top }=(\sigma _{1}(s,T_{i},\tilde{f}^{i}(s)),\ldots ,\sigma
_{d}(s,T_{i},\tilde{f}^{i}(s)))^{\top },  \notag \\
\tilde{S}_{I}(s,T_{i},\Delta ) &=&\left( \tilde{S}_{I_{1}}(s,T_{i},\Delta
),\ldots ,\tilde{S}_{I_{d}}(s,T_{i},\Delta )\right) ^{\top }  \notag
\end{eqnarray}%
and%
\begin{equation}
\tilde{S}_{I_{j}}(s,T_{i},\Delta )=\Delta \dsum_{k=\ell (s)}^{\kappa
(s,T_{i})}\gamma _{k}(s)\tilde{\sigma}_{k,j}(s).  \label{2.66}
\end{equation}%
We emphasize again that we extended the time interval from $t\in \lbrack
t_{0},t^{\ast }]$ to $t\in \lbrack t_{0},t^{\ast }\wedge T_{(i+1)\wedge N}].$

Assumptions~2.1-2.3 guarantee the existence of the unique strong solution of
(\ref{2.6})-(\ref{2.66}). Further, it is not difficult to show that they
also imply boundedness of exponential moments of $\tilde{f}^{i}(t),$ i.e.,
for a $c\in \mathbb{R}$ there is a constant $C>0$ such that (cf. (\ref%
{expmom})):%
\begin{equation}
E\exp (c|\tilde{f}^{i}(t)|)<C  \label{expmom2}
\end{equation}%
for all $t\in \left[ t_{0},t^{\ast }\right] \wedge T_{(i+1)\wedge N},$ $%
i=0,\ldots ,N.$ In connection with (\ref{expmom2}) we recall that due to
Assumption~2.3 the initial forward rate curve $f_{0}(T),\ t_{0}\leq T\leq
T^{\ast },$ is bounded by a finite constant. Hence, $\tilde{f}%
^{i}(0)=f_{0}^{i},$ $i=0,\ldots ,N,$ are bounded by the same constant.

In Section~\ref{sec:errMd}\ we prove (see Theorem~\ref{thm:msqi2f})
mean-square convergence of $\tilde{f}^{i}(t)$ to $f(t,T_{i})$ when $\Delta
\rightarrow 0.$ We note that the system (\ref{2.6})-(\ref{2.66}) plays only
an auxiliary role in our consideration. It is used as guidance in
constructing fully discrete numerical algorithms (i.e., discrete in both $T$
and $t)$ and also in proofs of their convergence.

\subsection{$t$-discretization\label{sec:tdis}}

In this section we discretize the finite system of coupled ordinary SDEs (%
\ref{2.6})-(\ref{2.66}) with respect to time $t$ and thus arrive at a fully
discrete method.

We introduce an equally-spaced grid for time $t$ with step ($t$-step) $%
h=(t^{\ast }-t_{0})/M$: 
\begin{equation*}
t_{0}<\cdots <t_{M}=t^{\ast },\ \ t_{k}=kh,\ \ k=0,\ldots ,M.
\end{equation*}%
In what follows we use the notation (cf. (\ref{indl}) and (\ref{indm})): 
\begin{equation}
\ell _{k}:=\ell (t_{k}),\ \ \ \varrho _{k}:=\varrho (t_{k}).
\label{indlrdis}
\end{equation}

We consider an approximation $\bar{f}_{k+1}^{i}$ of $\tilde{f}^{i}(t_{k+1})$
from (\ref{2.6}) (i.e., a full discretization of (\ref{3})-(\ref{3b}) in
both $T$ and $t)$ of the form

\begin{gather}
\bar{f}_{0}^{i}=f_{0}^{i},\text{ }i=0,\ldots ,N,  \label{2.7} \\
\bar{f}_{k+1}^{i}=\bar{f}_{k}^{i}+A^{i}(t_{k},T_{i};\bar{f}_{k}^{j},\text{ }%
j=\ell _{k+1},\ldots ,\kappa (t_{k+1},T_{i})\vee i;h;\xi _{k}),\text{ } 
\notag \\
i=\ell _{k+1},\ldots ,N,\ k=0,\ldots ,M,  \notag
\end{gather}%
where the form of the functions $A^{i}$ depends on the coefficients of (\ref%
{2.6})-(\ref{2.66}), i.e., on $\sigma $ and on a choice of the quadrature
rule $S_{I_{j}};$ $\kappa (t_{k},T_{i})$ is as in the quadrature (\ref{2.66}%
); $\xi _{k},\ k=0,\ldots ,M,\ $are some random vectors which have moments
of a sufficiently high order and $\xi _{k}$ for $k>0$ are independent of $%
\bar{f}_{j}^{i},$ $i=\ell _{j},\ldots N,$ $j=0,\ldots ,k,$ and of $\xi
_{0},\ldots ,\xi _{k-1}.$

To simplify the exposition of our theoretical analysis, in what follows we
consider the extended $\tilde{f}^{i}(t)$ and $\bar{f}_{k}^{i}.$ We put 
\begin{equation*}
\tilde{f}^{i}(t)=\tilde{f}^{i}(T_{i+1}),\ \ T_{(i+1)\wedge N}\wedge t^{\ast
}\leq t\leq t^{\ast },\ \ 0\leq i\leq \ell (t^{\ast })-1,
\end{equation*}%
and then the $N+1$-dimensional vector $\{\tilde{f}^{i}(t),$ $i=0,\ldots ,N\}$
is defined for all $t\in \lbrack t_{0},t^{\ast }].$ We put 
\begin{equation*}
\bar{f}_{k}^{i}=\bar{f}_{\mathbf{m}}^{i},\ \ k=\mathbf{m}+1,\ldots ,M,\ \
0\leq i\leq \ell (t^{\ast })-1,
\end{equation*}%
where $\mathbf{m}=\left\lceil \left( T_{i+1}-t_{0}\right) /h\right\rceil -1$
(we recall that $\left\lceil \cdot \right\rceil $ denotes the ceiling
function). Then the $N+1$-dimensional vector $\{\bar{f}_{k}^{i},$ $%
i=0,\ldots ,N\}$ is defined for all $k=0,\ldots ,M$. Let us emphasize that
we do not use the extension of $\bar{f}_{k}^{i}$ in numerical algorithms and
these extensions of $\tilde{f}^{i}(t)$ and $\bar{f}_{k}^{i}$ are done in
order to use the vector notation $\tilde{f}(t)$ and $\bar{f}_{k}$ without
need to adjust length of these vectors as $t$ and $k$ grow.

We assume that the $A^{i}$ in (\ref{2.7}) are such that $\bar{f}_{k}^{i}$
satisfy the following condition.\medskip

\noindent \textbf{Assumption 3.1 \ }\emph{For a} $c\in \mathbb{R}$ \emph{%
there is a constant} $C>0$ \emph{such that}%
\begin{equation}
E\exp (c|\bar{f}_{k}^{i}|)<C  \label{As31}
\end{equation}%
\emph{for all} $i=0,\ldots N,\ k=0,\ldots ,M.$ \medskip

This condition is satisfied by all sensible numerical schemes (i.e.,
sensible choices of $A^{i}$ in (\ref{2.7})) thanks to the uniform
boundedness of $\sigma _{i}(t,T,z)$ (see Assumption~2.1) and boundedness of
the initial condition (see Assumption~2.3 and also the comment after (\ref%
{expmom2})). In particular, it is satisfied by the weak Euler-type scheme (%
\ref{euler}) we use in the algorithms in Section~\ref{sec:alg}.

We also require that the numerical method (\ref{2.7}) for the SDEs (\ref{2.6}%
)-(\ref{2.66}) is of local weak order $q+1,$ i.e., that the following
assumption holds.\medskip

\noindent \textbf{Assumption 3.2. \ }\emph{We assume that the method} (\ref%
{2.7}) \emph{is such that for some} \emph{positive constant} $C$ \emph{%
independent of} $\Delta $ 
\begin{gather}
|E(\prod_{j=1}^{s}\delta \tilde{f}^{i_{j}}-\prod_{j=1}^{s}\delta \bar{f}%
^{i_{j}})|\leq Ch^{q+1},\;s=1,\ldots ,2q+1,  \label{Db03} \\
E\prod_{j=1}^{2q+2}|\delta \bar{f}^{i_{j}}|\leq Ch^{q+1},  \label{Db04}
\end{gather}%
\emph{where} 
\begin{equation*}
\delta \tilde{f}^{i}:=\tilde{f}_{t,x}^{i}(t+h)-x^{i},\ \ \delta \bar{f}^{i}:=%
\bar{f}_{t,x}^{i}(t+h)-x^{i},
\end{equation*}%
\emph{and} $\tilde{f}_{t,x}^{i}(t+h)$ \emph{is the solution of the SDEs} (%
\ref{2.6})\ \emph{with the initial condition }$x$ \emph{given at time }$t:$ $%
\tilde{f}_{t,x}^{i}(t)=x^{i},$ \emph{and} $\bar{f}_{t,x}^{i}(t+h)$ \emph{is
the one-step approximation} \emph{of }(\ref{2.6}) \emph{found according to} (%
\ref{2.7}) \emph{with} $\bar{f}_{t,x}^{i}(t)=x^{i}.$\medskip

Assumption~3.2 is similar to the one used in the standard theory of
numerical integration of SDEs in the weak sense (see, e.g. \cite{MT1}). As
we will see in Section~\ref{sec:err}, Assumptions~2.1-2.3 and~3.1-3.2
guarantee weak convergence of the numerical method (\ref{2.7}) to the
solution of the auxiliary system of SDEs (\ref{2.6}) with order $h^{q}.$

We note that $C$ in (\ref{Db03})-(\ref{Db04}) is independent of $x$ while in
the standard theory of numerical integration of SDEs one usually has $C$
depending on $x$ in such estimates (see \cite[p. 100]{MT1}). In our case it
is natural to put $C$ independent of $x$ since the coefficients of (\ref{2.6}%
) and their derivatives are uniformly bounded (see Assumptions~2.1-2.2). We
also emphasize that the constants $C$ in (\ref{Db03})-(\ref{Db04}) are
required not to depend on $\Delta $.

\begin{remark}
\label{rem:fict1}The numerical method $(\ref{2.7})$ contains the
approximation $\bar{f}_{k}^{\ell _{k}}$of the forward rate $f(t_{k},T_{\ell
_{k}})$ $($recall that $t_{k}\geq T_{\ell _{k}})$ which from the financial
point of view does not exist unless $t_{k}=T_{\ell _{k}}$. However, from
both theoretical and numerical points of view, it is not prohibiting to
consider the values $\bar{f}_{k}^{_{\ell _{k}}}$ which, as we will see later
in Section~\ref{sec:appyz}, is computationally beneficial. We may interpret
the points $(t_{k},T_{\ell _{k}})$ on our $(t,T)$-grid as fictitious nodes
(see also Remark~\ref{rem:fict2}).
\end{remark}

The approximation (\ref{2.7}) of the infinite-dimension stochastic equation (%
\ref{3})-(\ref{3b}) has two discretization steps: $T$-step $\Delta $ (i.e.,
step in maturity time) and $t$-step $h$ (i.e., step in time). We can say
that the $T$-step $\Delta $ controls the error of approximating (\ref{3})-(%
\ref{3b}) by (\ref{2.6})-(\ref{2.66}) while the $t$-step $h$ controls the
error of approximating (\ref{2.6})-(\ref{2.66}) by (\ref{2.7}). We will
later (see Remark~\ref{rem:reldh}) discuss how to choose $\Delta $ and $h$
in practice.

\subsection{Approximation of the price of an interest rate contract\label%
{sec:appyz}}

In the previous section we introduced an approximation $\bar{f}_{k}^{i}$ of
the solution to (\ref{3})-(\ref{3b}). Our aim is to evaluate the expectation
(\ref{7})-(\ref{9}), i.e., 
\begin{equation*}
F(t_{0},f_{0}\left( \cdot \right) ;s_{k},s_{i})=E\exp (-Y(s_{k}))G\left(
P(s_{k},s_{i})\right) .
\end{equation*}%
To evaluate $F$, one has to compute $Z(s_{k},s_{i})$ from (\ref{8}) and $%
Y(s_{k})$ from (\ref{9}). In this section we construct numerical
approximations for $Z(s_{k},s_{i})$ and $Y(s_{k}).$ For clarity of the
exposition, we assume in what follows that 
\begin{equation*}
s_{k}=t^{\ast }\text{\ \ \ and \ \ }s_{i}=T^{\ast }\ .
\end{equation*}

We approximate the maturity time integral from (\ref{9}) by a quadrature
rule $S_{Z}(t^{\ast },T^{\ast },\Delta ):$ 
\begin{equation}
Z(t^{\ast },T^{\ast })=\int_{t^{\ast }}^{T^{\ast }}f(t^{\ast },u)du\approx
S_{Z}(t^{\ast },T^{\ast },\Delta )=\Delta \dsum_{j=\varrho _{M}}^{N}\tilde{%
\gamma}_{j}f(t^{\ast },T_{j}),  \label{2.8}
\end{equation}%
where the weights $\tilde{\gamma}_{j}$ are chosen so that the quadrature
rule is of order $p>0,$ i.e., an inequality of the form (\ref{2.5}) holds: 
\begin{equation}
\left( E\left[ S_{Z}(t^{\ast },T^{\ast },\Delta )-Z(t^{\ast },T^{\ast })%
\right] ^{2}\right) ^{1/2}\leq C\Delta ^{p}.  \label{2.9}
\end{equation}%
The assumption we made at the beginning of Section~\ref{sec:Tdis} that there
is a sufficient number of nodes $T_{i}$ between $t^{\ast }$ and $T^{\ast }$
ensures that we can find a quadrature rule (\ref{2.8}) satisfying (\ref{2.9}%
). Some examples of quadratures $S_{Z}(t^{\ast },T^{\ast },\Delta )$ are
given in Section~\ref{sec:alg}.

In general, $T$-discretization and $t$-discretization have different steps $%
\Delta $ and $h,$ and approximate values of the short rate $%
r(t_{k})=f(t_{k},t_{k})$ (cf. (\ref{2})) are not directly available among $%
\bar{f}_{k}^{i}$ which are defined on the $(t,T)$-grid. Then to numerically
evaluate $Y(t^{\ast }),$ we need to construct an approximation of $%
f(t_{k},t_{k})$ based on the values $\bar{f}_{k}^{i},$ $i=\ell _{k},\ldots
,N $. To this end, let us first consider an approximation of the exact short
rate $r(t)=f(t,t)$ using the values of $f(t,T_{i}),$\ $i=\ell (t),\ldots ,N$%
. We recall that thanks to Assumptions~2.2-2.3 the solution $f(t,T)$ of (\ref%
{3})-(\ref{3b}) is sufficiently smooth in the last argument. We approximate $%
r(t)$ by $\pi (t)$ as 
\begin{eqnarray}
\pi (t) &=&\pi (t;f(t,T_{i}),\ i=\ell (t),\ldots ,\ell (t)+\theta )
\label{2.10} \\
&=&\dsum\limits_{l=0}^{\ell (t^{\ast })}\dsum\limits_{i=0}^{\theta }\lambda
_{i}(t)f(t,T_{l+i})\chi _{t\in \lbrack T_{l},T_{l+1})}  \notag \\
&=&\dsum\limits_{i=0}^{\theta }\lambda _{i}(t)f(t,T_{\ell (t)+i}),\ \ t\in
\lbrack t_{0},t^{\ast }],  \notag
\end{eqnarray}%
where $\lambda _{i}(t)$ are coefficients independent of $f,$ $|\lambda
_{i}(t)|$ are bounded by a constant independent of $\Delta $, $\theta $ is a
non-negative integer independent of $t$ and $\Delta ,$ and $\chi _{A}$ is
the indicator function of a set $A.$ We choose the number $\theta $ and the
coefficients $\lambda _{i}(t)$ so that the approximation (\ref{2.10}) is of
order $p:$ 
\begin{equation}
\left( E\left[ r(t)-\pi (t)\right] ^{2}\right) ^{1/2}\leq C\Delta ^{p},\text{
}p>0.  \label{2.11}
\end{equation}%
The form of (\ref{2.10}) covers both polynomial interpolation and
extrapolation. For interpolation, we approximate $r(t)$ using the values $%
f(t,T_{i}),$\ $i=\ell (t),\ldots ,\ell (t)+\theta .$ For extrapolation, the
coefficient $\lambda _{0}(t)=0$ and we approximate $r(t)$ using the values $%
f(t,T_{i}),$\ $i=\varrho (t),\ldots ,\varrho (t)+\theta -1.$ Some particular
examples of the approximation $\pi (t)$ are given in Section~\ref{sec:alg}.
Recall that in Section~\ref{sec:Tdis} we assumed that $\Delta $ is such that
there is a sufficient number of nodes $T_{i}$ between $t^{\ast }$ and $%
T^{\ast }$ which should, in particular, ensure that $\ell (t^{\ast })+\theta
\leq N.$

\begin{remark}
\label{rem:fict2}We note that we need fictitious points $(t_{k},T_{\ell
_{k}})$ on our $(t,T)$-grid (see also Remark~\ref{rem:fict1}) for the
interpolating form of $(\ref{2.10})$. The extrapolating form of $(\ref{2.10}%
) $ does not need the fictitious points as it is sufficient to compute $\bar{%
f}_{k}^{i}$ for $i=\varrho _{k},\ldots ,\varrho _{k}+\theta -1,$ \ $%
k=0,\ldots ,M,$ all of which have the usual financial meaning. However, we
reserve the possibility to use an interpolation (and, consequently, the
fictitious points) for simulating short rates since interpolation is usually
computationally preferable to extrapolation.
\end{remark}

Using the short rate approximation $\pi (s),$ we approximate the time
integral in (\ref{8}) as 
\begin{equation}
Y(t^{\ast })=\int_{t_{0}}^{t^{\ast }}r(s)ds\approx \int_{t_{0}}^{t^{\ast
}}\pi (s)ds\approx \tilde{Y}(t^{\ast }):=\int_{t_{0}}^{t^{\ast }}\tilde{\pi}%
(s)ds,  \label{2.12}
\end{equation}%
where $\tilde{\pi}(s)$ has the form of $\pi (s)$ from (\ref{2.10}) but with $%
\tilde{f}^{i}(t)$ instead of $f(t,T_{i}):$%
\begin{equation*}
\tilde{\pi}(s)=\pi (s;\tilde{f}^{i}(s),\ i=\ell (s),\ldots ,\ell (s)+\theta
).
\end{equation*}%
We extend the system of SDEs (\ref{2.6}) by adding to it the auxiliary
differential equation 
\begin{equation}
d\tilde{Y}=\pi (s;\tilde{f}^{i}(s),\ i=\ell (s),\ldots ,\ell (s)+\theta
)ds,\ \ \tilde{Y}(t_{0})=0.  \label{t5}
\end{equation}%
Recall that $\tilde{\pi}(s)$ for every $s\in \lbrack t_{0},t^{\ast }]$ is a
linear combination of $\tilde{f}^{i}(s),$ $i=\ell (s),\ldots ,\ell
(s)+\theta .$

Let $\tilde{Y}_{t,x,y}(s),$ $s\geq t,$ be the solution of (\ref{t5}) with
the initial condition $\tilde{Y}_{t,x,y}(t)=y$ and with $\tilde{f}^{i}(s)=%
\tilde{f}_{t,x}^{i}(s)$ (recall that $\tilde{f}_{t,x}^{i}(s)$ are defined in
Assumption~3.2). We observe that under $h\leq \alpha \Delta $ for some $%
\alpha >0:$ 
\begin{equation}
\tilde{Y}_{t,x,0}(t+h)=\dsum\limits_{l=\ell (t)}^{\ell
(t+h)}\int_{t}^{t+h}\dsum\limits_{i=0}^{\theta }\lambda _{i}(s)\tilde{f}%
_{t,x}^{l+i}(s)\chi _{s\in \lbrack T_{l},T_{l+1})}ds,  \label{nYY}
\end{equation}%
and for any positive integer $m$ 
\begin{equation}
E\left\vert \tilde{Y}_{t,x,0}(t+h)\right\vert ^{m}\leq Ch^{m}\left(
1+\dsum\limits_{l=\ell (t)}^{\ell (t+h)+\theta }|x^{l}|^{m}\right) ,
\label{estY}
\end{equation}%
where $C>0$ is a constant independent of $\Delta $ and $x.$ We note that the
condition $h\leq \alpha \Delta $ guarantees that the number $\ell (t+h)-\ell
(t)$ is independent of $\Delta ,$ which ensures that the constant $C$ in (%
\ref{estY}) is independent of $\Delta $ and the number of terms in the sum
on the right-hand side of (\ref{estY}) is also independent of $\Delta .$
This will be essential for proving convergence Theorem~\ref{prpt1}.

Now we extend the fully discrete approximation (\ref{2.7}) by adding to it
an approximation of (\ref{t5}):%
\begin{equation}
\bar{Y}_{0}=0,\text{ }\bar{Y}_{k+1}=\bar{Y}_{k}+A^{Y}(t_{k};\bar{f}_{k}^{j},%
\text{ }j=\ell _{k},\ldots ,\ell _{k+1}+\theta ;h),\text{ }k=0,\ldots ,M,
\label{t51}
\end{equation}%
where the form of $A^{Y}(t_{k};\bar{f}_{k}^{j},$ $j=\ell _{k},\ldots ,\ell
_{k+1}+\theta ;h)=A^{Y}(t_{k};h)$ depends on the form of $\pi (s)$ from (\ref%
{2.10}) and the accuracy required.

We replace Assumption~3.2 on the one-step approximation by the assumption
which is applicable to the extended system (\ref{2.6}), (\ref{t5}) and the
extended discretization (\ref{2.7}), (\ref{t51}). \medskip

\noindent \textbf{Assumption 3.2'}. $\ $\emph{Let }$h\leq \alpha \Delta $ 
\emph{for some} $\alpha >0.$ \emph{We assume that the method} (\ref{2.7}), (%
\ref{t51}) \emph{is such that for some} \emph{positive constant} $C$ \emph{%
independent of} $\Delta $%
\begin{gather}
\left\vert E\left( \delta \tilde{Y}^{m}\prod_{j=1}^{s-m}\delta \tilde{f}%
^{i_{j}}-\delta \bar{Y}^{m}\prod_{j=1}^{s-m}\delta \bar{f}^{i_{j}}\right)
\right\vert \leq Ch^{q+1}\left( 1+\dsum\limits_{l=\ell (t)}^{\ell
(t+h)+\theta }|x^{l}|^{m}\right) ,\;  \label{as321} \\
m=0,\ldots ,s,\ s=1,\ldots ,2q+1;  \notag \\
\left[ E\max_{0\leq m\leq 2q+2,\{i_{1},\ldots i_{2q+2-m}\}\in \{0,\ldots
,N\}}\left\vert \delta \bar{Y}^{m}\prod_{j=1}^{2q+2-m}\delta \bar{f}%
^{i_{j}}\right\vert ^{2}\right] ^{1/2}\leq Ch^{q+1}\left(
1+\dsum\limits_{l=\ell (t)}^{\ell (t+h)+\theta }|x^{l}|^{2q+2}\right) ,
\label{as322}
\end{gather}%
\emph{where} 
\begin{eqnarray*}
\delta \tilde{f}^{i} &=&\tilde{f}_{t,x}^{i}(t+h)-x^{i},\ \ \delta \bar{f}%
^{i}=\bar{f}_{t,x}^{i}(t+h)-x^{i},\ \  \\
\delta \tilde{Y} &=&\tilde{Y}_{t,x,y}(t+h)-y,\ \ \delta \bar{Y}=\bar{Y}%
_{t,x,y}(t+h)-y,
\end{eqnarray*}%
$\tilde{f}_{t,x}^{i}(t+h)$ \emph{and} $\bar{f}_{t,x}^{i}(t+h)$ \emph{are as
in Assumption~3.2 and} $\tilde{Y}_{t,x,y}(s),$ $s\geq t,$ \emph{is the
solution of} (\ref{t5}) \emph{with the initial condition} $\tilde{Y}%
_{t,x,y}(t)=y,$ \emph{and} $\bar{Y}_{t,x,y}(t+h)$ \emph{is its one-step
approximation} \emph{found\emph{\ }according to} (\ref{t51}) \emph{with} $%
\bar{Y}_{t,x,y}(t)=y.$\medskip

Note that the constants $C$ in (\ref{as321})-(\ref{as322}) do not depend on $%
x,$ $y,$ and $\Delta .$ The dependence of the estimates (\ref{as321})-(\ref%
{as322}) on $x$ is consistent with (\ref{estY}). The condition $h\leq \alpha
\Delta $ in Assumption~3.2' is not restrictive from the practical point of
view since we aim to be constructing efficient numerical algorithms for the
HJM model by allowing bigger $T$-steps $\Delta $ without losing accuracy. We
also note that this condition arises only when we need to approximate the
short rate (see also Remark~\ref{rem:hlessdelt}).

Further, we make the following assumption. \medskip

\noindent \textbf{Assumption 3.3}. \emph{For some }$c>0$\emph{\ and} $C>0$%
\begin{equation}
E\exp (c|\bar{Y}_{k}|)<C  \label{As311}
\end{equation}%
\emph{for all} $k=0,\ldots ,M.$\medskip

As a rule, the condition (\ref{As311}) immediately follows from
Assumption~3.1 which is the case, e.g., for the algorithms presented in
Section~\ref{sec:alg}.

Based on (\ref{2.8}), (\ref{2.12}) and using (\ref{2.7}), (\ref{t51}), we
arrive at the approximation $\bar{F}$ of $F$ from (\ref{7}): 
\begin{equation}
F(t_{0},f_{0}\left( \cdot \right) ;t^{\ast },T^{\ast })\approx \bar{F}%
(t_{0},f_{0};t^{\ast },T^{\ast })=E\exp (-\bar{Y}_{M})G\left( \bar{P}%
(t^{\ast },T^{\ast })\right) ,  \label{2.13}
\end{equation}%
where $\bar{Y}_{M}$ is from (\ref{t51});%
\begin{equation}
\bar{P}(t^{\ast },T^{\ast })=\exp \left( -\bar{S}_{Z}(t^{\ast },T^{\ast
},\Delta )\right) ,  \label{2.15}
\end{equation}%
$\bar{S}_{Z}(t^{\ast },T^{\ast },\Delta )$ is the quadrature rule of the
form (\ref{2.8}) with $f(t^{\ast },T_{j})$ replaced by $\bar{f}_{M}^{j}:$ 
\begin{equation}
\bar{S}_{Z}(t^{\ast },T^{\ast },\Delta )=\Delta \dsum_{j=\varrho _{M}}^{N}%
\tilde{\gamma}_{j}\ \bar{f}_{M}^{j};  \label{2.17}
\end{equation}%
and $f_{0}$ means the initial condition of (\ref{2.6}), which is the $N+1$%
-dimensional vector $(f_{0}^{0},\ldots ,f_{0}^{N})^{^{\top
}}=(f_{0}(T_{0}),\ldots ,f_{0}(T_{N}))^{^{\top }}.$

Finally, the expectation of the discounted payoff in (\ref{2.13}) is
approximated by the Monte Carlo method, i.e.,%
\begin{eqnarray}
F(t_{0},f_{0}\left( \cdot \right) ;t^{\ast },T^{\ast }) &\approx &\bar{F}%
(t_{0},f_{0};t^{\ast },T^{\ast })  \label{2.20} \\
&\approx &\hat{F}(t_{0},f_{0};t^{\ast },T^{\ast })=\frac{1}{L}%
\dsum\limits_{l=1}^{L}\exp (-\bar{Y}_{M}^{(l)})G\left( \bar{P}^{(l)}(t^{\ast
},T^{\ast })\right) ,  \notag
\end{eqnarray}%
where $\bar{Y}_{M}^{(l)},$ $\bar{P}^{(l)}$ are computed using independent
realizations $\bar{f}_{k}^{j,(l)},$ $j=\ell _{k},\ldots ,N,$ $k=1,\ldots ,M,$
of the random variables $\bar{f}_{k}^{j}.$

In (\ref{2.20}) the first approximate equality corresponds to the error of
numerical integration and the error in the second approximate equality comes
from the Monte Carlo technique. The numerical integration error is analyzed
in the next section. The Monte Carlo (i.e., statistical) error in (\ref{2.20}%
) is evaluated by%
\begin{eqnarray}
\bar{\rho}_{MC} &=&c\,\frac{\left[ Var\left\{ \exp (-\bar{Y}_{M})G\left( 
\bar{P}(t^{\ast },T^{\ast })\right) \right\} \right] ^{1/2}}{L^{1/2}}
\label{errMC} \\
&\approx &c\,\frac{\left[ Var\left\{ \exp (-Y(t^{\ast }))G\left( P(t^{\ast
},T^{\ast })\right) \right\} \right] ^{1/2}}{L^{1/2}},  \notag
\end{eqnarray}%
where, for example, the values $c=1,2,3$ correspond to the fiducial
probabilities $0.68,$\ $0.95,$\ $0.997,$ respectively. The Monte Carlo error
can be decreased by variance reduction techniques (see, e.g. \cite%
{Glasserman,GlassermanHJM,MT1,vari} and references therein). In this paper
we deal with the numerical integration error and numerical algorithms which
are effective with regard to $(t,T)$-discretization.

\section{Convergence theorems\label{sec:err}}

The aim of this section is to prove the convergence of the approximation $%
\bar{F}(t_{0},f_{0};t^{\ast },T^{\ast })$ to $F(t_{0},f_{0}\left( \cdot
\right) ;t^{\ast },T^{\ast })$ as $h\rightarrow 0$ and $\Delta \rightarrow
0. $

Denote by $\tilde{F}(t_{0},f_{0};t^{\ast },T^{\ast })$ the approximation of $%
F(t_{0},f_{0}\left( \cdot \right) ;t^{\ast },T^{\ast })$ from (\ref{7})
resulting from approximating the solution $f(t,T_{i})$ of (\ref{3})-(\ref{3b}%
) by $\tilde{f}^{i}(t)$ from (\ref{2.6}), i.e.,%
\begin{equation}
\tilde{F}(t_{0},f_{0};t^{\ast },T^{\ast })=E\exp (-\tilde{Y}(t^{\ast
}))G\left( \tilde{P}(t^{\ast },T^{\ast })\right) ,  \label{3.1}
\end{equation}%
where 
\begin{equation}
\tilde{P}(t^{\ast },T^{\ast })=\exp \left( -\tilde{S}_{Z}(t^{\ast },T^{\ast
},\Delta )\right) ,  \label{3.1b}
\end{equation}%
$\tilde{Y}(t)$ is from (\ref{t5}) and $\tilde{S}_{Z}(t^{\ast },T^{\ast
},\Delta )$ is the quadrature rule of the form (\ref{2.8}) with $f(t^{\ast
},T_{j})$ replaced by $\tilde{f}^{j}(t^{\ast }).$

The error $R$ of weak approximation of $F$ by $\bar{F}$ can be written as a
sum of two contributing terms: 
\begin{eqnarray}
R &=&F(t_{0},f_{0}\left( \cdot \right) ;t^{\ast },T^{\ast })-\bar{F}%
(t_{0},f_{0};t^{\ast },T^{\ast })  \label{3.2} \\
&=&\left[ F(t_{0},f_{0}\left( \cdot \right) ;t^{\ast },T^{\ast })-\tilde{F}%
(t_{0},f_{0};t^{\ast },T^{\ast })\right] +\left[ \tilde{F}%
(t_{0},f_{0};t^{\ast },T^{\ast })-\bar{F}(t_{0},f_{0};t^{\ast },T^{\ast })%
\right]  \notag \\
&:&=R_{1}+R_{2},  \notag
\end{eqnarray}%
where $R_{1}$ is the error due to $T$-discretization of (\ref{3})-(\ref{3b})
and $R_{2}$ is the error due to $t$-discretization of (\ref{2.6})-(\ref{2.66}%
). The first error, $R_{1},$ is analyzed in Section~\ref{sec:errMd} and the
second error, $R_{2},$ is analyzed in Section~\ref{sec:errtd}.

Note that in this section we shall use the letters $K,$ $C,$ and $c$ to
denote various constants which are independent of $\Delta \ $and $h.\medskip 
$

\subsection{$T$-discretization error\label{sec:errMd}}

In this section we analyze the error of the finite-dimensional approximation
(\ref{2.6})-(\ref{2.66}) for the infinite-dimensional stochastic equation (%
\ref{3})-(\ref{3b}). The plan of this section is as follows. First, we prove
(see Theorem~\ref{thm:msqi2f}) that the approximation (\ref{2.6})-(\ref{2.66}%
) has mean-square convergence of order $\Delta ^{p}.$ This result plays an
intermediate role for getting an estimate for the $T$-discretization error $%
R_{1}$ but, at the same time, it has its own theoretical value. Based on
Theorem~\ref{thm:msqi2f}, we prove (see Lemma~\ref{lem:msqY}) the
mean-square convergence of $\tilde{Y}$\ from (\ref{2.12}) to $Y$ from (\ref%
{8}). Finally, in Theorem~\ref{thm:weai2f} we prove that the weak-sense
error $R_{1}$ (see (\ref{3.2})) of (\ref{2.6})-(\ref{2.66}) is of order $%
\Delta ^{p}.$

\begin{theorem}
\label{thm:msqi2f}Suppose Assumptions~2.1-2.3 are satisfied. Then the
approximation $\tilde{f}^{i}(t)$ from $(\ref{2.6})$-$(\ref{2.66})$ converges
to $f(t,T_{i})$ from $(\ref{3})$-$(\ref{3b})$ as $\Delta \rightarrow 0$ with
the mean-square order $p$, i.e.,%
\begin{equation}
\left( E\left\vert \tilde{f}^{i}(t)-f(t,T_{i})\right\vert ^{2}\right)
^{1/2}\leq K\Delta ^{p},\ t\in \lbrack t_{0},t^{\ast }\wedge T_{(i+1)\wedge
N}],\text{ }i=0,\ldots ,N,  \label{3.3}
\end{equation}%
where $K>0$ is a constant independent of $\Delta ,$ $t,$ and $i.$
\end{theorem}

\noindent \textbf{Proof.} Denote by $\rho (t,T_{i})$ the error of the
approximation (\ref{2.6})-(\ref{2.66}):%
\begin{equation}
\rho (t,T_{i}):=\tilde{f}^{i}(t)-f(t,T_{i}),\text{ }t\in \lbrack
t_{0},t^{\ast }\wedge T_{(i+1)\wedge N}],\text{ }i=0,\ldots ,N.  \label{3.4}
\end{equation}%
Clearly,%
\begin{equation}
\rho (t_{0},T_{i})=0.  \label{3.45}
\end{equation}

Due to Assumption~2.2, $\sigma (s,T,z)$ is \ globally Lipschitz in $z$ whence%
\begin{equation}
\left\vert \tilde{\sigma}_{i}(s)-\sigma (s,T_{i})\right\vert =\left\vert
\sigma (s,T_{i},\tilde{f}^{i}(s))-\sigma (s,T_{i},f(s,T_{i}))\right\vert
\leq K\left\vert \rho (s,T_{i})\right\vert ,  \label{3.52}
\end{equation}%
and (cf. (\ref{2.444}) and (\ref{2.66}))%
\begin{eqnarray}
\left\vert \tilde{S}_{I}(s,T_{i},\Delta )-S_{I}(s,T_{i},\Delta )\right\vert
&=&\Delta \left\vert \dsum_{k=\ell (s)}^{\kappa (s,T_{i})}\gamma
_{k}(s)\left( \tilde{\sigma}_{k}(s)-\sigma (s,T_{k})\right) \right\vert
\label{3.53} \\
&\leq &K\Delta \dsum_{k=\ell (s)}^{\kappa (s,T_{i})}\left\vert \tilde{\sigma}%
_{k}(s)-\sigma (s,T_{k})\right\vert \leq K\Delta \dsum_{k=\ell (s)}^{\kappa
(s,T_{i})}\left\vert \rho (s,T_{k})\right\vert .  \notag
\end{eqnarray}

We have from (\ref{2.6})-(\ref{2.66}) and (\ref{3})-(\ref{3b}): 
\begin{eqnarray*}
\rho (t,T_{i}) &=&\int_{t_{0}}^{t}\left[ \tilde{\sigma}_{i}^{\top }(s)\tilde{%
S}_{I}(s,T_{i},\Delta )-\sigma ^{\top }(s,T_{i})I(s,T_{i})\right]
ds+\int_{t_{0}}^{t}\left[ \tilde{\sigma}_{i}(s)-\sigma (s,T_{i})\right]
^{\top }dW(s) \\
&=&\int_{t_{0}}^{t}\tilde{\sigma}_{i}^{\top }(s)\left[ \tilde{S}%
_{I}(s,T_{i},\Delta )-S_{I}(s,T_{i},\Delta )\right] ds \\
&&+\int_{t_{0}}^{t}\tilde{\sigma}_{i}^{\top }(s)\left[ S_{I}(s,T_{i},\Delta
)-I(s,T_{i})\right] ds \\
&&+\int_{t_{0}}^{t}\left[ \tilde{\sigma}_{i}(s)-\sigma (s,T_{i})\right]
^{\top }I(s,T_{i})ds+\int_{t_{0}}^{t}\left[ \tilde{\sigma}_{i}(s)-\sigma
(s,T_{i})\right] ^{\top }dW(s).
\end{eqnarray*}%
By Ito's formula, we obtain%
\begin{eqnarray}
\rho ^{2}(t,T_{i}) &=&\int_{t_{0}}^{t}2\rho (s,T_{i})\tilde{\sigma}%
_{i}^{\top }(s)\left[ \tilde{S}_{I}(s,T_{i},\Delta )-S_{I}(s,T_{i},\Delta )%
\right] ds  \label{3.5} \\
&&+\int_{t_{0}}^{t}2\rho (s,T_{i})\tilde{\sigma}_{i}(s)\left[
S_{I}(s,T_{i},\Delta )-I(s,T_{i})\right] ds  \notag \\
&&+\int_{t_{0}}^{t}2\rho (s,T_{i})\left[ \tilde{\sigma}_{i}(s)-\sigma
(s,T_{i})\right] ^{\top }I(s,T_{i})ds  \notag \\
&&+\int_{t_{0}}^{t}\left[ \tilde{\sigma}_{i}(s)-\sigma (s,T_{i})\right]
^{\top }\left[ \tilde{\sigma}_{i}(s)-\sigma (s,T_{i})\right] ds  \notag \\
&&+\int_{t_{0}}^{t}2\rho (s,T_{i})\left[ \tilde{\sigma}_{i}(s)-\sigma
(s,T_{i})\right] dW(s).  \notag
\end{eqnarray}%
Then 
\begin{eqnarray}
E\rho ^{2}(t,T_{i}) &=&2\int_{t_{0}}^{t}E\rho (s,T_{i})\tilde{\sigma}%
_{i}^{\top }(s)\left[ \tilde{S}_{I}(s,T_{i},\Delta )-S_{I}(s,T_{i},\Delta )%
\right] ds  \label{3.51} \\
&&+2\int_{t_{0}}^{t}E\rho (s,T_{i})\tilde{\sigma}_{i}^{\top }(s)\left[
S_{I}(s,T_{i},\Delta )-I(s,T_{i})\right] ds  \notag \\
&&+2\int_{t_{0}}^{t}E\rho (s,T_{i})\left[ \tilde{\sigma}_{i}(s)-\sigma
(s,T_{i})\right] ^{\top }I(s,T_{i})ds  \notag \\
&&+\int_{t_{0}}^{t}E\left[ \tilde{\sigma}_{i}(s)-\sigma (s,T_{i})\right]
^{\top }\left[ \tilde{\sigma}_{i}(s)-\sigma (s,T_{i})\right] ds.  \notag
\end{eqnarray}%
Using the boundedness of $\sigma (s,T,z)$ (see (\ref{4b})) and the
inequality (\ref{3.53}), the first term on the right-hand side of (\ref{3.51}%
) is estimated as 
\begin{eqnarray}
&&\left\vert 2\int_{t_{0}}^{t}E\rho (s,T_{i})\tilde{\sigma}_{i}^{\top }(s)%
\left[ \tilde{S}_{I}(s,T_{i},\Delta )-S_{I}(s,T_{i},\Delta )\right]
ds\right\vert  \label{3.55} \\
&\leq &K\Delta \left[ \int_{t_{0}}^{t}E\left\vert \rho (s,T_{i})\right\vert
\dsum_{k=\ell (s)}^{\kappa (s,T_{i})}\left\vert \rho (s,T_{k})\right\vert ds%
\right] .  \notag
\end{eqnarray}%
Using the boundedness of $\sigma (s,T,z),$ the inequality $2ab\leq
a^{2}+b^{2},$ and the condition (\ref{2.5}) for the quadrature rule $S_{I}$,
we obtain for the second term on right hand side of (\ref{3.51}):%
\begin{eqnarray}
&&\left\vert 2\int_{t_{0}}^{t}E\rho (s,T_{i})\tilde{\sigma}_{i}^{\top }(s)%
\left[ S_{I}(s,T_{i},\Delta )-I(s,T_{i})\right] ds\right\vert  \label{3.56}
\\
&\leq &K\int_{t_{0}}^{t}E|\rho (s,T_{i})|\left\vert S_{I}(s,T_{i},\Delta
)-I(s,T_{i})\right\vert ds  \notag \\
&\leq &K\int_{t_{0}}^{t}\left[ E\rho ^{2}(s,T_{i})+E\left\vert
S_{I}(s,T_{i},\Delta )-I(s,T_{i})\right\vert ^{2}\right] ds  \notag \\
&\leq &K\int_{t_{0}}^{t}\left[ E\rho ^{2}(s,T_{i})+\Delta ^{2p}\right] ds. 
\notag
\end{eqnarray}%
Using the inequality (\ref{3.52}) and the boundedness of $\sigma (s,T,z)$,
we get for the third term on the right-hand side of (\ref{3.51}):%
\begin{equation}
\left\vert 2\int_{t_{0}}^{t}E\rho (s,T_{i})\left[ \tilde{\sigma}%
_{i}(s)-\sigma (s,T_{i})\right] ^{\top }I(s,T_{i})ds\right\vert \leq
K\int_{t_{0}}^{t}E\rho ^{2}(s,T_{i})ds.  \label{3.57}
\end{equation}%
By the inequality (\ref{3.52}), the fourth term on the right-hand of (\ref%
{3.51}) is estimated as%
\begin{equation}
\int_{t_{0}}^{t}E\left[ \tilde{\sigma}_{i}(s)-\sigma (s,T_{i})\right] ^{\top
}\left[ \tilde{\sigma}_{i}(s)-\sigma (t,T_{i})\right] ds\leq
K\int_{t_{0}}^{t}E\rho ^{2}(s,T_{i})ds.  \label{3.6}
\end{equation}%
Substituting (\ref{3.55})-(\ref{3.6}) in (\ref{3.51}) and using the
inequality $2ab\leq a^{2}+b^{2}$, we obtain 
\begin{eqnarray}
E\rho ^{2}(t,T_{i}) &\leq &K\int_{t_{0}}^{t}\left\{ E\rho
^{2}(s,T_{i})+\Delta E\left[ \left\vert \rho (s,T_{i})\right\vert
\dsum_{k=\ell (s)}^{\kappa (s,T_{i})}\left\vert \rho (s,T_{k})\right\vert %
\right] +\Delta ^{2p}\right\} ds\ \ \ \ \   \label{3.66} \\
&\leq &K\int_{t_{0}}^{t}\left\{ \overset{\ }{\ }\left( 1+\Delta (\kappa
(s,T_{i})-\ell (s)+2)\right) E\rho ^{2}(s,T_{i})\right.  \notag \\
&&\left. +\Delta \dsum\limits_{k=\ell (s),\ k\neq i}^{\kappa (s,T_{i})}E\rho
^{2}(s,T_{k})+\Delta ^{2p}\right\} ds  \notag \\
&\leq &K\int_{t_{0}}^{t}\left\{ E\rho ^{2}(s,T_{i})+\Delta
\dsum\limits_{k=\ell (s),\ k\neq i}^{\kappa (s,T_{i})}E\rho
^{2}(s,T_{k})\right\} ds+K\Delta ^{2p},  \notag \\
t &\in &[t_{0},t^{\ast }\wedge T_{(i+1)\wedge N}],\text{\ \ }i=0,\ldots ,N. 
\notag
\end{eqnarray}%
We have used here that $\Delta (\kappa (s,T_{i})-\ell (s)+2)\leq T^{\ast
}-t_{0}.$

Introduce $\rho _{M}(t):=\max_{\ell (t)\leq i\leq N}E\rho ^{2}(t,T_{i}),$ $%
t\in \lbrack t_{0},t^{\ast }].$ Clearly (see (\ref{3.45})), $\rho
_{M}(t_{0})=0.$ Then we get from (\ref{3.66}): 
\begin{equation*}
\rho _{M}(t)\leq K\int_{t_{0}}^{t}\rho _{M}(s)ds+K\Delta ^{2p},
\end{equation*}%
whence (\ref{3.3}) follows by the Gronwall inequality. Theorem~\ref%
{thm:msqi2f} is proved. \ $\square \medskip $

Using Theorem~\ref{thm:msqi2f}, we prove the following lemma.

\begin{lemma}
\label{lem:msqY}Suppose Assumptions~2.1-2.3 are satisfied. The approximation 
$\tilde{Y}(t)$ from $(\ref{t5})$ converges to $Y(t)$ from $(\ref{8})$ as $%
\Delta \rightarrow 0$ with the mean-square order $p>0$, i.e., 
\begin{equation}
\left( E\left[ Y(t^{\ast })-\tilde{Y}(t^{\ast })\right] ^{2}\right)
^{1/2}\leq K\Delta ^{p},\text{ }  \label{3.12}
\end{equation}%
where $K>0$ is a constant independent of $\Delta .$
\end{lemma}

\noindent \textbf{Proof.} \ Consider the error of the approximation (\ref{t5}%
) for (\ref{8}) (see also (\ref{2.12})):%
\begin{equation}
Y(t^{\ast })-\tilde{Y}(t^{\ast })=\int_{t_{0}}^{t^{\ast
}}f(s,s)ds-\int_{t_{0}}^{t^{\ast }}\tilde{\pi}(s)ds.  \label{3.13}
\end{equation}%
We rearrange the right-hand side of (\ref{3.13}) to split this error into
the error due to approximation of the short rate $r(t)=f(t,t)$ by $\pi (t)$
and the error due to approximation of $f(t,T_{i})$ by $\tilde{f}^{i}(t):$%
\begin{equation}
Y(t^{\ast })-\tilde{Y}(t^{\ast })=\int_{t_{0}}^{t^{\ast }}\left( f(s,s)-\pi
(s)\right) ds+\int_{t_{0}}^{t^{\ast }}\left( \pi (s)-\tilde{\pi}(s)\right)
ds.  \label{3.131}
\end{equation}%
Due to the condition (\ref{2.11}) imposed on our choice of the approximation 
$\pi (s),$ we have 
\begin{equation}
E\left( \int_{t_{0}}^{t^{\ast }}\left( f(s,s)-\pi (s)\right) ds\right)
^{2}\leq K\Delta ^{2p}.  \label{3.132}
\end{equation}%
Recalling the form of the approximation $\pi (s)$ from (\ref{2.10}), we get 
\begin{eqnarray*}
&&E\left( \int_{t_{0}}^{t^{\ast }}\left( \pi (s)-\tilde{\pi}(s)\right)
ds\right) ^{2} \\
&=&E\left[ \int_{t_{0}}^{t^{\ast }}\sum_{l=0}^{\ell (t^{\ast
})}\dsum\limits_{i=0}^{\theta }\lambda _{i}(s)(f(s,T_{l+i})-\tilde{f}%
^{l+i}(s))\chi _{s\in \lbrack T_{l},T_{l+1})}ds\right] ^{2},
\end{eqnarray*}%
where $\lambda _{i}(s)$\ are bounded coefficients and the number $\theta $
is independent of $\Delta $. Then, using (\ref{3.3}), we obtain 
\begin{equation}
E\left( \int_{t_{0}}^{t^{\ast }}\left( \pi (s)-\tilde{\pi}(s)\right)
ds\right) ^{2}\leq K\Delta ^{2p}.  \label{3.133}
\end{equation}%
The relations (\ref{3.131})-(\ref{3.133}) imply the required error estimate (%
\ref{3.12}). \ $\square \medskip $

In the next theorem we obtain an estimate for the weak sense error $R_{1}$
from (\ref{3.2}).

\begin{theorem}
\label{thm:weai2f}Suppose Assumptions~2.1-2.3 are satisfied. Assume that the
payoff function $G(z)$ satisfies the global Lipschitz condition $(\ref{7a})$%
. Then the approximation $\tilde{F}(t_{0},f_{0};$ $t^{\ast },T^{\ast })$
from $(\ref{3.1})$ converges to $F(t_{0},f_{0}\left( \cdot \right) ;t^{\ast
},T^{\ast })$ from $(\ref{7})$, $(\ref{8})$-$(\ref{9})$ with order $p>0,$
i.e.,%
\begin{equation}
\left\vert F(t_{0},f_{0}\left( \cdot \right) ;t^{\ast },T^{\ast })-\tilde{F}%
(t_{0},f_{0};t^{\ast },T^{\ast })\right\vert \leq K\Delta ^{p},\text{ }
\label{3.14}
\end{equation}%
where $K>0$ is a constant independent of $\Delta .$
\end{theorem}

\noindent \textbf{Proof.} We have (see (\ref{7}), (\ref{8})-(\ref{9}) and (%
\ref{3.1})-(\ref{3.1b})): 
\begin{eqnarray}
&&R_{1}=F(t_{0},f_{0}\left( \cdot \right) ;t^{\ast },T^{\ast })-\tilde{F}%
(t_{0},f_{0};t^{\ast },T^{\ast })  \notag \\
&=&E\exp (-Y(t^{\ast }))G\left( P(t^{\ast },T^{\ast })\right) -E\exp (-%
\tilde{Y}(t^{\ast }))G\left( \tilde{P}(t^{\ast },T^{\ast })\right)  \notag \\
&=&E\left[ \exp \left( -Y(t^{\ast })\right) -\exp \left( -\tilde{Y}(t^{\ast
})\right) \right] G\left( \tilde{P}(t^{\ast },T^{\ast })\right)  \label{3.15}
\\
&&+E\left[ G\left( P(t^{\ast },T^{\ast })\right) -G\left( \tilde{P}(t^{\ast
},T^{\ast })\right) \right] \exp (-Y(t^{\ast })).  \notag
\end{eqnarray}

Consider the first term on the right-hand side of (\ref{3.15}). By the mean
value theorem, we get 
\begin{equation}
\exp \left( -Y(t^{\ast })\right) -\exp \left( -\tilde{Y}(t^{\ast })\right) =(%
\tilde{Y}(t^{\ast })-Y(t^{\ast }))\exp (\vartheta ),  \label{3.150}
\end{equation}%
where $\vartheta $ is a point between $-\tilde{Y}(t^{\ast })$ and $%
-Y(t^{\ast }).$ Due to the global Lipschitz condition (\ref{7a}) imposed on $%
G(z)$, we have (recall that $\tilde{S}_{Z}(t^{\ast },T^{\ast },\Delta )$ is
the quadrature rule of the form (\ref{2.8}) with $f(t^{\ast },T_{i})$
replaced by $\tilde{f}^{i}(t^{\ast }):$ 
\begin{eqnarray}
|G(\tilde{P}(t^{\ast },T^{\ast }))| &\leq &K\tilde{P}(t^{\ast },T^{\ast
})=K\exp \left( -\tilde{S}_{Z}(t^{\ast },T^{\ast },\Delta )\right)
\label{3.1500} \\
&=&K\exp \left( -\Delta \dsum_{j=\varrho _{M}}^{N}\tilde{\gamma}_{j}\ \tilde{%
f}^{j}(t^{\ast })\right) .  \notag
\end{eqnarray}%
Using (\ref{3.150}), (\ref{3.1500}), and the Cauchy--Bunyakovsky inequality
twice, we obtain 
\begin{eqnarray}
&&\left\vert E\left[ \exp \left( -Y(t^{\ast })\right) -\exp (-\tilde{Y}%
(t^{\ast }))\right] G(\tilde{P}(t^{\ast },T^{\ast }))\right\vert
\label{3.151} \\
&\leq &K\left\vert \left[ E(\tilde{Y}(t^{\ast })-Y(t^{\ast }))^{2}\right]
^{1/2}\left[ E\exp (4\vartheta )\right] ^{1/4}\left[ E\exp \left( -4\Delta
\dsum_{j=\varrho _{M}}^{N}\tilde{\gamma}_{j}\ \tilde{f}^{j}(t^{\ast
})\right) \right] ^{1/4}\right\vert .  \notag
\end{eqnarray}%
Thanks to (\ref{expmom}) and (\ref{expmom2}), exponential moments of $-%
\tilde{Y}(t^{\ast })$ and $-Y(t^{\ast })$ are bounded and, consequently, for
some $K>0$ we get $E\exp (4\vartheta )<K.$ Due to (\ref{expmom2}), we also
have 
\begin{eqnarray*}
E\exp \left( -4\Delta \dsum_{j=\varrho _{M}}^{N}\tilde{\gamma}_{j}\ \tilde{f}%
^{j}(t^{\ast })\right) &=&E\left[ \exp \left( -4T\dsum_{j=\varrho _{M}}^{N}%
\tilde{\gamma}_{j}\ \tilde{f}^{j}(t^{\ast })\right) \right] ^{1/N} \\
&\leq &\frac{1}{N}\dsum_{j=\varrho _{M}}^{N}E\exp \left( -4T\tilde{\gamma}%
_{j}\ \tilde{f}^{j}(t^{\ast })\right) <K.
\end{eqnarray*}%
Then (\ref{3.151}) together with (\ref{3.12}) implies 
\begin{equation}
|E\left[ \exp \left( -Y(t^{\ast })\right) -\exp (-\tilde{Y}(t^{\ast }))%
\right] G(\tilde{P}(t^{\ast },T^{\ast }))|\leq K\Delta ^{p}.  \label{3.152}
\end{equation}

Let us now consider the second term on the right-hand side of (\ref{3.15}).
Due to the global Lipschitz condition (\ref{7a}) imposed on $G(z)$, we have 
\begin{equation}
\left\vert G\left( P(t^{\ast },T^{\ast })\right) -G(\tilde{P}(t^{\ast
},T^{\ast }))\right\vert \leq K\cdot \left\vert P(t^{\ast },T^{\ast })-%
\tilde{P}(t^{\ast },T^{\ast })\right\vert .  \label{3.17}
\end{equation}%
Further, by the mean value theorem, we get 
\begin{eqnarray}
P(t^{\ast },T^{\ast })-\tilde{P}(t^{\ast },T^{\ast }) &=&\exp \left(
-Z\left( t^{\ast },T^{\ast }\right) \right) -\exp \left( -\tilde{S}%
_{Z}(t^{\ast },T^{\ast },\Delta )\right)  \notag \\
&=&\left( \tilde{S}_{Z}(t^{\ast },T^{\ast },\Delta )-Z\left( t^{\ast
},T^{\ast }\right) \right) \exp \left( \vartheta \right) ,  \label{3.18}
\end{eqnarray}%
where $\vartheta $ is between $-\tilde{S}_{Z}(t^{\ast },T^{\ast },\Delta )$
and $-Z\left( t^{\ast },T^{\ast }\right) $. Using (\ref{3.17}), (\ref{3.18}%
), and the Cauchy--Bunyakovsky inequality twice, we obtain 
\begin{eqnarray}
&&\left\vert E\left[ G\left( P(t^{\ast },T^{\ast })\right) -G(\tilde{P}%
(t^{\ast },T^{\ast }))\right] \exp (-Y(t^{\ast }))\right\vert  \label{3.181}
\\
&\leq &\left[ E\left( \tilde{S}_{Z}(t^{\ast },T^{\ast },\Delta )-Z\left(
t^{\ast },T^{\ast }\right) \right) ^{2}\right] ^{1/2}\left[ E\exp
(-4Y(t^{\ast }))\right] ^{1/4}\left[ E\exp (4\vartheta )\right] ^{1/4}. 
\notag
\end{eqnarray}%
It is clear that (\ref{expmom}) and (\ref{expmom2}) imply boundedness of the
exponential moments present in the right-hand side of (\ref{3.181}) and,
hence, 
\begin{eqnarray}
&&\left\vert E\left[ G\left( P(t^{\ast },T^{\ast })\right) -G\left( \tilde{P}%
(t^{\ast },T^{\ast })\right) \right] \exp (-Y(t^{\ast }))\right\vert
\label{3.nnn} \\
&\leq &K\left[ E\left( \tilde{S}_{Z}(t^{\ast },T^{\ast },\Delta )-Z\left(
t^{\ast },T^{\ast }\right) \right) ^{2}\right] ^{1/2}.  \notag
\end{eqnarray}%
We have%
\begin{eqnarray}
&&E\left( \tilde{S}_{Z}(t^{\ast },T^{\ast },\Delta )-Z\left( t^{\ast
},T^{\ast }\right) \right) ^{2}  \label{3.182} \\
&=&E\left( \tilde{S}_{Z}(t^{\ast },T^{\ast },\Delta )-S_{Z}(t^{\ast
},T^{\ast },\Delta )+S_{Z}(t^{\ast },T^{\ast },\Delta )-Z\left( t^{\ast
},T^{\ast }\right) \right) ^{2}  \notag \\
&\leq &2E\left[ \tilde{S}_{Z}(t^{\ast },T^{\ast },\Delta )-S_{Z}(t^{\ast
},T^{\ast },\Delta )\right] ^{2}+2E\left[ \tilde{S}_{Z}(t^{\ast },T^{\ast
},\Delta )-Z\left( t^{\ast },T^{\ast }\right) \right] ^{2}.  \notag
\end{eqnarray}%
Due to the condition (\ref{2.9}) imposed on the quadrature rule $%
S_{Z}(t^{\ast },T^{\ast },\Delta ),$ the second term on the right-hand side
of (\ref{3.182}) is bounded from above by $K\Delta ^{2p}.$ Using (\ref{3.3}%
), we obtain for the first term on the right-hand side of (\ref{3.182}) (cf.
(\ref{2.8})): 
\begin{eqnarray*}
2E\left[ \tilde{S}_{Z}(t^{\ast },T^{\ast },\Delta )-S_{Z}(t^{\ast },T^{\ast
},\Delta )\right] ^{2} &=&2\Delta ^{2}E\left[ \dsum_{j=\varrho _{M}}^{N}%
\tilde{\gamma}_{j}\ (\tilde{f}^{j}(t^{\ast })-f(t^{\ast },T_{j}))\right] ^{2}
\\
&\leq &K\Delta \dsum_{j=\varrho _{M}}^{N}E\left[ \tilde{f}^{j}(t^{\ast
})-f(t^{\ast },T_{j})\right] ^{2}, \\
&\leq &K\Delta (N-\varrho _{M}+1)\Delta ^{2p}\leq K\Delta ^{2p}.
\end{eqnarray*}%
Hence 
\begin{equation}
E\left( \tilde{S}_{Z}(t^{\ast },T^{\ast },\Delta )-Z\left( t^{\ast },T^{\ast
}\right) \right) ^{2}\leq K\Delta ^{2p}.  \label{3.184}
\end{equation}

The required estimate (\ref{3.14}) follows from (\ref{3.15}), (\ref{3.152}),
(\ref{3.nnn}), and (\ref{3.184}). Theorem~\ref{thm:weai2f} is proved. \ $%
\square $

\subsection{$t$-discretization error\label{sec:errtd}}

In this section we analyze the error $R_{2}$ (see (\ref{3.2})) due to $t$%
-discretization of (\ref{2.6})-(\ref{2.66}): 
\begin{equation*}
R_{2}=\tilde{F}(t_{0},f_{0};t^{\ast },T^{\ast })-\bar{F}(t_{0},f_{0};t^{\ast
},T^{\ast }).
\end{equation*}%
Then combining its estimate with the estimate (\ref{3.14}) for $R_{1}$ from
Theorem~\ref{thm:weai2f}, we prove convergence of the weak approximation $%
\bar{F}$ to $F$ (see (\ref{3.2})). In the analysis of $R_{2}$ the key is to
show that convergence of $\bar{F}(t_{0},f_{0};t^{\ast },T^{\ast })$ to $%
\tilde{F}(t_{0},f_{0};t^{\ast },T^{\ast })$ is uniform in $\Delta $, which
is the reason why we cannot just apply here the standard results of weak
convergence of numerical methods for SDEs (see, e.g. \cite{MT1}). The
convergence theorem is proved under the assumption that the pay-off function 
$G(z)$ in (\ref{7}) is sufficiently smooth. At the end of this section we
also discuss how this assumption can be relaxed.

To prove the convergence theorem (Theorem~\ref{prpt1}) of $\bar{F}%
(t_{0},f_{0};t^{\ast },T^{\ast })$ to $\tilde{F}(t_{0},f_{0};t^{\ast
},T^{\ast }),$ we need the following technical lemma. We will use the
multi-index notation: 
\begin{equation*}
\mathbf{i}=(i_{0},\ldots ,i_{N})
\end{equation*}%
with $i_{j}$ being nonnegative integers, $|\mathbf{i}|=i_{0}+\cdots +i_{N},$
and $\mathbf{i}!=i_{0}!\cdots i_{N}!.$

\begin{lemma}
\label{lemma}Let $\Lambda ^{m}$ be the $m^{th}$-order operator 
\begin{equation*}
\Lambda ^{m}=\Lambda _{\mu }^{m}=\sum_{|\mathbf{i}|=m}\mu ^{\mathbf{i}}\frac{%
\partial ^{m}}{\left( \partial x^{0}\right) ^{i_{0}}\cdots \left( \partial
x^{N}\right) ^{i_{N}}}
\end{equation*}%
with some $\mu ^{\mathbf{i}}.$ Suppose Assumptions~2.1 and~2.2 are
satisfied. Assume that the payoff function $G(z)$ has $m_{\ast }$ bounded
derivatives. Then for $m>0$ up to the order $m_{\ast }$%
\begin{equation}
\left\vert \Lambda ^{m}\tilde{F}(t,x;t^{\ast },T^{\ast })\right\vert \leq
K\mu _{Max}\exp (c\Delta |x|),  \label{lem0}
\end{equation}%
where $K>0$ and $c>0$ do not depend on $\Delta $ and $x\in \mathbb{R}^{N+1},$
and $\mu _{Max}:=\max_{|\mathbf{i}|=m}|\mu ^{\mathbf{i}}|$.
\end{lemma}

\begin{remark}
To help with intuitive understanding of this lemma, we remark that $\Lambda
^{m}\tilde{F}$ can be viewed as a Frechet derivative of the option price
with respect to the discretized initial forward rate curve.
\end{remark}

\noindent \textbf{Proof of Lemma~\ref{lemma}. } Recall the notation: $\tilde{%
f}_{t,x}^{j}(s),\ s\geq t,$ is the solution of the system of SDEs (\ref{2.6}%
)-(\ref{2.66}) with the initial condition at $t\geq t_{0}:$ $\tilde{f}%
_{t,x}^{j}(t)=x^{j}.$ We introduce a more detailed notation for $\tilde{S}%
_{Z}(t^{\ast },T^{\ast },\Delta )$ (cf. (\ref{2.8})): 
\begin{equation}
\tilde{S}_{Z}(t,x;t^{\ast },T^{\ast },\Delta )=\Delta \dsum_{j=\varrho
_{M}}^{N}\tilde{\gamma}_{j}\tilde{f}_{t,x}^{j}(t^{\ast }),  \label{tSznew}
\end{equation}%
which we can present as (cf. (\ref{2.6})) 
\begin{gather*}
\tilde{S}_{Z}(t,x;t^{\ast },T^{\ast },\Delta )=\Delta \dsum_{j=\varrho
_{M}}^{N}\tilde{\gamma}_{j}x^{j} \\
+\Delta \dsum_{j=\varrho _{M}}^{N}\tilde{\gamma}_{j}\left[ \int_{t}^{t^{\ast
}}\sigma ^{\top }(s,T_{j},\tilde{f}_{t,x}^{j}(s))\tilde{S}%
_{I}(t,x;s,T_{j},\Delta )ds+\int_{t}^{t^{\ast }}\sigma ^{\top }(s,T_{j},%
\tilde{f}_{t,x}^{j}(s))dW(s)\right] ,
\end{gather*}%
where (cf. (\ref{2.66})) 
\begin{equation}
\tilde{S}_{I}(t,x;s,T_{j},\Delta )=\Delta \dsum_{l=\ell (s)}^{\kappa
(s,T_{j})}\gamma _{l}(s)\sigma (s,T_{l},\tilde{f}_{t,x}^{l}(s)).
\label{SI_ext}
\end{equation}%
Then, thanks to Assumption~2.1, we obtain for any positive integer $m:$ 
\begin{gather}
E\tilde{P}^{m}(t^{\ast },T^{\ast })=E\exp \left( -m\tilde{S}_{Z}(t,x;t^{\ast
},T^{\ast },\Delta )\right)  \label{lem03} \\
=E\exp \left( -m\Delta \dsum_{j=\varrho _{M}}^{N}\tilde{\gamma}%
_{j}x^{j}\right.  \notag \\
\left. -m\Delta \dsum_{j=\varrho _{M}}^{N}\tilde{\gamma}_{j}\left[
\int_{t}^{t^{\ast }}\sigma ^{\top }(s,T_{j},\tilde{f}_{t,x}^{j}(s))\tilde{S}%
_{I}(t,x;s,T_{j},\Delta )ds+\int_{t}^{t^{\ast }}\sigma ^{\top }(s,T_{j},%
\tilde{f}_{t,x}^{j}(s))dW(s)\right] \right)  \notag \\
\leq K\exp \left( c\Delta \dsum_{j=\varrho _{M}}^{N}|x^{j}|\right) ,  \notag
\end{gather}%
where $K>0$ and $c>0$ do not depend on $\Delta .$

Further, recall that $\tilde{Y}_{t,x,y}(s),$ $s\geq t,$ is the solution of (%
\ref{t5}) with the initial condition $\tilde{Y}_{t,x,y}(t)=y$ and with $%
\tilde{f}^{i}(s)=\tilde{f}_{t,x}^{i}(s),$ i.e., 
\begin{eqnarray}
\tilde{Y}_{t,x,y}(s) &:&=y+\int_{t}^{s}\tilde{\pi}(s^{\prime })ds^{\prime }
\label{lem01} \\
&=&y+\int_{t}^{s}\pi (s^{\prime };\tilde{f}_{t,x}^{i}(s^{\prime }),\ i=\ell
(s^{\prime }),\ldots ,\ell (s^{\prime })+\theta )ds^{\prime }  \notag \\
&=&y+\dsum\limits_{l=\ell (t)}^{\ell (s)}\dsum\limits_{i=0}^{\theta
}\int_{t\vee T_{l}}^{s\wedge T_{l+1}}\lambda _{i}(s^{\prime })\tilde{f}%
_{t,x}^{l+i}(s^{\prime })\ ds^{\prime }  \notag \\
&=&y+\dsum\limits_{l=\ell (t)}^{\ell (s)}\dsum\limits_{m=l}^{l+\theta
}\int_{t\vee T_{l}}^{s\wedge T_{l+1}}\lambda _{m-l}(s^{\prime })\tilde{f}%
_{t,x}^{m}(s^{\prime })ds^{\prime },\ \ t\geq t_{0},\ \ s\geq t,  \notag
\end{eqnarray}%
where (cf. (\ref{2.10})) $\theta $ and $\lambda _{m-l}(s^{\prime })$ depend
on our choice of the accuracy order of short rate approximation, and $\theta 
$ does not depend on $\Delta ,\ $and $|\lambda _{m-l}(s^{\prime })|$ are
bounded by a constant independent of $\Delta .$

We also see that $\tilde{Y}_{t,x,y}(s)=y+\tilde{Y}_{t,x,0}(s).$ Using (\ref%
{lem01}), (\ref{2.6}), and Assumption~2.1, one can show that for any $%
\varkappa >0$ 
\begin{eqnarray}
E\left[ \exp (\varkappa |\tilde{Y}_{t,x,0}(t^{\ast })|)\right] &=&E\exp
\left( \varkappa \left\vert \dsum\limits_{l=\ell (t)}^{\ell (t^{\ast
})}\dsum\limits_{m=l}^{l+\theta }\int_{t\vee T_{l}}^{t^{\ast }\wedge
T_{l+1}}\lambda _{m-l}(s^{\prime })\tilde{f}_{t,x}^{m}(s^{\prime
})ds^{\prime }\right\vert \right)  \label{lem02} \\
&\leq &K\exp \left( c\Delta \dsum\limits_{l=\ell (t)}^{\ell (t^{\ast
})+\theta }|x^{l}|\right) ,  \notag
\end{eqnarray}%
where $K>0$ and $c>0$ do not depend on $\Delta .$

Using smoothness of $G(z)$, we obtain 
\begin{gather}
\Lambda ^{m}\tilde{F}(t,x;t^{\ast },T^{\ast })=E\Lambda ^{m}\exp (-\tilde{Y}%
_{t,x,0}(t^{\ast }))G(\tilde{P}(t^{\ast },T^{\ast }))  \label{lem5n} \\
=E\sum_{|\mathbf{i}|=m}\mu ^{\mathbf{i}}\frac{\partial ^{m}}{\left( \partial
x^{0}\right) ^{i_{0}}\cdots \left( \partial x^{N}\right) ^{i_{N}}}\exp (-%
\tilde{Y}_{t,x,0}(t^{\ast }))G(\tilde{P}(t^{\ast },T^{\ast }))  \notag \\
=E\exp (-\tilde{Y}_{t,x,0}(t^{\ast }))\sum_{k_{\ast }=0}^{m}\sum_{n_{\ast
}=0}^{m-k_{\ast }}\sum_{\alpha =0}^{m}\sum_{\beta =0}^{\alpha }\sum_{\bar{j}%
_{k_{\ast }}+\bar{l}_{n_{\ast }}=m}C(\alpha ,\beta ,j_{1},\ldots ,j_{k_{\ast
}},l_{1},\ldots ,l_{n_{\ast }})  \notag \\
\times \frac{d^{\alpha }}{dz^{\alpha }}G\left( \tilde{P}(t^{\ast },T^{\ast
})\right) \tilde{P}^{\beta }(t^{\ast },T^{\ast })  \notag \\
\times \sum_{i_{1},\ldots ,i_{\bar{j}_{k_{\ast }}},r_{1},\ldots ,r_{\bar{l}%
_{n_{\ast }}}=0}^{N}\mu ^{\mathbf{i}}\prod\limits_{k=1}^{k_{\ast }}\frac{%
\partial ^{j_{k}}}{\partial x^{i_{1+\bar{j}_{k-1}}}\cdots \partial x^{i_{%
\bar{j}_{k}}}}\tilde{Y}_{t,x,0}(t^{\ast })  \notag \\
\times \prod\limits_{n=1}^{n_{\ast }}\frac{\partial ^{l_{n}}}{\partial
x^{r_{1+\bar{l}_{n-1}}}\cdots \partial x^{r_{\bar{l}_{n}}}}\tilde{S}%
_{Z}(t,x;t^{\ast },T^{\ast },\Delta ),  \notag
\end{gather}%
where $C(\alpha ,\beta ,j_{1},\ldots ,j_{k_{\ast }},l_{1},\ldots ,l_{n_{\ast
}})\ $are constants independent of $N;$ $\bar{j}_{k}=\sum_{r=1}^{k}j_{r},$ $%
\bar{l}_{n}=\sum_{r=1}^{n}l_{r};$ the sum $\sum_{\bar{j}_{k_{\ast }}+\bar{l}%
_{n_{\ast }}=m}$ is taken over all positive integers $j_{1},\ldots
,j_{k_{\ast }}$ and $l_{1},\ldots ,l_{n_{\ast }}$ such that $j_{k}\leq
j_{k+1},$ $k=1,$ $\ldots ,k_{\ast }-1,$ $l_{n}\leq l_{n+1},$ $n=1,$ $\ldots
,n_{\ast }-1,$ and $\bar{j}_{k_{\ast }}+\bar{l}_{n_{\ast }}=m;$ and in the
right-hand side the multi-index $\mathbf{i}$ at $\mu ^{\mathbf{i}}$
corresponds to the values taken by $i_{1},\ldots ,i_{\bar{j}_{k_{\ast }}},$ $%
r_{1},\ldots ,r_{\bar{l}_{n_{\ast }}}.$

We have (cf. (\ref{tSznew})): 
\begin{equation}
\frac{\partial ^{l}}{\partial x^{i_{1}}\cdots \partial x^{i_{l}}}\tilde{S}%
_{Z}(t,x;t^{\ast },T^{\ast },\Delta )=\Delta \dsum_{q=\varrho _{M}}^{N}%
\tilde{\gamma}_{q}\frac{\partial ^{l}}{\partial x^{i_{1}}\cdots \partial
x^{i_{l}}}\tilde{f}_{t,x}^{q}(t^{\ast }).  \label{lem6n}
\end{equation}

Using the Cauchy-Bunyakovsky inequality, the assumed boundedness of
derivatives of $G(z),$ and the inequalities (\ref{lem03}) and (\ref{lem02}),
we obtain from (\ref{lem5n})-(\ref{lem6n}):%
\begin{gather}
|\Lambda ^{m}\tilde{F}(t,x;t^{\ast },T^{\ast })|\leq K\exp \left( c\Delta
\dsum_{j=\varrho _{M}}^{N}|x^{j}|\right)  \label{lem07n} \\
\times \sum_{k_{\ast }=0}^{m}\sum_{n_{\ast }=0}^{m-k_{\ast }}\sum_{\bar{j}%
_{k_{\ast }}+\bar{l}_{n_{\ast }}=m}\left( E\left[ \sum_{i_{1},\ldots ,i_{%
\bar{j}_{k_{\ast }}},r_{1},\ldots ,r_{\bar{l}_{n^{\ast }}}=0}^{N}\mu ^{%
\mathbf{i}}\ \prod\limits_{k=1}^{k_{\ast }}\frac{\partial ^{j_{k}}}{\partial
x^{i_{1+\bar{j}_{k-1}}}\cdots \partial x^{i_{\bar{j}_{k}}}}\tilde{Y}%
_{t,x,0}(t^{\ast })\right. \right.  \notag \\
\times \left. \left. \prod\limits_{n=1}^{n_{\ast }}\left( \Delta
\dsum_{q=\varrho _{M}}^{N}\tilde{\gamma}_{q}\frac{\partial ^{l_{n}}}{%
\partial x^{r_{1+\bar{l}_{n-1}}}\cdots \partial x^{r_{\bar{l}_{n}}}}\tilde{f}%
_{t,x}^{q}(t^{\ast })\right) \right] ^{2}\right) ^{1/2},  \notag
\end{gather}%
where $K>0$ and $c$ are independent of $\Delta $ and $x.$ Then, to complete
the proof of this lemma, it is sufficient to show that for any $0\leq
k_{\ast }\leq m$ and $0\leq n_{\ast }\leq m-k_{\ast },$ any combinations of $%
j_{1},\ldots ,j_{k^{\ast }}$ and $l_{1},\ldots ,l_{n^{\ast }}$ satisfying $%
\bar{j}_{k_{\ast }}+\bar{l}_{n^{\ast }}=m,$ and any combination of $%
q_{1},\ldots ,q_{n^{\ast }}$ with $\varrho _{M}\leq q_{i}\leq N:$%
\begin{eqnarray}
&&E\left[ \sum_{i_{1},\ldots ,i_{\bar{j}_{k_{\ast }}},r_{1},\ldots ,r_{\bar{l%
}_{n^{\ast }}}=0}^{N}\mu ^{\mathbf{i}}\ \prod\limits_{k=1}^{k_{\ast }}\frac{%
\partial ^{j_{k}}}{\partial x^{i_{1+\bar{j}_{k-1}}}\cdots \partial x^{i_{%
\bar{j}_{k}}}}\tilde{Y}_{t,x,0}(t^{\ast })\right.  \label{lem04n} \\
&&\times \left. \prod\limits_{n=1}^{n_{\ast }}\frac{\partial ^{l_{n}}}{%
\partial x^{r_{1+\bar{l}_{n-1}}}\cdots \partial x^{r_{\bar{l}_{n}}}}\tilde{f}%
_{t,x}^{q_{n}}(t^{\ast })\right] ^{2}\leq K\mu _{Max}^{2},  \notag
\end{eqnarray}%
where $K>0$ is independent of $\Delta $ and $x.$

We can obtain the following SDEs (see (\ref{t5})): 
\begin{eqnarray*}
d\frac{\partial ^{j}}{\partial x^{i_{1}}\cdots \partial x^{i_{j}}}\tilde{Y}%
_{t,x,0}(s) &=&\dsum\limits_{l=\ell (t)}^{\ell (t^{\ast
})}\dsum\limits_{r=l}^{l+\theta }\chi _{s\in \lbrack T_{l},T_{l+1})}\cdot
\lambda _{r-l}(s)\cdot \frac{\partial ^{j}}{\partial x^{i_{1}}\cdots
\partial x^{i_{j}}}\tilde{f}_{t,x}^{r}(s)ds, \\
\frac{\partial ^{j}}{\partial x^{i_{1}}\cdots \partial x^{i_{j}}}\tilde{Y}%
_{t,x,0}(t) &=&0,
\end{eqnarray*}%
and (see (\ref{2.6}))%
\begin{gather*}
d\frac{\partial ^{l}}{\partial x^{r_{1}}\cdots \partial x^{r_{l}}}\tilde{f}%
_{t,x}^{q}(s)=\sum_{\alpha =0}^{l}\sum_{\beta =0}^{l}\sum_{n_{\ast
}=0}^{l-1}\sum_{\tau _{\ast }=1}^{l-n_{\ast }}\sum_{\bar{l}_{n_{\ast }}+\bar{%
p}_{\tau _{\ast }}=l}C(\alpha ,\beta ,n_{\ast },\tau _{\ast }) \\
\times \sum_{\{k_{1},\ldots ,k_{l}\}=\{r_{1},\ldots ,r_{l}\}}\Delta
\dsum_{v=\ell (s)}^{\kappa (s,T_{q})}\gamma _{v}\ \frac{d^{\alpha }}{%
dz^{\alpha }}\sigma ^{\top }(s,T_{q},\tilde{f}_{t,x}^{q}(s))\frac{d^{\beta }%
}{dz^{\beta }}\sigma ^{\top }(s,T_{v},\tilde{f}_{t,x}^{v}(s)) \\
\times \prod\limits_{n=1}^{n_{\ast }}\frac{\partial ^{l_{n}}}{\partial
x^{k_{1+\bar{l}_{n-1}}}\cdots \partial x^{k_{_{\bar{l}_{n}}}}}\tilde{f}%
_{t,x}^{q}(s)\ \prod\limits_{\tau =1}^{\tau _{\ast }}\frac{\partial
^{p_{\tau }}}{\partial x^{k_{1+\bar{l}_{n_{\ast }}+\bar{p}_{\tau -1}}}\cdots
\partial x^{k_{\bar{l}_{n_{\ast }}+\bar{p}_{\tau }}}}\tilde{f}_{t,x}^{v}(s)\
ds \\
+\sum_{\alpha =1}^{l}\sum_{n_{\ast }=1}^{l}\sum_{\bar{l}_{n_{\ast
}}=l}C(\alpha ,n_{\ast })\sum_{\{k_{1},\ldots ,k_{l}\}=\{r_{1},\ldots
,r_{l}\}}\frac{d^{\alpha }}{dz^{\alpha }}\sigma ^{\top }(s,T_{q},\tilde{f}%
_{t,x}^{q}(s)) \\
\times \prod\limits_{n=1}^{n_{\ast }}\frac{\partial ^{l_{n}}}{\partial
x^{k_{1+\bar{l}_{n-1}}}\cdots \partial x^{k_{\bar{l}_{n}}}}\tilde{f}%
_{t,x}^{q}(s)\ dW(s), \\
\frac{\partial ^{l}}{\partial x^{r_{1}}\cdots \partial x^{r_{l}}}\tilde{f}%
_{t,x}^{q}(s)=\chi _{l=1},
\end{gather*}%
where $C(\alpha ,\beta ,n_{\ast },\tau _{\ast })$ and $C(\alpha ,n_{\ast })$
are constants independent of $N,$ and $\sum_{\{k_{1},\ldots
,k_{l}\}=\{r_{1},\ldots ,r_{l}\}}$ means summation over all possible
recombinations $\{k_{1},\ldots ,k_{l}\}$ of $r_{1},\ldots ,r_{l}$ (note that
the number of terms in this sum depends on $l$ but not on $N).$

To obtain (\ref{lem04n}), we first consider the case $m=1$ for which it is
sufficient to get an estimate for $E\left[ \sum_{i=0}^{N}\mu ^{i}\frac{%
\partial }{\partial x^{i}}\tilde{f}_{t,x}^{j}(s)\right] ^{2}.$ To this end,
introduce the process $\zeta _{t,x}(s)=(\zeta ^{0}(s),\ldots ,\zeta
^{N}(s))^{\top }$ with $\zeta ^{j}(s):=\sum_{i=0}^{N}\mu ^{i}\frac{\partial 
}{\partial x^{i}}\tilde{f}_{t,x}^{j}(s),$ $s\geq t,$ which satisfies the
following system of SDEs 
\begin{gather*}
d\zeta ^{j}=\frac{d}{dz}\sigma ^{\top }(s,T_{j},\tilde{f}_{t,x}^{j}(s))\cdot 
\tilde{S}_{I}(t,x;s,T_{j},\Delta )\cdot \zeta ^{j}\ ds \\
+\sigma ^{\top }(s,T_{j},\tilde{f}_{t,x}^{j}(s))\cdot \Delta \dsum_{l=\ell
(s)}^{\kappa (s,T_{j})}\frac{d}{dz}\sigma ^{\top }(s,T_{l},\tilde{f}%
_{t,x}^{l}(s))\cdot \zeta ^{l}\ ds+\frac{d}{dz}\sigma ^{\top }(s,T_{i},%
\tilde{f}_{t,x}^{j}(s))\cdot \zeta ^{j}\ dW(s), \\
\zeta ^{j}(t)=\mu ^{j},\ \ j=0,\ldots ,N.
\end{gather*}%
Then using Ito's formula and Assumptions~2.1 and~2.2, we obtain after some
straightforward calculations:%
\begin{equation*}
E\left[ \zeta _{t,x}^{j}(s)\right] ^{2}\leq K\left[ \mu ^{j}\right]
^{2}+K\int_{t}^{s}E\left[ \zeta ^{j}(s^{\prime })\right] ^{2}ds^{\prime
}+K\int_{t}^{s}\Delta \dsum_{l=\ell (s^{\prime })}^{\kappa (s,T_{j})}E\left[
\zeta ^{l}(s^{\prime })\right] ^{2}ds^{\prime }.
\end{equation*}%
Let $\mathcal{E}(s):=\max_{0\leq j\leq N}E\left[ \zeta _{t,x}^{j}(s)\right]
^{2}.$ Then 
\begin{equation*}
\mathcal{E}(s)\leq K\mu _{Max}^{2}+K\int_{t}^{s}\mathcal{E}(s^{\prime
})ds^{\prime },
\end{equation*}%
where $K>0$ does not depend on $\Delta $ and $x.$ Hence, by Gronwall's
inequality 
\begin{equation}
\mathcal{E}(s)\leq K\mu _{Max}^{2},\ \ t\leq s\leq t^{\ast }.  \label{lem4}
\end{equation}

Next, we consider the $(N+1)^{2}$-dimensional process 
\begin{equation*}
\zeta ^{j_{1},j_{2}}(s):=\sum_{i_{1},i_{2}=0}^{N}\mu ^{i_{1},i_{2}}\frac{%
\partial }{\partial x^{i_{1}}}\tilde{f}_{t,x}^{j_{1}}(s)\frac{\partial }{%
\partial x^{i_{2}}}\tilde{f}_{t,x}^{j_{2}}(s),\ j_{1},j_{2}=0,\ldots ,N,\
s\geq t.
\end{equation*}%
Using the same recipe as in the case of estimating $\max_{0\leq j\leq N}E%
\left[ \zeta _{t,x}^{j}(s)\right] ^{2},$ we get that 
\begin{equation}
\max_{0\leq j\leq N}E\left[ \zeta _{t,x}^{j_{1},j_{2}}(s)\right] ^{2}\leq
K\mu _{Max}^{2},  \label{lem08n}
\end{equation}%
where $K>0$ does not depend on $\Delta $ and $x.$ Using (\ref{lem08n}) and
repeating the same recipe again in the case of the processes 
\begin{eqnarray*}
_{2}\zeta _{t,x}^{j}(s) &:&=\sum_{i_{1},i_{2}=0}^{N}\mu ^{i_{1},i_{2}}\frac{%
\partial ^{2}}{\partial x^{i_{1}}\partial x^{i_{2}}}\tilde{f}_{t,x}^{j}(s),\
j=0,\ldots ,N, \\
\eta _{t,x}^{j}(s) &:&=\sum_{i_{1},i_{2}=0}^{N}\mu ^{i_{1},i_{2}}\frac{%
\partial }{\partial x^{i_{1}}}\tilde{Y}_{t,x,0}(s)\frac{\partial }{\partial
x^{i_{2}}}\tilde{f}_{t,x}^{j}(s),\ j=0,\ldots ,N,\ s\geq t,
\end{eqnarray*}%
we obtain 
\begin{eqnarray}
\max_{0\leq j\leq N}E\left[ _{2}\zeta _{t,x}^{j}(s)(s)\right] ^{2} &\leq
&K\mu _{Max}^{2},  \label{lem09n} \\
\max_{0\leq j\leq N}E\left[ \eta _{t,x}^{j}(s)\right] ^{2} &\leq &K\mu
_{Max}^{2},  \label{lem10n}
\end{eqnarray}%
where $K>0$ does not depend on $\Delta $ and $x.$ Using (\ref{lem09n}), it
is not difficult to get that for the process%
\begin{equation*}
_{2}\eta _{t,x}(s):=\sum_{i_{1},i_{2}=0}^{N}\mu ^{i_{1},i_{2}}\frac{\partial
^{2}}{\partial x^{i_{1}}\partial x^{i_{2}}}\tilde{Y}_{t,x,0}(s)
\end{equation*}%
the following estimate also holds%
\begin{equation}
E\left[ _{2}\eta _{t,x}(s)\right] ^{2}\leq K\mu _{Max}^{2}.  \label{lem11n}
\end{equation}%
It is clear that (\ref{lem08n})-(\ref{lem11n}) are sufficient for proving (%
\ref{lem04n}) with $m=2$. To show (\ref{lem04n}) for $m=3,$ we need to
obtain estimates for the second moments of the processes 
\begin{eqnarray*}
\zeta ^{j_{1},j_{2},j_{3}}(s) &=&\sum_{i_{1},i_{2},i_{3}=0}^{N}\mu
^{i_{1},i_{2},i_{3}}\frac{\partial }{\partial x^{i_{1}}}\tilde{f}%
_{t,x}^{j_{1}}(s)\frac{\partial }{\partial x^{i_{2}}}\tilde{f}%
_{t,x}^{j_{2}}(s)\frac{\partial }{\partial x^{i_{3}}}\tilde{f}%
_{t,x}^{j_{3}}(s),\ j_{1},j_{2},j_{3}=0,\ldots ,N, \\
_{2}\zeta _{t,x}^{j_{1},j_{2}}(s) &=&\sum_{i_{1},i_{2},i_{3}=0}^{N}\mu
^{i_{1},i_{2},i_{3}}\frac{\partial ^{2}}{\partial x^{i_{1}}\partial x^{i_{2}}%
}\tilde{f}_{t,x}^{j_{1}}(s)\frac{\partial }{\partial x^{i_{3}}}\tilde{f}%
_{t,x}^{j_{2}}(s),\ j_{1},j_{2}=0,\ldots ,N, \\
_{3}\zeta _{t,x}^{j}(s) &=&\sum_{i_{1},i_{2},i_{3}=0}^{N}\mu
^{i_{1},i_{2},i_{3}}\frac{\partial ^{3}}{\partial x^{i_{1}}\partial
x^{i_{2}}\partial x^{i_{3}}}\tilde{f}_{t,x}^{j}(s),\ j=0,\ldots ,N,\ s\geq t,
\end{eqnarray*}%
and 
\begin{eqnarray*}
\eta _{t,x}^{j_{1},j_{2}}(s) &=&\sum_{i_{1},i_{2},i_{3}=0}^{N}\mu
^{i_{1},i_{2},i_{3}}\frac{\partial }{\partial x^{i_{1}}}\tilde{f}%
_{t,x}^{j_{1}}(s)\frac{\partial }{\partial x^{i_{2}}}\tilde{f}%
_{t,x}^{j_{2}}(s)\frac{\partial }{\partial x^{i_{3}}}\tilde{Y}_{t,x,0}(s),\
j_{1},j_{2}=0,\ldots ,N, \\
_{2,1}\eta _{t,x}^{j}(s) &=&\sum_{i_{1},i_{2},i_{3}=0}^{N}\mu
^{i_{1},i_{2},i_{3}}\frac{\partial ^{2}}{\partial x^{i_{1}}\partial x^{i_{2}}%
}\tilde{f}_{t,x}^{j}(s)\frac{\partial }{\partial x^{i_{3}}}\tilde{Y}%
_{t,x,0}(s),\ j=0,\ldots ,N, \\
_{1,2}\eta _{t,x}^{j}(s) &=&\sum_{i_{1},i_{2},i_{3}=0}^{N}\mu
^{i_{1},i_{2},i_{3}}\frac{\partial }{\partial x^{i_{1}}}\tilde{f}%
_{t,x}^{j}(s)\frac{\partial ^{2}}{\partial x^{i_{2}}\partial x^{i_{3}}}%
\tilde{Y}_{t,x,0}(s),\ j=0,\ldots ,N,\  \\
_{3}\eta _{t,x}(s) &=&\sum_{i_{1},i_{2},i_{3}=0}^{N}\mu ^{i_{1},i_{2},i_{3}}%
\frac{\partial ^{3}}{\partial x^{i_{1}}\partial x^{i_{2}}\partial x^{i_{3}}}%
\tilde{Y}_{t,x,0}(s),\ s\geq t,
\end{eqnarray*}%
which can be done using the same recipe but with more laborious
calculations. In the case of an arbitrary $m$ one need to consider processes 
$\zeta ^{j_{1},\ldots ,j_{m}}(s),$ $_{2}\zeta _{t,x}^{j_{1},\ldots
,j_{m-1}}(s),$ $\ldots ,$ $_{m}\zeta _{t,x}^{j}(s),$ $\eta
_{t,x}^{j_{1},\ldots ,j_{m-1}}(s),$ $_{m-1,1}\eta _{t,x}^{j}(s),\ldots ,$ $%
_{m}\eta _{t,x}(s)$ defined in the same fashion as we did in the cases $m=2$
and $3.$ It is not difficult to see that employing the same recipe maxima of
their second moments will be again bounded by $K\mu _{Max}^{2},$ from which (%
\ref{lem04n}) follows for an arbitrary $m.$

The required inequality (\ref{lem0}) follows from (\ref{lem07n}) and (\ref%
{lem04n}). Lemma~\ref{lemma} is proved. \ $\square \medskip $

Using Lemma~\ref{lemma}, we now prove convergence of $\bar{F}%
(t_{0},f_{0};t^{\ast },T^{\ast })$ to $\tilde{F}(t_{0},f_{0};t^{\ast
},T^{\ast })$ in the case of smooth payoffs $G.$

\begin{theorem}
\label{prpt1}Let $h\leq \alpha \Delta $ for some $\alpha >0.$ Suppose
Assumptions~2.1-2.3 and Assumptions~3.1,~3.2', and~3.3 are satisfied. Assume
that the payoff function $G(z)$ has bounded derivatives up to a sufficiently
high order. Then the approximation $\bar{F}(t_{0},f_{0};t^{\ast },T^{\ast })$
defined by $(\ref{2.13})$-$(\ref{2.17}),$ $(\ref{2.7}),$ $(\ref{t51})$
converges to $\tilde{F}(t_{0},f_{0};t^{\ast },T^{\ast })$ from $(\ref{3.1})$
with order $q>0,$ i.e.,%
\begin{equation}
\left\vert \tilde{F}(t_{0},f_{0};t^{\ast },T^{\ast })-\bar{F}%
(t_{0},f_{0};t^{\ast },T^{\ast })\right\vert \leq Kh^{q},\text{ }  \label{t6}
\end{equation}%
where $K>0$ is a constant independent of $h$ and $\Delta .$
\end{theorem}

\noindent \textbf{Proof. } Using the standard technique (see \cite[p. 100]%
{MT1}), we can write the difference $R_{2}$ in the form 
\begin{eqnarray}
R_{2} &=&\tilde{F}(t_{0},f_{0};t^{\ast },T^{\ast })-\bar{F}%
(t_{0},f_{0};t^{\ast },T^{\ast })  \label{tn0} \\
&=&E\exp (-\tilde{Y}_{t_{0},f_{0},0}(t_{M}))G(\exp (-\tilde{S}%
_{Z}(t_{0},f_{0};t^{\ast },T^{\ast },\Delta )))  \notag \\
&&-E\exp (-\bar{Y}_{M})G(\exp (-\bar{S}_{Z}(t^{\ast },T^{\ast },\Delta ))) 
\notag
\end{eqnarray}%
\begin{eqnarray}
&=&\sum_{i=0}^{M-1}E\left[ \exp (-\tilde{Y}_{t_{i},\bar{f}_{i},\bar{Y}%
_{i}}(t_{M}))G(\exp (-\tilde{S}_{Z}(t_{i},\bar{f}_{i};t^{\ast },T^{\ast
},\Delta )))\right.  \notag \\
&&\left. -\exp (-\tilde{Y}_{t_{i+1},\bar{f}_{i+1},\bar{Y}_{i+1}}(t_{M}))G(%
\exp (-\tilde{S}_{Z}(t_{i+1},\bar{f}_{i+1};t^{\ast },T^{\ast },\Delta )))%
\right]  \notag \\
&=&\sum_{i=0}^{M-1}E\{\exp (-\tilde{Y}_{t_{i},\bar{f}_{i},\bar{Y}%
_{i}}(t_{i+1}))E[\exp (-\tilde{Y}_{t_{i+1},\tilde{f}_{t_{i},\bar{f}%
_{i}}(t_{i+1}),0}(t_{M}))  \notag \\
&&\times G(\exp (-\tilde{S}_{Z}(t_{i+1},\tilde{f}_{t_{i},\bar{f}%
_{i}}(t_{i+1});t^{\ast },T^{\ast },\Delta )))|\tilde{f}_{t_{i},\bar{f}%
_{i}}(t_{i+1})]  \notag \\
&&-\exp (-\bar{Y}_{i+1})E[\exp (-\tilde{Y}_{t_{i+1},\bar{f}%
_{i+1},0}(t_{M}))G(\exp (-\tilde{S}_{Z}(t_{i+1},\bar{f}_{i+1};t^{\ast
},T^{\ast },\Delta )))|\bar{f}_{i+1}]\}  \notag
\end{eqnarray}%
\begin{eqnarray*}
&=&\sum_{i=0}^{M-1}E\left\{ \exp (-\tilde{Y}_{t_{i},\bar{f}_{i},\bar{Y}%
_{i}}(t_{i+1}))\tilde{F}(t_{i+1},\tilde{f}_{t_{i},\bar{f}_{i}}(t_{i+1});t^{%
\ast },T^{\ast })-\exp (-\bar{Y}_{i+1})\tilde{F}(t_{i+1},\bar{f}%
_{i+1};t^{\ast },T^{\ast })\right\} \\
&=&\sum_{i=0}^{M-1}E\exp (-\bar{Y}_{i})E\left[ \exp (-\tilde{Y}_{t_{i},\bar{f%
}_{i},0}(t_{i+1}))\tilde{F}(t_{i+1},\tilde{f}_{t_{i},\bar{f}%
_{i}}(t_{i+1});t^{\ast },T^{\ast })\right. \\
&&\left. -\exp (-\bar{Y}_{t_{i},\bar{f}_{i}^{\cdot },0}(t_{i+1}))\tilde{F}%
(t_{i+1},\bar{f}_{t_{i},\bar{f}_{i}}(t_{i+1});t^{\ast },T^{\ast })\left\vert 
\bar{f}_{i}\right. \right] \\
&=&\sum_{i=0}^{M-1}E\exp (-\bar{Y}_{i})\rho (t_{i},\bar{f}_{i}),
\end{eqnarray*}%
where 
\begin{eqnarray}
\rho (t,x) &=&E[\exp (-\tilde{Y}_{t,x,0}(t+h))\tilde{F}(t+h,\tilde{f}%
_{t,x}(t+h);t^{\ast },T^{\ast })  \label{tn1} \\
&&-\exp (-\bar{Y}_{t,x,0}(t+h))\tilde{F}(t+h,\bar{f}_{t,x}(t+h);t^{\ast
},T^{\ast })].  \notag
\end{eqnarray}%
Now we write the Taylor expansion of the terms under expectation in (\ref%
{tn1}) in powers of $\delta \tilde{Y}=-\tilde{Y}_{t,x,0}(t+h)$ and $\delta 
\tilde{f}^{i}=\tilde{f}_{t,x}^{i}(t+h)-x^{i}$ and in powers of $\delta \bar{Y%
}=-\bar{Y}_{t,x,0}(t+h)$ and $\delta \bar{f}_{t,x}^{i}=\bar{f}%
_{t,x}^{i}(t+h)-x^{i}.$ As a result, we obtain 
\begin{eqnarray}
&&\exp (-\tilde{Y}_{t,x,0}(t+h))\tilde{F}(t+h,\tilde{f}_{t,x}(t+h);t^{\ast
},T^{\ast })  \label{tn2} \\
&=&\tilde{F}(t+h,x;t^{\ast },T^{\ast })  \notag \\
&&+\sum_{|\mathbf{i}|+k=1}^{2q+1}\frac{1}{\mathbf{i}!k!}\frac{\partial ^{|%
\mathbf{i}|}}{\left( \partial x^{0}\right) ^{i_{0}}\cdots \left( \partial
x^{N}\right) ^{i_{N}}}\tilde{F}(t+h,x;t^{\ast },T^{\ast })\left( \delta 
\tilde{f}^{0}\right) ^{i_{0}}\cdots \left( \delta \tilde{f}^{N}\right)
^{i_{N}}\delta \tilde{Y}^{k}  \notag \\
&&+\sum_{|\mathbf{i}|+k=2q+2}\frac{1}{\mathbf{i}!k!}\frac{\partial ^{|%
\mathbf{i}|}}{\left( \partial x^{0}\right) ^{i_{0}}\cdots \left( \partial
x^{N}\right) ^{i_{N}}}\tilde{F}(t+h,x+\tilde{\chi}(\tilde{f}%
_{t,x}(t+h)-x);t^{\ast },T^{\ast })  \notag \\
&&\times \exp (-\tilde{\theta}\tilde{Y}_{t,x,0}(t+h))\times \left( \delta 
\tilde{f}^{0}\right) ^{i_{0}}\cdots \left( \delta \tilde{f}^{N}\right)
^{i_{N}}\delta \tilde{Y}^{k},  \notag
\end{eqnarray}%
where $\tilde{\chi}$ and $\tilde{\theta}$ are from $[0,1]$.

Further, 
\begin{eqnarray}
&&\exp (-\bar{Y}_{t,x,0}(t+h))\tilde{F}(t+h,\bar{f}_{t,x}(t+h);t^{\ast
},T^{\ast })  \label{tn3} \\
&=&\tilde{F}(t+h,x;t^{\ast },T^{\ast })  \notag \\
&&+\sum_{|\mathbf{i}|+k=1}^{2q+1}\frac{1}{\mathbf{i}!k!}\frac{\partial ^{|%
\mathbf{i}|}}{\left( \partial x^{0}\right) ^{i_{0}}\cdots \left( \partial
x^{N}\right) ^{i_{N}}}\tilde{F}(t+h,x;t^{\ast },T^{\ast })\left( \delta \bar{%
f}^{0}\right) ^{i_{0}}\cdots \left( \delta \bar{f}^{N}\right) ^{i_{N}}\delta 
\bar{Y}^{k}  \notag \\
&&+\sum_{|\mathbf{i}|+k=2q+2}\frac{1}{\mathbf{i}!k!}\frac{\partial ^{|%
\mathbf{i}|}}{\left( \partial x^{0}\right) ^{i_{0}}\cdots \left( \partial
x^{N}\right) ^{i_{N}}}\tilde{F}(t+h,x+\bar{\chi}(\bar{f}_{t,x}(t+h)-x);t^{%
\ast },T^{\ast })  \notag \\
&&\times \exp (-\bar{\theta}\bar{Y}_{t,x,0}(t+h))\left( \delta \bar{f}%
^{0}\right) ^{i_{0}}\cdots \left( \delta \bar{f}^{N}\right) ^{i_{N}}\delta 
\bar{Y}^{k},  \notag
\end{eqnarray}%
with $\bar{\chi}$ and $\bar{\theta}$ being from $[0,1]$.

It is not difficult to check (see also (\ref{estY})) that under the assumed
condition\emph{\ }$h\leq \alpha \Delta ,$ $\alpha >0,$ the following
inequality holds:%
\begin{equation}
\left[ E\max_{0\leq m\leq 2q+2,\{i_{1},\ldots i_{2q+2-m}\}\in \{0,\ldots
,N\}}\left\vert \delta \tilde{Y}^{m}\prod_{j=1}^{2q+2-m}\delta \tilde{f}%
^{i_{j}}\right\vert ^{2}\right] ^{1/2}\leq Ch^{q+1}\left(
1+\dsum\limits_{l=\ell (t)}^{\ell (t+h)+\theta }|x^{l}|^{2q+2}\right) ,
\label{Db06}
\end{equation}%
where $C>0$ is independent of $\Delta .$ We note that the number of
components $x^{l}$ appearing in the right-hand side of (\ref{Db06}) is not
larger than $1+\theta ,$ which does not depend on $\Delta .$

Using Lemma~\ref{lemma}, the inequalities (\ref{Db06}), (\ref{lem02}) and (%
\ref{expmom2}), and the Cauchy-Bunyakovsky inequality, we obtain 
\begin{eqnarray}
&&|E\sum_{|\mathbf{i}|+k=2q+2}\frac{1}{\mathbf{i}!k!}\frac{\partial ^{|%
\mathbf{i}|}}{\left( \partial x^{0}\right) ^{i_{0}}\cdots \left( \partial
x^{N}\right) ^{i_{N}}}\tilde{F}(t+h,x+\tilde{\chi}(\tilde{f}%
_{t,x}(t+h)-x);t^{\ast },T^{\ast })  \label{tn37} \\
&&\times \exp (-\tilde{\theta}\tilde{Y}_{t,x,0}(t+h))\left( \delta \tilde{f}%
^{0}\right) ^{i_{0}}\cdots \left( \delta \tilde{f}^{N}\right) ^{i_{N}}\delta 
\tilde{Y}^{k}|  \notag
\end{eqnarray}%
\begin{eqnarray*}
&\leq &E\exp (|\tilde{Y}_{t,x,0}(t+h)|)\times \sum_{k=0}^{2q+2}\left\vert
\sum_{|\mathbf{i}|=2q+2-k}\frac{1}{\mathbf{i}!k!}\left( \delta \tilde{f}%
^{0}\right) ^{i_{0}}\cdots \left( \delta \tilde{f}^{N}\right) ^{i_{N}}\delta 
\tilde{Y}^{k}\right. \\
&&\left. \times \frac{\partial ^{|\mathbf{i}|}}{\left( \partial x^{0}\right)
^{i_{0}}\cdots \left( \partial x^{N}\right) ^{i_{N}}}\tilde{F}(t+h,x+\tilde{%
\chi}(\tilde{f}_{t,x}(t+h)-x);t^{\ast },T^{\ast })\right\vert \\
&\leq &KE\left[ \exp (|\tilde{Y}_{t,x,0}(t+h)|)\max_{|\mathbf{i}%
|+k=2q+2}|\left( \delta \tilde{f}^{0}\right) ^{i_{0}}\cdots \left( \delta 
\tilde{f}^{N}\right) ^{i_{N}}\delta \tilde{Y}^{k}|\right. \\
&&\left. \times \exp (c\Delta |x+\tilde{\chi}(\tilde{f}_{t,x}(t+h)-x)|)%
\right] \\
&\leq &K\exp (c\Delta |x|)\left( 1+\dsum\limits_{l=\ell (t)}^{\ell
(t+h)+\theta }|x^{l}|^{2q+2}\right) h^{q+1},
\end{eqnarray*}%
where $K>0$ and $c>0$ independent of $\Delta ,$ $h,$ and $x.$

Analogously, using Lemma~\ref{lemma}, the inequality (\ref{as322}) from
Assumption~3.2', and Assumptions~3.1 and~3.3, we get%
\begin{gather}
|E\sum_{|\mathbf{i}|+k=2q+2}\frac{1}{\mathbf{i}!k!}\frac{\partial ^{|\mathbf{%
i}|}}{\left( \partial x^{0}\right) ^{i_{0}}\cdots \left( \partial
x^{N}\right) ^{i_{N}}}\tilde{F}(t+h,x+\bar{\chi}(\bar{f}_{t,x}(t+h)-x);t^{%
\ast },T^{\ast })  \label{tn38} \\
\times \exp (-\bar{\theta}\bar{Y}_{t,x,0}(t+h))\left( \delta \bar{f}%
^{0}\right) ^{i_{0}}\cdots \left( \delta \bar{f}^{N}\right) ^{i_{N}}|\  
\notag \\
\leq K\exp (c\Delta |x|)\left( 1+\dsum\limits_{l=\ell (t)}^{\ell
(t+h)+\theta }|x^{l}|^{2q+2}\right) h^{q+1}.  \notag
\end{gather}

We obtain from (\ref{tn1})-(\ref{tn3}) and (\ref{tn37}), (\ref{tn38}): 
\begin{eqnarray}
|\rho (t,x)| &\leq &\sum_{k=0}^{2q}\left\vert \sum_{|\mathbf{i}%
|=1}^{2q+1-k}\mu _{k}^{\mathbf{i}}\frac{\partial ^{|\mathbf{i}|}}{\left(
\partial x^{0}\right) ^{i_{0}}\cdots \left( \partial x^{N}\right) ^{i_{N}}}%
\tilde{F}(t+h,x;t^{\ast },T^{\ast })\right\vert  \label{tn4} \\
&&+K\exp (c\Delta |x|)\left( 1+\dsum\limits_{l=\ell (t)}^{\ell (t+h)+\theta
}|x^{l}|^{2q+2}\right) h^{q+1},  \notag
\end{eqnarray}%
with $K>0$ and $c>0$ independent of $\Delta $ and 
\begin{equation*}
\mu _{k}^{\mathbf{i}}=\frac{1}{\mathbf{i}!k!}\left[ E\left( \delta \tilde{f}%
^{0}\right) ^{i_{0}}\cdots \left( \delta \tilde{f}^{N}\right) ^{i_{N}}\delta 
\tilde{Y}^{k}-E\left( \delta \bar{f}^{0}\right) ^{i_{0}}\cdots \left( \delta 
\bar{f}^{N}\right) ^{i_{N}}\delta \bar{Y}^{k}\right] .
\end{equation*}%
Applying Lemma~\ref{lemma} and using the inequality (\ref{as321}) from
Assumption~3.2', we obtain from (\ref{tn4}): 
\begin{equation}
|\rho (t,x)|\leq K\exp (c\Delta |x|)\left( 1+\dsum\limits_{l=\ell (t)}^{\ell
(t+h)+\theta }|x^{l}|^{2q+2}\right) h^{q+1},  \label{tn6}
\end{equation}%
where $K>0$ and $c>0$ do not depend on $\Delta $ and $x.$

Substituting (\ref{tn6}) in (\ref{tn0}) and using Assumptions~3.1 and~3.3
and the Cauchy-Bunyakovsky inequality, we arrive at the required (\ref{t6}).
Theorem~\ref{prpt1} is proved. \ $\square $

\begin{remark}
\label{rem:hlessdelt}As it follows from the proof, in Theorem~\ref{prpt1}
the condition $h\leq \alpha \Delta $ is used only for estimating the parts
of the error involving the approximate discounting factor $\tilde{Y}%
_{t,x,0}(s)$. If pricing an interest rate derivative does not require a
discounting factor (e.g., when one uses the forward measure pricing, cf.
Remark~\ref{rem:forwmeas}) then a theorem analogous to Theorem~\ref{prpt1}
can be proved under Assumptions~2.1-~2.3 and Assumptions~3.1 and~3.2 without
the restriction on $h.$
\end{remark}

Theorems~\ref{thm:weai2f}\ and~\ref{prpt1} imply the following result.

\begin{theorem}
\label{thm_t1}Under the conditions of Theorems~\ref{thm:msqi2f} and~\ref%
{prpt1}, the approximation $\bar{F}(t_{0},f_{0};$ $t^{\ast },T^{\ast })$
defined by $(\ref{2.13})$-$(\ref{2.17}),$ $(\ref{2.7}),$ $(\ref{t51})$
converges to $F(t_{0},f_{0}\left( \cdot \right) ;t^{\ast },T^{\ast })$ from $%
(\ref{7})$-$(\ref{9})$, $(\ref{3})$-$(\ref{3b})$ with order $p>0$ in $\Delta 
$ and with order $q>0$ in $h,$ i.e.,%
\begin{equation}
\left\vert F(t_{0},f_{0}\left( \cdot \right) ;t^{\ast },T^{\ast })-\bar{F}%
(t_{0},f_{0};t^{\ast },T^{\ast })\right\vert \leq K(\Delta ^{p}+h^{q}),
\label{t7}
\end{equation}%
where $K>0$ is a constant independent of $\Delta $ and $h.$
\end{theorem}

\begin{remark}
\label{rem:reldh}(Relationship between $\Delta $ and $h$) A higher order $p,$
i.e., a higher order of an approximation $\tilde{F}(t_{0},f_{0};t^{\ast
},T^{\ast })$ of $F(t_{0},f_{0}\left( \cdot \right) ;t^{\ast },T^{\ast }),$
can be achieved by using a higher-order quadrature rules in $(\ref{2.444})$
and $(\ref{2.8})$ and higher-order interpolation or extrapolation in $(\ref%
{2.10})$. For this purpose, we can use a large arsenal of effective
quadrature rules and interpolation/extrapolations methods from the
deterministic numerical analysis which are directly applicable here $($see
Section~\ref{sec:alg}$)$. To achieve a higher order $q,$ we need a
higher-order weak-sense numerical scheme for $(\ref{2.6})$-$(\ref{2.66})$.
As it is known $($see, e.g. \cite{MT1}$)$, this is a harder task, and, due
to complexity of stochastic schemes, one usually restricts themselves to
using weak methods of orders $1$ or $2.$ As a result, in practice we will
take $p\geq q.$ Then, to balance the two errors in $(\ref{t7})$, we choose $%
\Delta =\alpha h^{q/p}$ for some $\alpha >0$ to obtain the overall error to
be of order $O(h^{q}).$ In other words, by increasing the order $p$ we can
take larger $T$-discretization steps $\Delta $ and, consequently,
significantly improve computational efficiency of HJM simulation which, in
particular, is illustrated in our numerical experiments in Section~\ref%
{sec:exp}.
\end{remark}

According to the motivation examples considered in Section~\ref{sec:hjm},
the payoff $G(z)$ is usually globally Lipschitz (see (\ref{7a})) but not
sufficiently smooth function as it is required in Theorem~\ref{prpt1} and,
consequently, in Theorem~\ref{thm_t1}. Let us discuss two ways how one can
deal with this theoretical difficulty.

First, as it was noted in, e.g. \cite{MT2005}, we can approximate the payoff
function $G(z)$ by a smooth function $\breve{G}(z).$ Denote by $\varepsilon $
an error of this approximation. The proposed numerical method can be applied
to the smooth approximating function $\breve{G}(z)$ and Theorems~\ref{prpt1}
and~\ref{thm_t1} remain valid for $F$ with $\breve{G}$ instead of $G.$ In
this case, the overall error in evaluating the price of an interest rate
contract consists of the numerical integration errors estimated in Theorem~%
\ref{thm_t1} and the error $\varepsilon $ of the smoothening of $G.$

Second, one can exploit the result of \cite{BT} which in application to our
problem means that if the transition Markov function for the process $\tilde{%
f}(t)$ is sufficiently smooth and $\bar{f}_{k}^{i}$ is simulated by the
strong Euler scheme then $\bar{F}(t_{0},f_{0};t^{\ast },T^{\ast })$
converges to $\tilde{F}(t_{0},f_{0};t^{\ast },T^{\ast })$ with order one in $%
h$ even for nonsmooth $G.$

We remark that the computational practice (see our numerical experiments in
Section~\ref{sec:exp}) suggests that the error estimates of Theorems~\ref%
{prpt1}\ and~\ref{thm_t1} are valid for the weak Euler-type scheme (see (\ref%
{euler}) below) in the case of nonsmooth $G(z)$. Further, it is natural to
expect that for higher-order weak schemes the error estimates of Theorems~%
\ref{prpt1}\ and~\ref{thm_t1} are also valid for nonsmooth payoffs $G(z).$
We note that to answer on these theoretical questions related to
nonsmoothness of $G(z)$ further development of the general theory of
numerical integration of ordinary SDEs is required which is outside the
scope of the present paper.

\section{Numerical algorithms\label{sec:alg}}

In this section we provide some particular examples of the generic numerical
method introduced in Section~\ref{sec:met}. For simplicity of the
presentation, we restrict ourselves in this section to the case of $T$-step
being not larger than the $t$-step, i.e.,%
\begin{equation}
h\leq \Delta ,  \label{4.0}
\end{equation}%
which is a stronger condition than the one assumed in Theorems~\ref{prpt1}\
and~\ref{thm_t1}: $h\leq \alpha \Delta ,\ \alpha >0.$ This requirement is
not particular restricting since our aim is to construct efficient
algorithms for the HJM model by allowing bigger $T$-steps $\Delta $ without
losing accuracy as it was discussed in the Introduction and Remark~\ref%
{rem:reldh}. We note that there is no difficulty in constructing algorithms
imposing $h\leq \alpha \Delta $ for some $\alpha >0$ instead of (\ref{4.0}).
The condition (\ref{4.0}) ensures that there cannot be more than one node $%
T_{i}$ in any interval $[t_{k},t_{k+1})$ hence there are only two cases
possible: either $\ell _{k+1}=\ell _{k}$ or $\ell _{k+1}=\ell _{k}+1.$ This
is used in constructing numerical algorithms of this section.

We need the following new notation in this section:%
\begin{equation}
\Delta _{i,k}:=T_{i}-t_{k}  \label{4.1}
\end{equation}%
and 
\begin{equation*}
t_{k+1/2}=\frac{t_{k}+t_{k+1}}{2}\ .
\end{equation*}

In the paper we limit the illustration (see also Remark~\ref{rem:high}) of
the generic numerical method from Section~\ref{sec:met} to considering only
the weak Euler-type scheme (i.e., with $q=1)$ as a numerical approximation
of the SDEs (\ref{2.6}), (\ref{t5}), i.e., as an approximation of the $t$%
-dynamics. In this case the extended discretization (\ref{2.7}), (\ref{t51})
takes the form 
\begin{gather}
\bar{f}_{0}^{i}=f_{0}^{i},\text{ }i=0,\ldots ,N,\text{ \ }\bar{Y}_{0}=0,
\label{euler} \\
\bar{f}_{k+1}^{i}=\bar{f}_{k}^{i}+\tsum\limits_{j=1}^{d}\bar{\sigma}%
_{i,j}(t_{k})\mathbb{\bar{S}}_{I_{j}}(t_{k},T_{i};\Delta
,h)+h^{1/2}\tsum\limits_{j=1}^{d}\bar{\sigma}_{i,j}(t_{k})\xi _{j,k+1}, 
\notag \\
i=\ell _{k+1},\ldots ,N,  \notag \\
\bar{Y}_{k+1}=\bar{Y}_{k}+A^{Y}(t_{k};\bar{f}_{k}^{j},\text{ }j=\ell
_{k},\ldots ,\ell (t^{\ast })+\theta ;h),\text{ }k=0,\ldots ,M-1,  \notag
\end{gather}%
where $\xi _{j,k+1}$ are independent random variables distributed by the law 
$P(\xi =\pm 1)=1/2,$\ 
\begin{equation*}
(\bar{\sigma}_{i,1}(t_{k}),\ldots ,\bar{\sigma}_{i,d}(t_{k}))^{\top
}=(\sigma _{1}(t_{k},T_{i},\bar{f}_{k}^{i}),\ldots ,\sigma _{d}(t_{k},T_{i},%
\bar{f}_{k}^{i}))^{\top },
\end{equation*}%
$\mathbb{\bar{S}}_{I_{j}}(t_{k},T_{i};\Delta ,h)$ depends on our choice of
the quadrature rule (\ref{2.444}), and $A^{Y}$ is as in (\ref{t51}) and
depends on the choice of approximation for the short rate (\ref{2.10}).

In the remaining part of this section, we give three algorithms based on
rectangle ($p=1)$, trapezoid ($p=2)$, and Simpson ($p=4)$ quadrature rules $%
S_{I_{j}}(t_{k},T_{i},\Delta )$ accompanied by short rate approximations of
the corresponding orders. In all these cases it is not difficult to check
that (\ref{euler}) satisfies Assumption~3.1 and that $\bar{Y}_{k}$ satisfy
Assumption~3.3.

\subsection{Algorithm of order $O(\Delta +h)$\label{order1}}

The application of the composite rectangle rule to approximate the integrals 
$I_{j}(t_{k},T_{i})$ in (\ref{2.3}) and $Z(t^{\ast },T^{\ast })$ in (\ref{8}%
) yields 
\begin{eqnarray}
\ \ \ \ \ \mathbb{\bar{S}}_{I_{j}}(t_{k},T_{\ell _{k+1}};\Delta ,h)
&=&h\Delta _{\ell _{k+1},k}\bar{\sigma}_{\ell _{k+1},j}(t_{k}),
\label{4.1.1} \\
\mathbb{\bar{S}}_{I_{j}}(t_{k},T_{\varrho _{k+1}};\Delta ,h) &=&\left\{ 
\begin{array}{c}
h\Delta _{\varrho _{k+1},k}\bar{\sigma}_{\varrho _{k+1},j}(t_{k}),\text{ \
if\ }T_{\ell _{k+1}}<t_{k}, \\ 
\mathbb{\bar{S}}_{I_{j}}(t_{k},T_{\ell _{k+1}};\Delta ,h)+h\Delta \bar{\sigma%
}_{\varrho _{k+1},j}(t_{k}),\ \text{\ otherwise,}%
\end{array}%
\right.  \notag \\
\mathbb{\bar{S}}_{I_{j}}(t_{k},T_{i};\Delta ,h) &=&\mathbb{\bar{S}}%
_{I_{j}}(t_{k},T_{\varrho _{k+1}};\Delta ,h)+h\Delta \dsum_{m=\varrho
_{k+1}+1}^{i}\bar{\sigma}_{m,j}(t_{k}),\text{ }i=\varrho _{k+1}+1,\ldots ,N,
\notag \\
j &=&1,\ldots ,d,  \notag
\end{eqnarray}%
\begin{equation}
\bar{S}_{Z}(t^{\ast },T^{\ast },\Delta )=\bar{f}_{M}^{\varrho _{M}}\Delta
_{\varrho _{M},M}+\Delta \dsum_{m=\varrho _{M+1}}^{N}\bar{f}_{M}^{m}.
\label{4.1.2}
\end{equation}%
By straightforward calculations one can show that the used rectangle rule
satisfies the order conditions (\ref{2.5}) and (\ref{2.9}) with $p=1$. We
pay attention that we incorporated two cases in (\ref{4.1.1}): when $\ell
_{k+1}=\ell _{k}$ and hence $T_{\ell _{k+1}}\leq t_{k}$ and when (see also (%
\ref{4.0})) $\ell _{k+1}=\ell _{k}+1$ and hence $T_{\ell _{k+1}}>t_{k}.$

We use the piecewise approximation of the short rate (cf. (\ref{2.10})): 
\begin{equation}
\pi (t)=\dsum\limits_{l=0}^{\ell (t^{\ast })}f(t,T_{l})\chi _{t\in \lbrack
T_{l},T_{l+1})},\ \ t\in \lbrack t_{0},t^{\ast }].  \label{4.1.31}
\end{equation}%
The approximation (\ref{4.1.31}) obviously satisfies the order condition (%
\ref{2.11}) with $p=1$. To satisfy Assumption~3.2' with $q=1$, we, in
particular, need to approximate the integral $\tilde{Y}_{t,x,0}(t+h)$ in (%
\ref{nYY}) by $\bar{Y}_{t,x,0}(t+h)$ from (\ref{t51}) with local order $%
O(h^{2}).$ In the case of (\ref{4.1.31}) the coefficient in the right-hand
side of (\ref{t5}) $\pi (s;x^{i},\ i=\ell (s))=\dsum\limits_{l=0}^{\ell
(t^{\ast })}x^{l}\chi _{s\in \lbrack T_{l},T_{l+1})}=x^{\ell (s)}$ is only
piece-wise smooth. Further, according to the condition (\ref{4.0}), we can
have two cases: either an open interval $(t_{k},t_{k+1})$ does not contain
any node $T_{i}$ of the $T$-grid or it contains a single node $T_{\varrho
_{k}}.$ In the former case we can approximate the integral $\tilde{Y}%
_{t,x,0}(t+h)$ in (\ref{nYY}) by the left rectangle rule and we have $%
A^{Y}(t_{k};\bar{f}_{k}^{j},$ $j=\ell _{k},\ell _{k+1};h)=h\bar{f}_{k}^{\ell
_{k}}$ with the local error of order $O(h^{2})$ as needed. In the second
case to achieve the local error of order $O(h^{2})$ despite lack of
smoothness of $\pi (s;x^{i},\ i=\ell (s))$, we split the integral $\tilde{Y}%
_{t,x,0}(t+h)=\tilde{Y}_{t,x,0}(T_{\ell _{k}+1})+\tilde{Y}_{T_{\varrho
_{k}},f(T_{\ell _{k}+1}),0}(t+h)$ and approximate the first integral by the
left-rectangle rule and the second by the right-rectangle rule: $A^{Y}(t_{k};%
\bar{f}_{k}^{j},$ $j=\ell _{k},\ell _{k+1};h)=\Delta _{\ell _{k+1},k}\bar{f}%
_{k}^{\ell _{k}}-\Delta _{\ell _{k}+1,k+1}\bar{f}_{k+1}^{\ell _{k+1}}.$
Thus, 
\begin{equation}
A^{Y}(t_{k};\bar{f}_{k}^{j},j=\ell _{k},\ell _{k+1};h)=\left( h\wedge \Delta
_{\ell _{k+1},k}\right) \bar{f}_{k}^{\ell _{k}}-\left( 0\wedge \Delta _{\ell
_{k}+1,k+1}\right) \bar{f}_{k+1}^{\ell _{k+1}}.  \label{4.1.3}
\end{equation}%
We note that despite the use of $\bar{f}_{k+1}^{\ell _{k+1}}$ in the
right-hand side of (\ref{4.1.3}) the method does not require to resolve any
implicitness.

Assumption~3.2' with $q=1$ can be checked for the scheme (\ref{euler}), (\ref%
{4.1.1}), (\ref{4.1.3}) following the standard, routine way (see, e.g. \cite[%
Chap. 2]{MT1}).

The algorithm based on (\ref{euler}) and (\ref{4.1.1}), (\ref{4.1.2}), (\ref%
{4.1.3}), we will call \textbf{Algorithm~5.1} for the option price (\ref{7}%
)-(\ref{9}). According to Theorem~\ref{thm_t1}, this algorithm is of order $%
O(\Delta +h)$, which under the condition (see (\ref{4.0})) $\Delta =\alpha
h, $ $\alpha \geq 1,$ resulting in $O(h).$ We also note that in the case $%
\Delta =h$ the short rate is readily available on the grid and its
approximation is not needed. Algorithm~5.1 with $\Delta =h$ is analogous to
the numerical methods for the HJM model considered in \cite%
{HJM1990,Jarrow2002,Glasserman}. As it is shown in our numerical experiments
(see Section~\ref{sec:exp}), Algorithm~5.1 is less efficient than the new
algorithms (Algorithms~5.2 and~5.3) which we propose in the next two
sections.

\begin{remark}
If we replace $A^{Y}$ in $(\ref{4.1.3})$ by 
\begin{equation}
A^{Y}(t_{k};h)=h\bar{f}_{k}^{\ell _{k}}  \label{roughY}
\end{equation}%
then Assumption~3.2' with $q=1$ is not satisfied and we cannot guarantee
closeness of $\bar{Y}_{k}$ and $\tilde{Y}(t_{k}).$ Nevertheless, $\bar{Y}%
_{k} $ from $(\ref{roughY})$ still apparently approximates $Y(t_{k})$ so
that the overall algorithm for computing the option price $(\ref{7})$-$(\ref%
{9})$ remains of weak order $O(\Delta +h).$ This can be justified by some
nonrigorous arguments and this was also demonstrated in our numerical
experiments. To obtain such a result rigorously, we need to conduct
convergence proof without using the intermediate finite-dimensional SDEs $(%
\ref{2.6}),$ $(\ref{t5})$. We do not pursue this direction in the paper. At
the same time, we note that in all our numerical tests the scheme using $%
A^{Y}$ from $(\ref{4.1.3})$ gave more accurate results than the scheme with $%
A^{Y}$ from $(\ref{roughY})$ in the cases when the $T_{i}$ nodes do not
belong to the $t$-grid. Otherwise $A^{Y}$ in $(\ref{4.1.3})$ and $A^{Y}$ in $%
(\ref{roughY})$ obviously coincide.
\end{remark}

\subsection{Algorithm of order $O(\Delta ^{2}+h)$\label{order2}}

In this section we use quadrature rules (\ref{2.444}), (\ref{2.8}) and a
short rate approximation (\ref{2.10}) of order $O(\Delta ^{2}).$

We aim at applying the standard composite trapezoid rule to the integrals $%
I_{j}(s,T_{i})$ in (\ref{2.3}) and (\ref{2.5a}). The trapezoid rule requires
that each of the integration subintervals $[T_{\ell (s)},s],$ $[s,T_{\varrho
_{(s)}}],$ $[T_{\varrho _{(s)}},T_{\varrho _{(s)}+1}],$ $\ldots ,$ $%
[T_{i-1},T_{i}]$ span at least two nodes on the $T$-grid. However, the
integration intervals $[T_{\ell (s)},s]\ $and $[s,T_{\varrho _{(s)}}]$
usually contain just a single node on the $T$-grid: $T_{\ell (s)}$ and $%
T_{\varrho (s)},$ respectively. We resolve this issue by applying the right
and left rectangle rules on these two intervals, respectively. Thus, the
quadrature rule $S_{I_{j}}(s,T_{i},\Delta )$ takes the form for $s\in
\lbrack t_{0},t^{\ast }],$ $i=\ell (s),\ldots ,T^{\ast }:$ 
\begin{eqnarray}
S_{I_{j}}(s,T_{\ell (s)},\Delta )=\left( T_{\ell (s)}-s\right) \sigma
_{j}(s,T_{\ell (s)}),  \label{puretrap} \\
S_{I_{j}}(s,T_{i},\Delta )=(T_{\varrho (s)}-s)\sigma _{j}(s,T_{\varrho (s)})+%
\frac{\Delta }{2}\dsum_{m=\varrho (s)}^{i-1}\left[ \sigma
_{j}(s,T_{m})+\sigma _{j}(s,T_{m+1})\right] \ \ \text{for }s\leq T_{i}, 
\notag \\
j=1,\ldots ,d.  \notag
\end{eqnarray}%
This quadrature rule satisfies the order condition (\ref{2.5}) with $p=2$.
To this end, we recall that left and right rectangle rules have local order
two and we use them here on one or two integration steps only while the
trapezoid rule has local order three and the composite trapezoid rule is of
order two.

To ensure that (\ref{euler}) satisfies Assumption~3.2' with $q=1$, we, in
particular, need to approximate the integral $\int_{t_{k}}^{t_{k+1}}\tilde{S}%
_{I_{j}}(s,T_{i},\Delta )ds$ by $\mathbb{\bar{S}}_{I_{j}}(t_{k},T_{i};\Delta
,h)$ on a single step with weak order $O(h^{2}).$ If the node $T_{\ell
_{k+1}}$ is not between $t_{k}$ and $t_{k+1}$ (due to (\ref{4.0}) it cannot
be more than one $T$-node in $(t_{k},t_{k+1}))$, it is sufficient to
approximate the integral by the left rectangle rule and put $\mathbb{\bar{S}}%
_{I_{j}}(t_{k},T_{i};\Delta ,h)=h\bar{S}_{I_{j}}(t_{k},T_{i},\Delta ),$
where $\bar{S}_{I_{j}}(t_{k},T_{i},\Delta )$ is of the form (\ref{puretrap})
but with $\bar{\sigma}_{m,j}(t_{k})$ instead of $\sigma _{j}(s,T_{m}).$
However, if $T_{\ell _{k+1}}>t_{k}$ then due to (\ref{puretrap}) we apply
one integration rule on $[t_{k},T_{\ell _{k+1}}]$ and the other on $[T_{\ell
_{k+1}},T_{\varrho _{k+1}}],$ which causes loss of smoothness of the
integrand $\tilde{S}_{I_{j}}(s,T_{i},\Delta ).$ To reach the required order $%
O(h^{2}),$ we construct the approximation using the following guidance:%
\begin{eqnarray}
\ \ \ \ \ \ \ \ \ \ \ \ \ \int_{t_{k}}^{t_{k+1}}\tilde{S}_{I_{j}}(s,T_{i},%
\Delta )ds &=&\int_{t_{k}}^{T_{\ell _{k+1}}}\tilde{S}_{I_{j}}(s,T_{i},\Delta
)ds+\int_{T_{\ell _{k+1}}}^{t_{k+1}}\tilde{S}_{I_{j}}(s,T_{i},\Delta )ds
\label{5.1ex} \\
&=&\int_{t_{k}}^{T_{\ell _{k+1}}}\left[ (T_{\ell _{k+1}}-s)\tilde{\sigma}%
_{\ell _{k+1},j}(s)+\frac{\Delta }{2}\tilde{\sigma}_{\ell _{k+1},j}(s)+\frac{%
\Delta }{2}\tilde{\sigma}_{\varrho _{k+1},j}(s)\right] ds  \notag \\
&&+\int_{T_{\ell _{k+1}}}^{t_{k+1}}(T_{\varrho _{k+1}}-s)\tilde{\sigma}%
_{\varrho _{k+1},j}(s)  \notag \\
&&+\frac{\Delta }{2}\int_{t_{k}}^{t_{k+1}}\dsum_{m=\varrho _{k+1}}^{i-1}%
\left[ \tilde{\sigma}_{m,j}(s)+\tilde{\sigma}_{m+1,j}(s)\right]  \notag \\
&\approx &(T_{\ell _{k+1}}-t_{k})\frac{T_{\ell _{k+1}}-t_{k}+\Delta }{2}%
\tilde{\sigma}_{\ell _{k+1},j}(t_{k})  \notag \\
&&+\frac{\Delta }{2}[t_{k+2}-T_{\ell _{k+1}}]\tilde{\sigma}_{\varrho
_{k+1},j}(t_{k})  \notag \\
&&+h\frac{\Delta }{2}\dsum_{m=\varrho _{k+1}}^{i-1}\left[ \tilde{\sigma}%
_{m,j}(t_{k}) +\tilde{\sigma}_{m+1,j}(t_{k})\right] .  \notag
\end{eqnarray}%
As a result $\mathbb{\bar{S}}_{I_{j}}(t_{k},T_{i},\Delta )$ in (\ref{euler})
is taken of the form: 
\begin{gather}
\mathbb{\bar{S}}_{I_{j}}(t_{k},T_{\ell _{k+1}};\Delta ,h)=h\Delta _{\ell
_{k+1},k}\bar{\sigma}_{\ell _{k+1},j}(t_{k}),  \label{5.1} \\
\mathbb{\bar{S}}_{I_{j}}(t_{k},T_{\varrho _{k+1}};\Delta ,h)=\left\{ 
\begin{array}{l}
h\Delta _{\varrho _{k+1},k}\bar{\sigma}_{\varrho _{k+1},j}(t_{k}),\text{ \
if\ }T_{\ell _{k+1}}\leq t_{k}, \\ 
\Delta _{\ell _{k+1},k}\ \frac{\Delta _{\varrho _{k+1},k}}{2}\bar{\sigma}%
_{\ell _{k+1},j}(t_{k})-\Delta _{\ell _{k+1},k+2}\ \frac{\Delta }{2}\bar{%
\sigma}_{\varrho _{k+1},j}(t_{k}),\text{\ otherwise,}%
\end{array}%
\right.  \notag \\
\mathbb{\bar{S}}_{I_{j}}(t_{k},T_{i};\Delta ,h)=\mathbb{\bar{S}}%
_{I_{j}}(t_{k},T_{\varrho _{k+1}};\Delta ,h)+h\frac{\Delta }{2}\left( \bar{%
\sigma}_{\varrho _{k+1},j}(t_{k})+2\sum_{m=\varrho _{k+1}+1}^{i-1}\bar{\sigma%
}_{m,j}(t_{k})+\bar{\sigma}_{i,j}(t_{k})\right) ,  \notag \\
i=\varrho _{k+1}+1,\ldots ,N,\ \ j=1,\ldots ,d.  \notag
\end{gather}

By a similar reasoning used to derive (\ref{puretrap}), we obtain the
corresponding quadrature rule $S_{Z}(t^{\ast },T^{\ast },\Delta )$ (see (\ref%
{2.8})). Namely, we apply the right-rectangle rule on the integration
interval $\left[ t_{M},T_{\varrho _{M}}\right] $ and the composite trapezoid
rule on the rest of the integration interval, i.e.,%
\begin{equation}
\bar{S}_{Z}(t^{\ast },T^{\ast },\Delta )=\bar{f}_{M}^{\varrho _{M}}\Delta
_{\varrho _{M},M}+\frac{\Delta }{2}\left( \bar{f}_{M}^{\varrho
_{M}}+2\sum_{j=\varrho _{M}+1}^{N-1}\bar{f}_{M}^{j}+\bar{f}_{M}^{N}\right) .
\label{5.4}
\end{equation}%
It is not difficult to show that the combination of rectangle and trapezoid
rules used for deriving (\ref{5.4}) satisfies the order condition (\ref{2.9}%
) with $p=2$.

We use linear interpolation for the short rate in (\ref{2.10}): 
\begin{equation}
\pi (t)=\dsum\limits_{l=0}^{\ell (t^{\ast })}\left[ \frac{t-T_{l}}{\Delta }%
f(t,T_{l+1})+\frac{T_{l+1}-t}{\Delta }f(t,T_{l})\right] \chi _{t\in \lbrack
T_{l},T_{l+1})},\ t\in \lbrack t_{0},t^{\ast }].  \label{5.50}
\end{equation}%
The approximation (\ref{5.50}) obviously satisfies the order condition (\ref%
{2.11}) with $p=2$. As in the case of Algorithm~5.1, the coefficient in the
right-hand side of (\ref{t5}) is also only piece-wise smooth here. Consider
first the case when the node $T_{\ell _{k+1}}$ is not between $t_{k}$ and $%
t_{k+1}.$ The application of the left-rectangle rule to the integral $%
\int_{t_{k}}^{t_{k+1}}\tilde{\pi}(s)ds$ has the error $O(h^{2}/\Delta ),$
i.e., it does not lead to a uniform error estimate $O(h^{2})$ required by
Assumption~3.2'. To ensure that the estimates $O(h^{2})$ in Assumptions~3.2'
are uniform in $\Delta ,$ we use the following guidance: 
\begin{eqnarray*}
\int_{t_{k}}^{t_{k+1}}\tilde{\pi}(s)ds &=&\int_{t_{k}}^{t_{k+1}}\left[ \frac{%
s-T_{\ell _{k+1}}}{\Delta }\tilde{f}^{\varrho _{k+1}}(s)+\frac{T_{\varrho
_{k+1}}-s}{\Delta }\tilde{f}^{\ell _{k+1}}(s)\right] ds \\
&\approx &\tilde{f}^{\varrho _{k+1}}(t_{k})\int_{t_{k}}^{t_{k+1}}\frac{%
s-T_{\ell _{k+1}}}{\Delta }ds+\tilde{f}^{\ell
_{k+1}}(t_{k})\int_{t_{k}}^{t_{k+1}}\frac{T_{\varrho _{k+1}}-s}{\Delta }ds \\
&=&\tilde{f}^{\varrho _{k+1}}(t_{k})h\frac{t_{k+1/2}-T_{\ell _{k+1}}}{\Delta 
}+\tilde{f}^{\ell _{k+1}}(t_{k})h\frac{T_{\varrho _{k+1}}-t_{k+1/2}}{\Delta }%
.
\end{eqnarray*}%
So, in this case we put $A^{Y}(t_{k};h)=h\left[ \frac{\Delta _{\rho
_{k+1},k+1/2}}{\Delta }\bar{f}_{k}^{\ell _{k+1}}-\frac{\Delta _{\ell
_{k+1},k+1/2}}{\Delta }\bar{f}_{k}^{\varrho _{k+1}}\right] .$ In the other
case, i.e., if $T_{\ell _{k+1}}>t_{k},$ we split the integral $%
\int_{t_{k}}^{t_{k+1}}\tilde{\pi}(s)ds=\int_{t_{k}}^{T_{\ell _{k+1}}}\tilde{%
\pi}(s)ds+\int_{T_{\ell _{k+1}}}^{t_{k+1}}\tilde{\pi}(s)ds$ and approximate
each of them separately as we did in constructing (\ref{4.1.3}). As a
result, we arrive at%
\begin{equation}
A^{Y}(t_{k};h)=\left\{ 
\begin{array}{l}
h\left[ \frac{\Delta _{\rho _{k+1},k+1/2}}{\Delta }\bar{f}_{k}^{\ell _{k+1}}-%
\frac{\Delta _{\ell _{k+1},k+1/2}}{\Delta }\bar{f}_{k}^{\varrho _{k+1}}%
\right] \text{ \ \ if \ }T_{\ell _{k+1}}\leq t_{k}, \\ 
\Delta _{\ell _{k+1},k}\left[ \frac{\Delta _{\ell _{k+1},k}}{2\Delta }\bar{f}%
_{k}^{\ell _{k}}-\frac{\Delta _{\ell _{k}-1,k}}{2\Delta }\bar{f}_{k}^{\ell
_{k+1}}\right] \ \  \\ 
-\Delta _{\ell _{k+1},k+1}\left[ \frac{\Delta _{\varrho _{k+1}+1,k+1}}{%
2\Delta }\bar{f}_{k+1}^{\ell _{k+1}}-\frac{\Delta _{\ell _{k+1},k+1}}{%
2\Delta }\bar{f}_{k+1}^{\varrho _{k+1}}\right] \ \ \text{otherwise.}%
\end{array}%
\right.  \label{5.5}
\end{equation}

Assumption~3.2' with $q=1$ can be checked for the scheme (\ref{euler}), (\ref%
{5.1}), (\ref{5.5}) following the standard way.

The algorithm based on (\ref{euler}) and (\ref{5.1}), (\ref{5.4}), (\ref{5.5}%
) we will call \textbf{Algorithm~5.2} for the option price (\ref{7})-(\ref{9}%
). According to Theorem~\ref{thm_t1}, this algorithm is of order $O(\Delta
^{2}+h).$ In practice (see Remark~\ref{rem:reldh}) we choose $\Delta =\alpha 
\sqrt{h}$ with $\alpha >0$ such that (\ref{4.0}) is satisfied, which results
in the algorithm's accuracy $O(h).$ In our experiments (see Section~\ref%
{sec:exp}) Algorithm~5.2 outperformed Algorithm~5.1.

\subsection{Algorithm of order $O(\Delta ^{4}+h)$\label{order4}}

At the beginning of Section~\ref{sec:Tdis} we made the assumption that there
is a sufficient number of nodes $T_{i}$ between $t^{\ast }$ and $T^{\ast }\ $%
which ensures that we have enough nodes on the $T$-grid for using the
quadrature rules (\ref{2.444}) and (\ref{2.8}) and the short rate
approximations (\ref{2.10}) of the required accuracy. This assumption gives
an unnecessary restriction for using higher-order algorithms in practice and
we now demonstrate how it can be relaxed. To this end, we introduce $%
N^{^{\prime }}$ instead of $N$ in the method (\ref{2.7}) as the number of
discretization nodes on $T$-grid: 
\begin{equation}
N^{^{\prime }}:=N\vee \max_{0<i\leq N}\kappa (t^{\ast },T_{i})\vee \left(
\ell (t^{\ast })+\theta \right) ,  \label{Nprime}
\end{equation}%
where $\kappa (t^{\ast },T_{i})$ and $\theta $ are as in (\ref{2.444}) and (%
\ref{2.10}), respectively. Also, in (\ref{2.8}) we can put $N^{^{\prime }}$
instead of $N$ and if required increase $N^{^{\prime }}$ further to be able
to approximate the integral $Z(t^{\ast },T^{\ast })$ on the left-hand side
of (\ref{2.8}) with the prescribed accuracy. As a result, we avoid the
restriction on how close $t^{\ast }$ can be to $T^{\ast }.$ It is clear that
this extension of the $T$-grid by a fixed number of nodes in the case of
large $\Delta $ does not influence our theoretical results.

Without re-writing the Euler-type scheme (\ref{euler}), we will assume in
this section that we run it for $i=\ell _{k+1},\ldots ,N^{^{\prime }}$
instead of $i=\ell _{k+1},\ldots ,N.$

We are aiming at constructing an algorithm of order $O(\Delta ^{4}+h)$ and
would like to exploit the standard composite Simpson rule for approximation
of the integrals $I_{j}(s,T_{i})=\int_{s}^{T_{i}}\sigma _{j}(s,u)du$ from (%
\ref{2.3}) and (\ref{2.5a}). The Simpson rule needs three nodes per
integration step. But the integrals $I_{j}(s,T_{\ell (s)}),$ $%
I_{j}(s,T_{\varrho (s)}),$ and $I_{j}(s,T_{\varrho (s)+1})$ are over the
intervals which have just one or two nodes on the $T$-grid under (\ref{4.0}).

We first consider the integrals $I_{j}(s,T_{i})=\int_{s}^{T_{i}}\sigma
_{j}(s,u)du$ with $T_{i}=T_{\ell (s)},$ $T_{\varrho (s)},$ and $T_{\varrho
(s)+1},$ which we approximate by quadrature rules $S_{I_{j}}(s,T_{i},\Delta
) $ of the form%
\begin{equation}
S_{I_{j}}(s,T_{i},\Delta )=\left( T_{i}-s\right) \left[ \beta _{1}^{i}\sigma
_{j}(s,T_{\ell (s)})+\beta _{2}^{i}\sigma _{j}(s,T_{\varrho (s)})+\beta
_{3}^{i}\sigma _{j}(s,T_{\varrho (s)+1})\right] ,\text{ }  \label{4.3}
\end{equation}%
where the coefficients $\beta _{1}^{i},$ $\beta _{2}^{i}$, $\beta _{3}^{i}$
depend on the value of $T_{i}.$ We require that (\ref{4.3}) is of order $4,$
i.e., that (\ref{2.5}) is satisfied for these three integrals with $p=4.$
One can show that the following sets of coefficients satisfy this order
requirement:%
\begin{gather}
\beta _{1}^{\ell (s)}=\frac{5}{12}+\frac{5}{12}\frac{T_{\varrho (s)}-s}{%
\Delta }+\frac{1}{6}\frac{\left( T_{\varrho (s)}-s\right) ^{2}}{\Delta ^{2}}%
,\   \label{4.6} \\
\beta _{2}^{\ell (s)}=\frac{2}{3}-\frac{1}{3}\frac{T_{\varrho (s)}-s}{\Delta 
}-\frac{1}{3}\frac{\left( T_{\varrho (s)}-s\right) ^{2}}{\Delta ^{2}},\ \
\beta _{3}^{\ell (s)}=-\frac{1}{12}-\frac{1}{12}\frac{T_{\varrho (s)}-s}{%
\Delta }+\frac{1}{6}\frac{\left( T_{\varrho (s)}-s\right) ^{2}}{\Delta ^{2}};
\notag
\end{gather}%
\begin{eqnarray}
\beta _{1}^{\varrho (s)} &=&\frac{1}{4}\frac{T_{\varrho (s)}-s}{\Delta }+%
\frac{1}{6}\frac{\left( T_{\varrho (s)}-s\right) ^{2}}{\Delta ^{2}},\ \beta
_{2}^{\varrho (s)}=1-\frac{1}{3}\frac{\left( T_{\varrho (s)}-s\right) ^{2}}{%
\Delta ^{2}},\   \label{4.4} \\
\beta _{3}^{\varrho (s)} &=&-\frac{1}{4}\frac{T_{\varrho (s)}-s}{\Delta }+%
\frac{1}{6}\frac{\left( T_{\varrho (s)}-s\right) ^{2}}{\Delta ^{2}};  \notag
\end{eqnarray}%
\begin{gather}
\beta _{1}^{\varrho (s)+1}=-\frac{1}{12}+\frac{1}{12}\frac{T_{\varrho (s)}-s%
}{\Delta }+\frac{1}{6}\frac{\left( T_{\varrho (s)}-s\right) ^{2}}{\Delta ^{2}%
};\   \label{4.5} \\
\beta _{2}^{\varrho (s)+1}=\frac{2}{3}+\frac{1}{3}\frac{T_{\varrho (s)}-s}{%
\Delta }-\frac{1}{3}\frac{\left( T_{\varrho (s)}-s\right) ^{2}}{\Delta ^{2}}%
;\ \beta _{3}^{\varrho (s)+1}=\frac{5}{12}-\frac{5}{12}\frac{T_{\varrho
(s)}-s}{\Delta }+\frac{1}{6}\frac{\left( T_{\varrho (s)}-s\right) ^{2}}{%
\Delta ^{2}}.  \notag
\end{gather}

Further, for $\varrho (s)+1<i\leq N$ we write $I_{j}(s,T_{i})=I_{j}(s,T_{%
\varrho (s)})+I_{j}(T_{\varrho (s)},T_{i};s)$ with $I_{j}(T_{\varrho
(s)},T_{i};s):=\int_{T_{\varrho (s)}}^{T_{i}}\sigma _{j}(s,u)du;$ and we
approximate the integral $I_{j}(T_{\varrho (s)},T_{i};s),\ \varrho
(s)+1<i\leq N^{^{\prime }},$ by the composite Simpson rule $%
S_{I_{j}}(T_{\varrho (s)},T_{i},\Delta ;s)$ if its integration interval
spans an odd number of maturity time nodes: 
\begin{eqnarray}
S_{I_{j}}(T_{\varrho (s)},T_{i},\Delta ;s) &=&\frac{\Delta }{3}\left( \sigma
_{j}(s,T_{\varrho (s)})+2\tsum\limits_{l=1}^{(i-\varrho (s))/2-1}\sigma
_{j}(s,T_{\varrho (s)+2l})\right.  \label{4.7} \\
&&\left. +4\tsum\limits_{l=1}^{(i-\varrho (s))/2}\sigma _{j}(s,T_{\varrho
(s)+2l-1})+\sigma _{j}(s,T_{i})\right) ,  \notag
\end{eqnarray}%
and otherwise we apply the Simpson's 3/8 rule for the last four nodes:%
\begin{eqnarray}
S_{I_{j}}(T_{\varrho (s)},T_{i},\Delta ;s) &=&\frac{\Delta }{3}\left( \sigma
_{j}(s,T_{\varrho (s)})+2\tsum\limits_{l=1}^{(i-\varrho (s)-1)/2-2}\sigma
_{j}(s,T_{\varrho (s)+2l})\right.  \notag \\
&&\left. +4\tsum\limits_{l=1}^{(i-\varrho (s)-1)/2-1}\sigma
_{j}(s,T_{\varrho (s)+2l-1})+\sigma _{j}(s,T_{i-3})\right)  \label{4.7a} \\
&&+\frac{3\Delta }{8}\left( \sigma _{j}(s,T_{i-3})+3\sigma
_{j}(s,T_{i-2})+3\sigma _{j}(s,T_{i-1})+\sigma _{j}(s,T_{i})\right) .  \notag
\end{eqnarray}%
By straightforward calculations one can show that the quadrature rule (\ref%
{4.3}), (\ref{4.7}), and (\ref{4.7a}) satisfies the order condition (\ref%
{2.5}) with $p=4.$

To obtain $\mathbb{\bar{S}}_{I_{j}}(t_{k},T_{i};\Delta ,h)$ based on (\ref%
{4.3}), (\ref{4.7}), and (\ref{4.7a}), we need again to consider the two
cases: when $\ell _{k+1}=\ell _{k}$ and hence $T_{\ell _{k+1}}\leq t_{k}$
and when (see also (\ref{4.0})) $\ell _{k+1}=\ell _{k}+1$ and hence $T_{\ell
_{k+1}}>t_{k}.$ If $\ell _{k+1}=\ell _{k},$ then we put $\mathbb{\bar{S}}%
_{I_{j}}(t_{k},T_{i};\Delta ,h):=h\bar{S}_{I_{j}}(t_{k},T_{i},\Delta ;t_{k})$
with $\bar{S}_{I_{j}}(t_{k},T_{i},\Delta ;t_{k})$ having the form (\ref{4.3}%
), (\ref{4.7})-(\ref{4.7a}) with $\bar{\sigma}_{l,j}(t_{k})$ instead of $%
\sigma _{j}(t_{k},T_{l}).$ Otherwise, we split the integral 
\begin{equation*}
\int_{t_{k}}^{t_{k+1}}\tilde{S}_{I_{j}}(s,T_{i},\Delta
)ds=\int_{t_{k}}^{T_{\ell _{k+1}}}\tilde{S}_{I_{j}}(s,T_{i},\Delta
)ds+\int_{T_{\ell _{k+1}}}^{t_{k+1}}\tilde{S}_{I_{j}}(s,T_{i},\Delta )ds
\end{equation*}%
and approximate each of them to obtain the required $\mathbb{\bar{S}}%
_{I_{j}}(t_{k},T_{i};\Delta ,h)$ analogously to how we have proceeded in
constructing Algorithm~5.2. We do not write the corresponding expressions of 
$\mathbb{\bar{S}}_{I_{j}}(t_{k},T_{i};\Delta ,h)$ here though there is no
difficulty to restore them.

Using (\ref{4.3}), (\ref{4.7}), and (\ref{4.7a}), we construct the
quadrature rule $S_{Z}(t^{\ast },T^{\ast },\Delta )$ (see (\ref{2.8})) and
arrive at 
\begin{equation}
\bar{S}_{Z}(t^{\ast },T_{i},\Delta )=\Delta _{i,M}\left[ \beta _{1}^{i}\bar{f%
}_{M}^{\ell _{M}}+\beta _{2}^{i}\bar{f}_{M}^{\varrho _{M}}+\beta _{3}^{i}%
\bar{f}_{M}^{\varrho _{M}+1}\right] ;\text{ }  \label{4.91}
\end{equation}%
\begin{equation}
\bar{S}_{Z}(T_{\varrho _{M}},T^{\ast },\Delta )=\frac{\Delta }{3}\left( \bar{%
f}_{M}^{\varrho _{M}}+2\tsum\limits_{l=1}^{(N-\varrho _{M}-1)/2}\bar{f}%
_{M}^{\varrho _{M}+2l}+4\tsum\limits_{l=1}^{(N-\varrho _{M}+1)/2}\bar{f}%
_{M}^{\varrho _{M}+2l-1}+\bar{f}_{M}^{N}\right) ;  \label{4.92}
\end{equation}%
\begin{eqnarray}
\bar{S}_{Z}(T_{\varrho _{M}},T^{\ast },\Delta ) &=&\frac{\Delta }{3}\left( 
\bar{f}_{M}^{\varrho _{M}}+2\tsum\limits_{l=1}^{(N-\varrho _{M})/2-2}\bar{f}%
_{M}^{\varrho _{M}+2l}+4\tsum\limits_{l=1}^{(N-\varrho _{M})/2-1}\bar{f}%
_{M}^{\varrho _{M}+2l-1}+\bar{f}_{M}^{N-3}\right)  \notag \\
&&+\frac{3\Delta }{8}\left( \bar{f}_{M}^{N-3}+3\bar{f}_{M}^{N-2}+3\bar{f}%
_{M}^{N-1}+\bar{f}_{M}^{N}\right) .  \label{4.93}
\end{eqnarray}%
Then we define $\bar{S}_{Z}(t^{\ast },T^{\ast },\Delta )$ to be used in the
algorithm as 
\begin{equation}
\bar{S}_{Z}(t^{\ast },T^{\ast },\Delta )=\left\{ 
\begin{array}{l}
\text{(\ref{4.91}), (\ref{4.4}) with }i=\varrho _{M}\ \ \text{if }N=\varrho
_{M}, \\ 
\text{(\ref{4.91}), (\ref{4.5}) with }i=\varrho _{M}+1\ \ \text{if }%
N=\varrho _{M}+1, \\ 
\bar{S}_{Z}(t_{M},T_{\varrho _{M}},\Delta )+\bar{S}_{Z}(T_{\varrho
_{M}},T_{N},\Delta )\ \ \text{if }N>\varrho _{M}+1,%
\end{array}%
\right.  \label{4.9}
\end{equation}%
where $\bar{S}_{Z}(t_{M},T_{\varrho _{M}},\Delta )$ is from (\ref{4.91}), (%
\ref{4.5}) with $i=\varrho _{M}$ and 
\begin{equation*}
\bar{S}_{Z}(T_{\varrho _{M}},T_{N},\Delta )=\left\{ 
\begin{array}{l}
\text{(\ref{4.92}) if }N-\varrho _{M}+1\text{ is odd,} \\ 
\text{(\ref{4.93}) if }N-\varrho _{M}+1\text{ is even.}%
\end{array}%
\right.
\end{equation*}%
It is not difficult to show that the quadrature rules used for deriving (\ref%
{4.9}) satisfy the order condition (\ref{2.9}) with $p=4$.

For the short rate approximation $\pi (t)$ (see (\ref{2.10})), we use cubic
polynomial interpolation which obviously satisfies the order condition (\ref%
{2.11}) with $p=4$:%
\begin{equation}
\pi (t)=\dsum\limits_{j=0}^{3}L_{j}(t)f(t,T_{\ell (t)+j}),  \label{4.10}
\end{equation}%
where 
\begin{equation*}
L_{j}(t)=\dprod\limits_{\substack{ i=0  \\ i\neq j}}^{3}\frac{t-T_{\ell
(t)+i}}{T_{\ell (t)+j}-T_{\ell (t)+i}}.
\end{equation*}%
To obtain the corresponding $A^{Y}(t_{k};h)=A^{Y}(t_{k};\bar{f}%
_{k}^{j},j=\ell _{k},\ldots ,\ell _{k+1}+3;h)$, we follow similar guidance
as the one used to obtain (\ref{5.5}). We do not write the expression of $%
A^{Y}(t_{k};h)$ here but there is no difficulty to restore it.

The algorithm presented in this section, we will call \textbf{Algorithm~5.3}
for the option price (\ref{7})-(\ref{9}). Assumption~3.2' with $q=1$ can be
checked for this algorithm following the standard way. According to Theorem~%
\ref{thm_t1}, Algorithm~5.3 is of order $O(\Delta ^{4}+h).$ In practice (see
Remark~\ref{rem:reldh}) we will choose $\Delta =\alpha \sqrt[4]{h}$ with $%
\alpha >0$ such that (\ref{4.0}) holds, which results in the algorithm's
accuracy $O(h).$

\begin{remark}
\label{algcomplexity}$($Complexity of the algorithms$)$ Let us estimate
computational complexity of the algorithms considered in this section. The
number of operations in these algorithms is of order $O(MN).$ Then running
times of Algorithm~5.1 with $\Delta =\alpha h$, Algorithm~5.2 with $\Delta
=\alpha \sqrt{h}$, and\ Algorithm~5.3 with $\Delta =\alpha \sqrt[4]{h}$ are
proportional to $M^{2},$ $M\sqrt{M},$ and $M\sqrt[4]{M},$ respectively.
Also, it should be taken into account that Algorithms~5.2 and~5.3 require
approximately twice and four times number of operations per $t$-step,
respectively, than Algorithm~5.1. Hence one can expect that in reaching a
similar accuracy Algorithm~5.3 is approximately $M^{3/4}/4$ faster than
Algorithm~5.1 and Algorithm~5.2 is $\sqrt{M}/2$ faster than Algorithm~5.1.
This is confirmed in our numerical experiments $($see Section~\ref{sec:exp}$%
) $.
\end{remark}

\begin{remark}
\label{rem:high}If in Algorithms~5.2 and~5.3 we substitute the Euler scheme $%
(\ref{euler})$ by a second-order $($i.e., $q=2)$ weak scheme $($see examples
of such schemes in, e.g. \cite{MT1}$),$ then these modified algorithms (they
should satisfy Assumption~3.2' with $q=2$) will become of order $%
O(h^{2}+\Delta ^{2})$ and $O(h^{2}+\Delta ^{4}),$ respectively. Choosing $%
\Delta =h$ and $\Delta =\alpha \sqrt{h}$ in the modified Algorithms~5.2
and~5.3, respectively, their accuracy becomes of order $O(h^{2}).$
\end{remark}

\section{Mean-square method and its convergence\label{sec:msq}}

In most of the financial applications weak numerical methods, which we have
considered in the previous sections, are sufficient. At the same time,
mean-square methods can be useful for simulating scenarios. Also,
mean-square convergence of fully discrete approximations for the HJM model
is of theoretical interest. In this section we consider a mean-square method
for (\ref{2.6})-(\ref{2.66}) and prove its convergence.

We consider an approximation $\bar{f}_{k+1}^{i}$ of $\tilde{f}^{i}(t_{k+1})$
from (\ref{2.6}) (i.e., a full discretization of (\ref{3})-(\ref{3b}) in
both $T$ and $t)$ of the form%
\begin{gather}
\bar{f}_{0}^{i}=f_{0}^{i},\text{ }i=0,\ldots ,N,  \label{msq1} \\
\bar{f}_{k+1}^{i}=\bar{f}_{k}^{i}+  \notag \\
+A^{i}(t_{k},T_{i};\bar{f}_{k}^{j},\text{ }j=\ell _{k+1},\ldots ,\kappa
(t_{k+1},T_{i})\vee i;h;W_{l}(s)-W_{l}(t_{k}),\;l=1,\ldots ,d,\;t_{k}\leq
s\leq t_{k+1}),  \notag \\
i=\ell _{k+1},\ldots ,N,\ k=0,\ldots ,M,  \notag
\end{gather}%
where the form of function $A^{i}$ depends on the coefficients of (\ref{2.6}%
)-(\ref{2.66}), i.e., on $\sigma $ and on choice of the quadrature rule $%
S_{I_{j}};$ $\kappa (t_{k},T_{i})$ is as in the quadrature (\ref{2.66}).
Note that in this section we use the same notation $\bar{f}_{k}^{i}$ for the
mean-square approximation as the one we use for weak approximations in all
the other sections of this paper. Since mean-square approximations of (\ref%
{2.6}) are considered in this section only, this abuse of notation does not
lead to any confusion.

As before, we put 
\begin{equation*}
\bar{f}_{k}^{i}=\bar{f}_{\mathbf{m}}^{i},\ \ \ \ k=\mathbf{m}+1,\ldots ,M,\
0\leq i\leq \ell (t^{\ast })-1,
\end{equation*}%
where $\mathbf{m}=\left\lceil \left( T_{i+1}-t_{0}\right) /h\right\rceil -1.$
Then the $N+1$-dimensional vector $\{\bar{f}_{k}^{i},$ $i=0,\ldots ,N\}$ is
defined for all $k=0,\ldots ,M.$

We impose the following assumption on the one-step approximation $\bar{f}%
_{t,x}^{i}(t+h)$ of the method (\ref{msq1}) for the solution $\tilde{f}%
_{t,x}^{i}(t+h)$ of (\ref{2.6})\ with the initial condition $x$ given at
time $t:$ $\tilde{f}_{t,x}^{i}(t)=x^{i}$.\medskip

\noindent \textbf{Assumption 6.1 } \emph{Let} 
\begin{equation}
q_{2}\geq \frac{1}{2}\,,\;q_{1}\geq q_{2}+\frac{1}{2}\,.  \label{Ba07}
\end{equation}%
\emph{Suppose the one-step approximation }$\bar{f}_{t,x}^{i}(t+h)$\emph{\
has order of accuracy }$q_{1}$\emph{\ for expectation of the deviation and
order of accuracy }$q_{2}$\emph{\ for the mean-square deviation; more
precisely, for arbitrary} $t_{0}\leq t\leq t^{\ast }-h,$\ $x\in \mathbb{R}%
^{N+1}$ \emph{the following inequalities hold:} 
\begin{equation}
|E(\tilde{f}_{t,x}^{i}(t+h)-\bar{f}_{t,x}^{i}(t+h))|\leq Ch^{q_{1}}\,,
\label{Ba05}
\end{equation}%
\begin{gather}
\left[ E|\tilde{f}_{t,x}^{i}(t+h)-\bar{f}_{t,x}^{i}(t+h))|^{2}\right]
^{1/2}\leq Ch^{q_{2}},  \label{Ba06} \\
i=0,\ldots N,  \notag
\end{gather}%
\emph{where }$C>0$ \emph{is a constant independent of }$h,$ $\Delta ,$ and $%
x.$ \medskip

Assumption~6.1 is analogous to the conditions of the fundamental theorem of
mean-square convergence \cite[p. 4]{MT1}. We note that $C$ in (\ref{Ba05})-(%
\ref{Ba06}) are independent of $x$ while in the fundamental theorem such $C$
depend on $x$. In our case it is natural to put $C$ independent of $x$ since
the coefficients of (\ref{2.6}) and their derivatives are uniformly bounded
(see Assumptions~2.1-2.2). We also emphasize that the constants $C$ in (\ref%
{Ba05})-(\ref{Ba06}) do not depend on $\Delta $.

Under the stated assumptions we will prove mean-square convergence of $\bar{f%
}_{k}^{i}$ to $\tilde{f}^{i}(t_{k})$ uniform in $\Delta $ in order then to
prove mean-square convergence of $\bar{f}_{k}^{i}$ to $f(t_{k},T_{i})$
exploiting in addition Theorem~\ref{thm:msqi2f}. We cannot use here the
fundamental theorem of mean-square convergence \cite[p. 4]{MT1} since we
need to show that the convergence is uniform in $\Delta .$

\begin{theorem}
\label{thm:msq6}Suppose Assumptions~2.1-2.3 and Assumption~6.1 are
satisfied. Then for any $M,$ $N$ and $k=0,1,\ldots ,M,$ $i=0,1,\ldots ,N$
the following inequality holds: 
\begin{equation}
\left[ E|\tilde{f}^{i}(t_{k})-\bar{f}_{k}^{i}|^{2}\right] ^{1/2}\leq
Kh^{q_{2}-1/2}\,,  \label{Ba08}
\end{equation}%
i.e., the order of mean-square accuracy of the method $(\ref{msq1})$ for $(%
\ref{2.6})$ is $q=q_{2}-1/2.$
\end{theorem}

\noindent \textbf{Proof. } We have (cf . \cite[pp. 7-8]{MT1}) 
\begin{gather}
\tilde{f}^{i}(t_{k+1})-\bar{f}_{k+1}^{i}=\tilde{f}%
_{t_{0},f_{0}}^{i}(t_{k+1})-\bar{f}_{t_{0},f_{0}}^{i}(t_{k+1})=\tilde{f}%
_{t_{k},\tilde{f}(t_{k})}^{i}(t_{k+1})-\bar{f}_{t_{k},\bar{f}%
_{k}}^{i}(t_{k+1})  \label{Ba27} \\
=(\tilde{f}_{t_{k},\tilde{f}(t_{k})}^{i}(t_{k+1})-\tilde{f}_{t_{k},\bar{f}%
_{k}}^{i}(t_{k+1}))+(\tilde{f}_{t_{k},\bar{f}_{k}}^{i}(t_{k+1})-\bar{f}%
_{t_{k},\bar{f}_{k}}^{i}(t_{k+1}))\,,  \notag
\end{gather}%
where the first difference in the right-hand side of (\ref{Ba27}) is the
error of the solution arising due to the error in the initial data at time $%
t_{k},$ accumulated over $k$ steps, and the second difference is the
one-step error at the $(k+1)$-step. Taking the square of both sides of (\ref%
{Ba27}), we obtain 
\begin{gather}
R_{i,k+1}^{2}:=E|\tilde{f}^{i}(t_{k+1})-\bar{f}_{k+1}^{i}|^{2}  \label{Ba28}
\\
=EE(|\tilde{f}_{t_{k},\tilde{f}(t_{k})}^{i}(t_{k+1})-\tilde{f}_{t_{k},\bar{f}%
_{k}}^{i}(t_{k+1})|^{2}|\mathcal{F}_{t_{k}})  \notag \\
+EE(|\tilde{f}_{t_{k},\bar{f}_{k}}^{i}(t_{k+1})-\bar{f}_{t_{k},\bar{f}%
_{k}}^{i}(t_{k+1})|^{2}|\mathcal{F}_{t_{k}})\,  \notag \\
+2EE((\tilde{f}_{t_{k},\tilde{f}(t_{k})}^{i}(t_{k+1})-\tilde{f}_{t_{k},\bar{f%
}_{k}}^{i}(t_{k+1}))(\tilde{f}_{t_{k},\bar{f}_{k}}^{i}(t_{k+1})-\bar{f}%
_{t_{k},\bar{f}_{k}}^{i}(t_{k+1}))|\mathcal{F}_{t_{k}})\,\,.  \notag
\end{gather}

Due to the condition (\ref{Ba06}), we get for the second term on the
right-hand side of (\ref{Ba28}): 
\begin{equation}
|EE(|\tilde{f}_{t_{k},\bar{f}_{k}}^{i}(t_{k+1})-\bar{f}_{t_{k},\bar{f}%
_{k}}^{i}(t_{k+1})|^{2}|\mathcal{F}_{t_{k}})|\leq Ch^{2q_{2}}.  \label{msq11}
\end{equation}

Let us estimate the first term on the right-hand side of (\ref{Ba28}). Ito's
formula implies that 
\begin{eqnarray*}
&&\varepsilon _{i}^{2}(t_{k+1}):=E|\tilde{f}_{t_{k},\tilde{f}%
(t_{k})}^{i}(t_{k+1})-\tilde{f}_{t_{k},\bar{f}_{k}}^{i}(t_{k+1})|^{2} \\
&=&E|\tilde{f}^{i}(t_{k})-\bar{f}_{k}^{i}|^{2} \\
&&+2E\int_{t_{k}}^{t_{k+1}}(\tilde{f}_{t_{k},\tilde{f}(t_{k})}^{i}(s)-\tilde{%
f}_{t_{k},\bar{f}_{k}}^{i}(s)) \\
&&\times \lbrack \sigma ^{\top }(s,T_{i},\tilde{f}_{t_{k},\tilde{f}%
(t_{k})}^{i}(s))\tilde{S}_{I}(t_{k},\tilde{f}(t_{k});s,T_{i},\Delta )-\sigma
^{\top }(s,T_{i},\tilde{f}_{t_{k},\bar{f}_{k}}^{i}(s))\tilde{S}_{I}(t_{k},%
\bar{f}_{k};s,T_{i},\Delta )]ds \\
&&+E\int_{t_{k}}^{t_{k+1}}|\sigma (s,T_{i},\tilde{f}_{t_{k},\tilde{f}%
(t_{k})}^{i}(s))-\sigma (s,T_{i},\tilde{f}_{t_{k},\bar{f}%
_{k}}^{i}(s))|^{2}ds.
\end{eqnarray*}%
Then, recalling that $\sigma (s,T,z)$ is\ globally Lipschitz in $z$ due to
Assumption~2.2 and the form of $\tilde{S}_{I}(t_{k},\bar{f}%
_{k};s,T_{i},\Delta )$ (see (\ref{SI_ext})), we obtain 
\begin{eqnarray*}
\varepsilon _{i}^{2}(t_{k+1}) &\leq &E|\tilde{f}^{i}(t_{k})-\bar{f}%
_{k}^{i}|^{2}+K\int_{t_{k}}^{t_{k+1}}E|\tilde{f}_{t_{k},\tilde{f}%
(t_{k})}^{i}(s)-\tilde{f}_{t_{k},\bar{f}_{k}}^{i}(s)|^{2}ds \\
&&+K\Delta \int_{t_{k}}^{t_{k+1}}\dsum_{l=\ell (s)}^{\kappa (s,T_{i})}E|%
\tilde{f}_{t_{k},\tilde{f}(t_{k})}^{l}(s)-\tilde{f}_{t_{k},\bar{f}%
_{k}}^{l}(s)|^{2}ds,
\end{eqnarray*}%
where $K>0$ does not depend on $\Delta $. Introduce $\varepsilon
_{Max}^{2}(s):=\max_{_{0\leq i\leq M}}\varepsilon _{i}^{2}(s).$ Then 
\begin{equation*}
\varepsilon _{Max}^{2}(t_{k+1})\leq \max_{_{0\leq i\leq M}}E|\tilde{f}%
^{i}(t_{k})-\bar{f}_{k}^{i}|^{2}+K\int_{t_{k}}^{t_{k+1}}\varepsilon
_{Max}^{2}(s)ds
\end{equation*}%
which implies that for all $0\leq i\leq M$ and all sufficiently small $h>0:$%
\begin{equation}
\varepsilon _{i}^{2}(t_{k+1})\leq e^{Kh}\max_{_{0\leq i\leq M}}E|\tilde{f}%
^{i}(t_{k})-\bar{f}_{k}^{i}|^{2}\leq \max_{_{0\leq i\leq M}}E|\tilde{f}%
^{i}(t_{k})-\bar{f}_{k}^{i}|^{2}\cdot (1+Kh),  \label{msq12}
\end{equation}%
where $K>0$ does not depend on $\Delta $ and $h.$

Now let us estimate the third term on the right-hand side of (\ref{Ba28}).
We have (cf. Lemma~1.1.3 in \cite[p. 5]{MT1}): 
\begin{equation}
\tilde{f}_{t_{k},\tilde{f}(t_{k})}^{i}(t_{k+1})-\tilde{f}_{t_{k},\bar{f}%
_{k}}^{i}(t_{k+1})=\tilde{f}^{i}(t_{k})-\bar{f}_{k}^{i}+Z^{i},  \label{msq13}
\end{equation}%
where 
\begin{eqnarray*}
Z^{i} &=&\int_{t_{k}}^{t_{k+1}}[\sigma ^{\top }(s,T_{i},\tilde{f}_{t_{k},%
\tilde{f}(t_{k})}^{i}(s))\tilde{S}_{I}(t_{k},\tilde{f}(t_{k});s,T_{i},\Delta
) \\
&&-\sigma ^{\top }(s,T_{i},\tilde{f}_{t_{k},\bar{f}_{k}}^{i}(s))\tilde{S}%
_{I}(t_{k},\bar{f}_{k};s,T_{i},\Delta )]ds \\
&&+\int_{t_{k}}^{t_{k+1}}[\sigma ^{\top }(s,T_{i},\tilde{f}_{t_{k},\tilde{f}%
(t_{k})}^{i}(s))-\sigma ^{\top }(s,T_{i},\tilde{f}_{t_{k},\bar{f}%
_{k}}^{i}(s))]dW(s).
\end{eqnarray*}%
Using (\ref{msq12}), it is not difficult to get 
\begin{equation}
E\left( Z^{i}\right) ^{2}\leq Kh\cdot \max_{_{0\leq i\leq M}}E|\tilde{f}%
^{i}(t_{k})-\bar{f}_{k}^{i}|^{2},  \label{msq14}
\end{equation}%
where $K>0$ does not depend on $\Delta $ and $h.$ Using (\ref{msq13}), (\ref%
{Ba05}), (\ref{Ba06}), (\ref{msq14}), and (\ref{Ba07}), we obtain 
\begin{eqnarray}
&&|EE((\tilde{f}_{t_{k},\tilde{f}(t_{k})}^{i}(t_{k+1})-\tilde{f}_{t_{k},\bar{%
f}_{k}}^{i}(t_{k+1}))(\tilde{f}_{t_{k},\bar{f}_{k}}^{i}(t_{k+1})-\bar{f}%
_{t_{k},\bar{f}_{k}}^{i}(t_{k+1}))|\mathcal{F}_{t_{k}})|  \label{msq15} \\
&\leq &|E(\tilde{f}^{i}(t_{k})-\bar{f}_{k}^{i})E(\tilde{f}_{t_{k},\bar{f}%
_{k}}^{i}(t_{k+1})-\bar{f}_{t_{k},\bar{f}_{k}}^{i}(t_{k+1}))|\mathcal{F}%
_{t_{k}})|  \notag \\
&&+|EZ^{i}\cdot (\tilde{f}_{t_{k},\bar{f}_{k}}^{i}(t_{k+1})-\bar{f}_{t_{k},%
\bar{f}_{k}}^{i}(t_{k+1}))|  \notag \\
&\leq &(E|\tilde{f}^{i}(t_{k})-\bar{f}_{k}^{i}|^{2})^{1/2}\cdot
Kh^{q_{1}}+\left( E\left( Z^{i}\right) ^{2}\right) ^{1/2}\left( E(\tilde{f}%
_{t_{k},\bar{f}_{k}}^{i}(t_{k+1})-\bar{f}_{t_{k},\bar{f}%
_{k}}^{i}(t_{k+1})^{2}\right) ^{1/2}  \notag \\
&\leq &Kh^{q_{1}}(E|\tilde{f}^{i}(t_{k})-\bar{f}%
_{k}^{i}|^{2})^{1/2}+Kh^{q_{2}+1/2}\cdot \left( \max_{_{0\leq i\leq M}}E|%
\tilde{f}^{i}(t_{k})-\bar{f}_{k}^{i}|^{2}\right) ^{1/2}  \notag \\
&\leq &Kh^{q_{2}+1/2}\cdot \left( \max_{_{0\leq i\leq M}}E|\tilde{f}%
^{i}(t_{k})-\bar{f}_{k}^{i}|^{2}\right) ^{1/2},  \notag
\end{eqnarray}%
where $K>0$ does not depend on $\Delta $ and $h.$

Let $R_{Max,k}^{2}:=\max_{_{0\leq i\leq M}}R_{i,k}^{2}.$ Then it follows
from (\ref{Ba28}), (\ref{msq11}), (\ref{msq12}) and (\ref{msq15}) that%
\begin{equation*}
R_{Max,k+1}^{2}\leq R_{Max,k}^{2}\cdot
(1+Kh)+Kh^{q_{2}+1/2}R_{Max,k}+Ch^{2q_{2}}.
\end{equation*}%
Using the elementary relation 
\begin{equation*}
h^{q_{2}+1/2}R_{Max,k}\leq \frac{R_{Max,k}^{2}h}{2}+\frac{h^{2q_{2}}}{2}\,,
\end{equation*}%
we get 
\begin{equation*}
R_{Max,k+1}^{2}\leq R_{Max,k}^{2}\cdot (1+Kh)+Ch^{2q_{2}}
\end{equation*}%
whence (\ref{Ba08}) follows taking into account Lemma~1.1.6 from \cite[p. 7]%
{MT1} and the fact that $R_{Max,0}^{2}=0.$ Theorem~\ref{thm:msq6} is proved.
\ $\square \medskip $

Theorems~\ref{thm:msqi2f}\ and~\ref{thm:msq6} imply the following result.

\begin{theorem}
\label{thm:msq62}Assume that the conditions of Theorems~\ref{thm:msqi2f} and~%
\ref{thm:msq6} hold. Then for any $M,$ $N$ and $i=0,1,\ldots ,N,$ $%
k=0,1,\ldots ,\left\lceil \left( T_{i}-t_{0}\right) /h\right\rceil -1$ the
mean-square error is estimated as 
\begin{equation}
\left[ E|f(t_{k},T_{i})-\bar{f}_{k}^{i}|^{2}\right] ^{1/2}\leq K\cdot
(\Delta ^{p}+h^{q_{2}-1/2})\,,  \label{msq99}
\end{equation}%
where $K>0$ is a constant independent of $\Delta $ and $h.$
\end{theorem}

\begin{example}
To illustrate the results of this section, let us present a mean-square
algorithm for $(\ref{3})$-$(\ref{3b})$ based on the mean-square Euler-type
scheme: 
\begin{gather}
\bar{f}_{0}^{i}=f_{0}^{i},\text{ }i=0,\ldots ,N,\text{ \ }\bar{Y}_{0}=0,
\label{msqE1} \\
\bar{f}_{k+1}^{i}=\bar{f}_{k}^{i}+\tsum\limits_{j=1}^{d}\bar{\sigma}%
_{i,j}(t_{k})\mathbb{\bar{S}}_{I_{j}}(t_{k},T_{i};\Delta
,h)+h^{1/2}\tsum\limits_{j=1}^{d}\bar{\sigma}_{i,j}(t_{k})\xi _{j,k+1}, 
\notag \\
i=\ell _{k+1},\ldots ,N,  \notag
\end{gather}%
where $\xi _{j,k+1}$ are independent Gaussian random variables with zero
mean and unit variance,\ 
\begin{equation*}
(\bar{\sigma}_{i,1}(t_{k}),\ldots ,\bar{\sigma}_{i,d}(t_{k}))^{\top
}=(\sigma _{1}(t_{k},T_{i},\bar{f}_{k}^{i}),\ldots ,\sigma _{d}(t_{k},T_{i},%
\bar{f}_{k}^{i}))^{\top },
\end{equation*}%
and $\mathbb{\bar{S}}_{I_{j}}(t_{k},T_{i};\Delta ,h)$ depends on our choice
of the quadrature rule $(\ref{2.444})$. If $\mathbb{\bar{S}}%
_{I_{j}}(t_{k},T_{i};\Delta ,h)$ is taken from $(\ref{4.1.1})$ or $(\ref{5.1}%
)$ or from Algorithm~5.3 then $p=1,$ $p=2$ or $p=4,$ respectively, and $%
q_{1}=2$ and $q_{2}=1$ under $h\leq \Delta .$ The overall error of these
algorithms are $O(\Delta +h^{1/2}),$ $O(\Delta ^{2}+h^{1/2}),$ and $O(\Delta
^{4}+h^{1/2}),$ respectively.
\end{example}

\section{Numerical examples\label{sec:exp}}

In this section we demonstrate accuracy and convergence properties of the
algorithms from Section~\ref{sec:alg}. We also compare computational costs
of the algorithms. This comparison illustrates that the algorithms with
higher-order quadrature rules are more efficient.

For illustration, we price an interest rate caplet which is an interest rate
derivative providing protection against an increase in an interest rate for
a single period. Suppose a caplet is set at time $t^{\ast }$ with payment
date at $T^{\ast }$ and has the unit nominal value and a strike $K.$ The
arbitrage price of the caplet is given by (\ref{5}) with $t^{\ast }=s_{k}$
and $T^{\ast }=s_{i}.$ The caplet parameters chosen for the experiments are $%
t^{\ast }=1.0,$ $T^{\ast }=6.0,\ K=0.03.$

A particular model within the HJM framework (\ref{3})-(\ref{3b}) is
specified by a choice of the volatility function and initial forward rate
curve. Here we consider two examples: a one-factor model with deterministic
exponential volatility function (Vasicek model, see, e.g. \cite%
{BR96,BrigoMercurio}) and a two-factor model with proportional volatility
function (see, e.g. \cite{HJM1992,Glasserman,NBS07,Gupta}). The former one
admits a closed-form formula for the caplet price.

The algorithms were implemented using C++ with GCC 3.4.3 compiler. The
experiments were run on ALICE HPC Computer nodes of the University of
Leicester, each with dual quad-core 2.67GHz Intel Xeon X5550 processor, 12
GB RAM, and OS 64-bit Scientific Linux 5.4.

\subsection{Vasicek model}

We consider the one-factor HJM model (\ref{3})-(\ref{3b}) with the
deterministic volatility function given by%
\begin{equation}
\sigma (t,T)=\sigma \exp (-\kappa (T-t)),  \label{7.1.1}
\end{equation}%
and the initial forward curve defined as%
\begin{equation}
f_{0}(T)=\exp (-\kappa (T-t_{0}))r_{0}+\left( 1-\exp (-\kappa
(T-t_{0}))\right) \vartheta -\frac{\sigma ^{2}}{2\kappa ^{2}}\left( 1-\exp
(-\kappa (T-t_{0}))\right) ^{2},  \label{7.1.2}
\end{equation}%
where $\sigma ,$ $\kappa ,$ $r_{0},$ and $\vartheta $ are given positive
constants.

%TCIMACRO{\TeXButton{B}{\begin{table}[h] \centering}}%
%BeginExpansion
\begin{table}[h] \centering%
%EndExpansion
%TCIMACRO{%
%\TeXButton{tab1}{\caption{{\it Algorithm~5.1 for the Vasicek model.}
%Performance of Algorithm~5.1
%with $\Delta=h$ in the case of  the Vasicek model (\ref{7.1.1}), (\ref{7.1.2}) with parameters $\sigma=0.02$, $r_{0}=0.05$, $\kappa=1$ and $\theta=1$
%for pricing a unit nominal caplet with parameters $t_0=0$, $t^{*}=1.0$, $T^{\ast}=6.0,~K=0.03.$ $L$ is the number 
%of independent runs in the Monte Carlo simulation (see (\ref{2.20})). 
%\label{tab1}}} }%
%BeginExpansion
\caption{{\it Algorithm~5.1 for the Vasicek model.}
Performance of Algorithm~5.1
with $\Delta=h$ in the case of  the Vasicek model (\ref{7.1.1}), (\ref{7.1.2}) with parameters $\sigma=0.02$, $r_{0}=0.05$, $\kappa=1$ and $\theta=1$
for pricing a unit nominal caplet with parameters $t_0=0$, $t^{*}=1.0$, $T^{\ast}=6.0,~K=0.03.$ $L$ is the number 
of independent runs in the Monte Carlo simulation (see (\ref{2.20})). 
\label{tab1}}
%EndExpansion
\setlength{\tabcolsep}{2pt} 
\begin{tabular}{cccc}
\hline
$h$ & $L$ & $error$ & $\ \ \ \ CPU~time,$ $min\ \ \ \ $ \\ \hline
\multicolumn{1}{l}{$\ \ \ \ 0.2\ \ \ \ $} & $\ \ \ \ 10^{7}\ \ \ \ $ & $%
4.22\times 10^{-2}\pm 2.80\times 10^{-6}$ & $4.00\times 10^{-1}$ \\ \hline
\multicolumn{1}{l}{$\ \ \ \ 0.1\ \ \ \ $} & $\ \ \ \ 10^{7}\ \ \ \ $ & $%
2.04\times 10^{-2}\pm 3.06\times 10^{-6}$ & $9.00\times 10^{-1}$ \\ \hline
\multicolumn{1}{l}{$\ \ \ \ 0.05\ \ \ \ $} & $\ \ \ \ 10^{7}\ \ \ \ $ & $%
1.00\times 10^{-2}\pm 3.19\times 10^{-6}$ & $2.45\times 10^{0}$ \\ \hline
\multicolumn{1}{l}{$\ \ \ \ 0.025\ \ \ \ $} & $\ \ \ \ 10^{7}\ \ \ \ $ & $%
4.98\times 10^{-3}\pm 3.26\times 10^{-6}$ & $7.73\times 10^{0}$ \\ \hline
\multicolumn{1}{l}{$\ \ \ \ 0.0125\ \ \ \ $} & $\ \ \ \ 10^{9}\ \ \ \ $ & $%
2.48\times 10^{-3}\pm 3.29\times 10^{-7}$ & $2.30\times 10^{3}$ \\ \hline
$\ \ \ \ 0.00625\ \ \ \ $ & $\ \ \ \ 10^{9}\ \ \ \ $ & $1.24\times
10^{-3}\pm 3.31\times 10^{-7}$ & $8.32\times 10^{3}$ \\ \hline
\end{tabular}%
%TCIMACRO{\TeXButton{E}{\end{table}}}%
%BeginExpansion
\end{table}%
%EndExpansion

%TCIMACRO{\TeXButton{B}{\begin{table}[h] \centering}}%
%BeginExpansion
\begin{table}[h] \centering%
%EndExpansion
%TCIMACRO{%
%\TeXButton{tab2}{\caption{{\it Algorithm~5.2 for the Vasicek model.}
%Performance of Algorithm~5.2
%with $\Delta=\sqrt{h}$ in the case of the Vasicek model (\ref{7.1.1}), (\ref{7.1.2}) with 
%the same parameters as in Table~\ref{tab1}.
%\label{tab2}}} }%
%BeginExpansion
\caption{{\it Algorithm~5.2 for the Vasicek model.}
Performance of Algorithm~5.2
with $\Delta=\sqrt{h}$ in the case of the Vasicek model (\ref{7.1.1}), (\ref{7.1.2}) with 
the same parameters as in Table~\ref{tab1}.
\label{tab2}}
%EndExpansion
\smallskip \setlength{\tabcolsep}{2pt} 
\begin{tabular}{cccc}
\hline
$h$ & $L$ & $error$ & $\ \ \ \ CPU~time,$ $min\ \ \ \ $ \\ \hline
\multicolumn{1}{l}{$\ \ \ \ 0.2\ \ \ \ $} & $\ \ \ \ 10^{7}\ \ \ \ $ & $%
6.53\times 10^{-3}\pm 2.72\times 10^{-6}$ & $3.00\times 10^{-1}$ \\ \hline
\multicolumn{1}{l}{$\ \ \ \ 0.1\ \ \ \ $} & $\ \ \ \ 10^{7}\ \ \ \ $ & $%
3.32\times 10^{-3}\pm 2.99\times 10^{-6}$ & $5.33\times 10^{-1}$ \\ \hline
\multicolumn{1}{l}{$\ \ \ \ 0.05\ \ \ \ $} & $\ \ \ \ 10^{7}\ \ \ \ $ & $%
1.65\times 10^{-3}\pm 3.11\times 10^{-6}$ & $1.03\times 10^{0}$ \\ \hline
\multicolumn{1}{l}{$\ \ \ \ 0.025\ \ \ \ $} & $\ \ \ \ 10^{7}\ \ \ \ $ & $%
8.29\times 10^{-4}\pm 3.22\times 10^{-6}$ & $2.13\times 10^{0}$ \\ \hline
\multicolumn{1}{l}{$\ \ \ \ 0.0125\ \ \ \ $} & $\ \ \ \ 10^{9}\ \ \ \ $ & $%
4.13\times 10^{-4}\pm 3.27\times 10^{-7}$ & $4.65\times 10^{2}$ \\ \hline
$0.00625$ & $\ \ \ \ 10^{9}\ \ \ \ $ & $2.07\times 10^{-4}\pm 3.30\times
10^{-7}$ & $1.09\times 10^{3}$ \\ \hline
\end{tabular}%
%TCIMACRO{\TeXButton{E}{\end{table}}}%
%BeginExpansion
\end{table}%
%EndExpansion

%TCIMACRO{\TeXButton{B}{\begin{table}[h] \centering}}%
%BeginExpansion
\begin{table}[h] \centering%
%EndExpansion
%TCIMACRO{%
%\TeXButton{tab3}{\caption{{\it Algorithm~5.3 for the Vasicek model.}
%Performance of Algorithm~5.3
%with $\Delta=\alpha \sqrt[4]{h}$ in the case of the Vasicek model (\ref{7.1.1}), (\ref{7.1.2}) with 
%the same parameters as in Table~\ref{tab1}.
%\label{tab3}}} }%
%BeginExpansion
\caption{{\it Algorithm~5.3 for the Vasicek model.}
Performance of Algorithm~5.3
with $\Delta=\alpha \sqrt[4]{h}$ in the case of the Vasicek model (\ref{7.1.1}), (\ref{7.1.2}) with 
the same parameters as in Table~\ref{tab1}.
\label{tab3}}
%EndExpansion
\smallskip \setlength{\tabcolsep}{2pt} 
\begin{tabular}{ccccc}
\hline
$h$ & $L$ & $error$ & $\ \ \ \ CPU~time,$ $min\ \ \ \ $ & $\ \ \ \ \alpha \
\ \ \ $ \\ \hline
\multicolumn{1}{l}{$\ \ \ \ 0.2\ \ \ \ $} & $\ \ \ \ 10^{7}\ \ \ \ $ & $%
1.25\times 10^{-3}\pm 2.53\times 10^{-6}$ & $2.59\times 10^{-1}$ & $%
9.97\times 10^{-1}$ \\ \hline
\multicolumn{1}{l}{$\ \ \ \ 0.1\ \ \ \ $} & $\ \ \ \ 10^{7}\ \ \ \ $ & $%
6.28\times 10^{-4}\pm 2.87\times 10^{-6}$ & $4.10\times 10^{-1}$ & $%
9.70\times 10^{-1}$ \\ \hline
\multicolumn{1}{l}{$\ \ \ \ 0.05\ \ \ \ $} & $\ \ \ \ 10^{7}\ \ \ \ $ & $%
3.18\times 10^{-4}\pm 3.09\times 10^{-6}$ & $8.11\times 10^{-1}$ & $%
9.76\times 10^{-1}$ \\ \hline
\multicolumn{1}{l}{$\ \ \ \ 0.025\ \ \ \ $} & $\ \ \ \ 10^{7}\ \ \ \ $ & $%
1.56\times 10^{-4}\pm 3.20\times 10^{-6}$ & $1.59\times 10^{0}$ & $%
9.43\times 10^{-1}$ \\ \hline
\multicolumn{1}{l}{$\ \ \ \ 0.0125\ \ \ \ $} & $\ \ \ \ 10^{9}\ \ \ \ $ & $%
9.62\times 10^{-5}\pm 3.26\times 10^{-7}$ & $3.13\times 10^{2}$ & $%
9.97\times 10^{-1}$ \\ \hline
\multicolumn{1}{l}{$\ \ \ \ 0.00625\ \ \ \ $} & $\ \ \ \ 10^{9}\ \ \ \ $ & $%
4.71\times 10^{-5}\pm 3.30\times 10^{-7}$ & $5.96\times 10^{2}$ & $%
9.70\times 10^{-1}$ \\ \hline
\end{tabular}%
%TCIMACRO{\TeXButton{E}{\end{table}}}%
%BeginExpansion
\end{table}%
%EndExpansion

It is known (see, e.g. \cite{BR96,BrigoMercurio,Rebo}) that a caplet
corresponds to a put option on a zero-coupon bond. In \cite{Jamshidian}
analytic expressions for the European option prices on zero-coupon and
coupon bearing bonds under the Vasicek model are derived. In particular, the
price of the caplet set at time $t^{\ast }$ with payment date at $T^{\ast }$%
, unit nominal value and strike $K$ is given by%
\begin{equation}
F(t_{0},f_{0}\left( \cdot \right) ;t^{\ast },T^{\ast })=P(t_{0},t^{\ast
})\Phi (-c_{P}+\sigma _{P})-(1+K(T^{\ast }-t^{\ast }))P(t_{0},T^{\ast })\Phi
(-c_{P}),  \label{Vascapprice}
\end{equation}%
where $\Phi (\cdot )$ denotes the standard normal cumulative distribution
function and 
\begin{eqnarray*}
\sigma _{P} &=&\frac{\sigma }{\kappa }\sqrt{\frac{1-\exp (-2\kappa (t^{\ast
}-t_{0}))}{2\kappa }}\left[ 1-\exp (-2\kappa (T^{\ast }-t^{\ast }))\right] ,
\\
&&c_{P}=\frac{1}{\sigma _{P}}\ln \frac{(1+K(T^{\ast }-t^{\ast
}))P(t_{0},T^{\ast })}{P(t_{0},t^{\ast })}+\frac{\sigma _{P}}{2}.
\end{eqnarray*}

The values of parameters chosen in the experiments are $t_{0}=0,$ $\sigma
=0.02,$ $r_{0}=0.05,$ $\kappa =1,$ and $\vartheta =1.$ The values $\kappa =1$
and $\vartheta =1$ are rather unrealistic from the financial point of view
and are chosen for illustrative purposes. Under a more realistic choice of
parameters, simulations are done with a particular time step $h$ (see Table~%
\ref{tab4}).

The results of the experiments with Algorithm~5.1 of order $O(\Delta +h)$,
Algorithm~5.2 of order $O(\Delta ^{2}+h)$, and Algorithm~5.3 of order $%
O(\Delta ^{4}+h)$ are presented in Tables~\ref{tab1}, \ref{tab2}, and~\ref%
{tab3}, respectively. For Algorithms~5.1 and~5.2, we set $\Delta =h$ and $%
\Delta =\sqrt{h},$ respectively. For Algorithm~5.3, we set $\Delta =\alpha 
\sqrt[4]{h}$ with $\alpha >0$ so that $T$-grid remains equally spaced: 
\begin{equation}
\alpha =\frac{1}{\sqrt[4]{h}}\left\{ 
\begin{array}{l}
\frac{T^{\ast }-t_{0}}{\left\lfloor \frac{T^{\ast }-t_{0}}{\sqrt[4]{h}}%
\right\rfloor +1},\text{ if \ }\frac{T^{\ast }-t_{0}}{\sqrt[4]{h}}%
-\left\lfloor \frac{T^{\ast }-t_{0}}{\sqrt[4]{h}}\right\rfloor \geq \frac{1}{%
2}, \\ 
\frac{T^{\ast }-t_{0}}{\left\lfloor \frac{T^{\ast }-t_{0}}{\sqrt[4]{h}}%
\right\rfloor },\text{ otherwise,}%
\end{array}%
\right.  \label{alpha3}
\end{equation}%
where $\left\lfloor \cdot \right\rfloor $ denotes the integer part of a real
number. It is clear that $\alpha \approx 1.$

As a result, the errors of all three algorithms become of order $O(h)$. In
the tables, the values before \textquotedblleft $\pm $\textquotedblright\
are estimates of the bias computed as the difference between the exact
caplet price (\ref{Vascapprice}) and its sampled approximation (see (\ref%
{2.20})), while the values after \textquotedblleft $\pm $\textquotedblright\
give half of the size of the confidence interval for the corresponding
estimator with probability $0.95$. The number of Monte Carlo runs $L$ is
chosen here so that the Monte Carlo error is small in comparison with the
bias. It is not difficult to see from the tables that the experimentally
observed convergence rate is in agreement with the theoretical first order
convergence in $h$. We note that in the analysis of convergence of
Algorithm~5.3 one has to take into account not only values of $h$ but also
of $\alpha .$ As expected, the experiments demonstrate that Algorithm~5.3 is
the most computationally efficient among the three algorithms tested and
also Algorithm~5.2 outperforms Algorithm~5.1. As it follows from Tables~\ref%
{tab1}-\ref{tab4}, for a fixed time step $h$ the ratios of running times of
the considered algorithms is in agreement with the theoretical prediction
(see Remark~\ref{algcomplexity}).

%TCIMACRO{\TeXButton{B}{\begin{table}[h] \centering}}%
%BeginExpansion
\begin{table}[h] \centering%
%EndExpansion
%TCIMACRO{%
%\TeXButton{tab4}{\caption{{\it Vasicek model.} Performance of the algorithms  (\ref{order1})-(\ref{order4}) with $h=2^{-6}$ 
%and  $L=10^{9}$ 
%in the case of the Vasicek model (\ref{7.1.1}), (\ref{7.1.2}) with
%parameters $\sigma=0.02$, $r_{0}=0.05$, $\kappa=0.178$ and $\theta=0.086$
%for pricing a unit nominal caplet with parameters  $t_0=0$, $t^{*}=1.0$, $T^{\ast}=6.0,~K=0.03.$
%\label{tab4}}} }%
%BeginExpansion
\caption{{\it Vasicek model.} Performance of the algorithms  (\ref{order1})-(\ref{order4}) with $h=2^{-6}$ 
and  $L=10^{9}$ 
in the case of the Vasicek model (\ref{7.1.1}), (\ref{7.1.2}) with
parameters $\sigma=0.02$, $r_{0}=0.05$, $\kappa=0.178$ and $\theta=0.086$
for pricing a unit nominal caplet with parameters  $t_0=0$, $t^{*}=1.0$, $T^{\ast}=6.0,~K=0.03.$
\label{tab4}}
%EndExpansion
\smallskip \setlength{\tabcolsep}{2pt} 
\begin{tabular}{ccccc}
\hline
& $h$ & $L$ & $error$ & $CPU$ $time,$ $min$ \\ \hline
\multicolumn{1}{l}{\ Algorithm 5.1$\ $} & \multicolumn{1}{l}{$\ \ \ 0.1\ \ \ 
$} & $\ \ 10^{9}\ \ $ & $\ \ -5.38\times 10^{-4}\pm 2.90\times 10^{-6}\ \ $
& $\ \ 9.79\times 10^{1}\ $ \\ \hline
\multicolumn{1}{l}{\ Algorithm 5.2$\ $} & \multicolumn{1}{l}{$\ \ \ 0.1\ \ \ 
$} & $10^{9}$ & $1.75\times 10^{-4}\pm 2.86\times 10^{-6}$ & $5.51\times
10^{1}$ \\ \hline
\multicolumn{1}{l}{$\ $Algorithm 5.3$\ $} & \multicolumn{1}{l}{$\ \ \ 0.1\ \
\ $} & $10^{9}$ & $7.81\times 10^{-5}\pm 2.89\times 10^{-6}$ & $4.10\times
10^{1}$ \\ \hline
\end{tabular}
%TCIMACRO{\TeXButton{E}{\end{table}}}%
%BeginExpansion
\end{table}%
%EndExpansion

In Table~\ref{tab4} we present results for $h=0.1$ and $L=10^{9}$ in the
case of the more realistic choice of parameters $\kappa =0.178$ and $%
\vartheta =0.086$ of the Vasicek model. With these parameters, the bias is
very small, and if one would like to analyze it, e.g. for $h=2^{-2}\times
5^{-1},$ then the number of Monte Carlo runs has to be increased up to $%
10^{11}$ in order to make the Monte Carlo error sufficiently smaller than
the bias. We see from Table~\ref{tab4} that Algorithm~5.3 is more than twice
faster than Algorithm~5.1 in producing the results of a similar accuracy.

\subsection{Proportional volatility model}

Here we choose the volatility functions of the form%
\begin{equation}
\sigma _{j}(t,T)=\sigma _{j}\exp (-\kappa _{j}(T-t))\min \left(
f(t,T),\Gamma \right) ,  \label{7.2.1}
\end{equation}%
where $\sigma _{j}$ and $\kappa _{j}$ are positive constants and $\Gamma $
is a large positive number introduced to cap the proportional volatility in
order to avoid an explosion of the forward-rate process (cf. Assumption~2.1
and also Remark~\ref{Rem:as2}). The volatility specification of the form (%
\ref{7.2.1}) yields an approximately lognormal distribution of forward rates.

%TCIMACRO{\TeXButton{B}{\begin{table}[h] \centering}}%
%BeginExpansion
\begin{table}[h] \centering%
%EndExpansion
%TCIMACRO{%
%\TeXButton{tab5}{\caption{{\it Algorithm~5.1 for the Proportional volatility model.}
%Performance of Algorithm~5.1
%with $\Delta=h$ in the case of the proportional volatility model (\ref{7.2.1})
%with parameters (\ref{7.2.2}) and with initial forward curve (\ref{7.2.3})
%for pricing a unit nominal caplet with parameters  $t_0=0$, $t^{*}=1.0$, $T^{\ast}=6.0,~K=0.03.$
%\label{tab5}}}}%
%BeginExpansion
\caption{{\it Algorithm~5.1 for the Proportional volatility model.}
Performance of Algorithm~5.1
with $\Delta=h$ in the case of the proportional volatility model (\ref{7.2.1})
with parameters (\ref{7.2.2}) and with initial forward curve (\ref{7.2.3})
for pricing a unit nominal caplet with parameters  $t_0=0$, $t^{*}=1.0$, $T^{\ast}=6.0,~K=0.03.$
\label{tab5}}%
%EndExpansion
\smallskip\ \setlength{\tabcolsep}{2pt} 
\begin{tabular}{cccc}
\hline
$h$ & $L$ & $error$ & $\ \ \ \ CPU~time,$ $min\ \ \ \ $ \\ \hline
\multicolumn{1}{l}{$\ \ \ \ 0.2\ \ \ \ $} & $\ \ \ \ 10^{9}\ \ \ \ $ & $%
5.80\times 10^{-4}\pm 2.57\times 10^{-6}$ & $\ \ \ \ 6.99\times 10^{1}\ \ \
\ $ \\ \hline
\multicolumn{1}{l}{$\ \ \ \ 0.125\ \ \ \ $} & $\ \ \ \ 10^{9}\ \ \ \ $ & $%
3.64\times 10^{-4}\pm 2.57\times 10^{-6}$ & $\ \ \ \ 1.21\times 10^{2}\ \ \
\ $ \\ \hline
\multicolumn{1}{l}{$\ \ \ \ 0.1\ \ \ \ $} & $\ \ \ \ 10^{9}\ \ \ \ $ & $%
2.92\times 10^{-4}\pm 2.57\times 10^{-6}$ & $\ \ \ \ 1.67\times 10^{2}\ \ \
\ $ \\ \hline
\multicolumn{1}{l}{$\ \ \ \ 0.05\ \ \ \ $} & $\ \ \ \ 10^{9}\ \ \ \ $ & $%
1.48\times 10^{-4}\pm 2.56\times 10^{-6}$ & $\ \ \ \ 4.84\times 10^{2}\ \ \
\ $ \\ \hline
\end{tabular}%
%TCIMACRO{\TeXButton{E}{\end{table}}}%
%BeginExpansion
\end{table}%
%EndExpansion

%TCIMACRO{\TeXButton{B}{\begin{table}[h] \centering}}%
%BeginExpansion
\begin{table}[h] \centering%
%EndExpansion
%TCIMACRO{%
%\TeXButton{tab6}{\caption{{\it Algorithm~5.2  for the Proportional volatility model.}
%Performance of Algorithm~5.2
%with $\Delta=\sqrt{h}$ in the case of the proportional volatility model (\ref{7.2.1})
%with the same parameters as in Table~\ref{tab5}.
%\label{tab6}}} }%
%BeginExpansion
\caption{{\it Algorithm~5.2  for the Proportional volatility model.}
Performance of Algorithm~5.2
with $\Delta=\sqrt{h}$ in the case of the proportional volatility model (\ref{7.2.1})
with the same parameters as in Table~\ref{tab5}.
\label{tab6}}
%EndExpansion
\smallskip \setlength{\tabcolsep}{2pt} 
\begin{tabular}{cccc}
\hline
$h$ & $L$ & $error$ & $\ \ \ \ CPU~time,$ $min\ \ \ \ $ \\ \hline
\multicolumn{1}{l}{$\ \ \ \ 0.2\ \ \ \ $} & $\ \ \ \ 10^{9}\ \ \ \ $ & $%
1.50\times 10^{-4}\pm 2.56\times 10^{-6}$ & $5.70\times 10^{1}$ \\ \hline
\multicolumn{1}{l}{$\ \ \ \ 0.125\ \ \ \ $} & $\ \ \ \ 10^{9}\ \ \ \ $ & $%
8.91\times 10^{-5}\pm 2.56\times 10^{-6}$ & $7.88\times 10^{1}$ \\ \hline
\multicolumn{1}{l}{$\ \ \ \ 0.1\ \ \ \ $} & $\ \ \ \ 10^{9}\ \ \ \ $ & $%
8.01\times 10^{-5}\pm 2.56\times 10^{-6}$ & $1.03\times 10^{2}$ \\ \hline
\multicolumn{1}{l}{$\ \ \ \ 0.05\ \ \ \ $} & $\ \ \ \ 10^{9}\ \ \ \ $ & $%
3.47\times 10^{-5}\pm 2.56\times 10^{-6}$ & $2.16\times 10^{2}$ \\ \hline
\end{tabular}%
%TCIMACRO{\TeXButton{E}{\end{table}}}%
%BeginExpansion
\end{table}%
%EndExpansion

%TCIMACRO{\TeXButton{B}{\begin{table}[h] \centering}}%
%BeginExpansion
\begin{table}[h] \centering%
%EndExpansion
\caption{{\it Algorithm~5.3  for the Proportional volatility model.}
Performance of Algorithm~5.3
with $\Delta=\alpha \sqrt[4]{h}$ in the case of the proportional volatility model (\ref{7.2.1})
 with  the same parameters as in Table~\ref{tab5}.
\label{tab7}} \smallskip \setlength{\tabcolsep}{2pt} 
\begin{tabular}{ccccc}
\hline
$h$ & $L$ & $error$ & $\ \ \ \ CPU~time,$ $min\ \ \ \ $ & $\ \ \ \ \alpha \
\ \ \ $ \\ \hline
\multicolumn{1}{l}{$\ \ \ \ 0.2\ \ \ \ $} & $\ \ \ \ 10^{9}\ \ \ \ $ & $%
7.04\times 10^{-5}\pm 2.57\times 10^{-6}$ & $5.32\times 10^{1}$ & $%
9.97\times 10^{-1}$ \\ \hline
\multicolumn{1}{l}{$\ \ \ \ 0.125\ \ \ \ $} & $\ \ \ \ 10^{9}\ \ \ \ $ & $%
4.59\times 10^{-5}\pm 2.57\times 10^{-6}$ & $6.34\times 10^{1}$ & $%
9.17\times 10^{-1}$ \\ \hline
\multicolumn{1}{l}{$\ \ \ \ 0.1\ \ \ \ $} & $\ \ \ \ 10^{9}\ \ \ \ $ & $%
3.66\times 10^{-5}\pm 2.57\times 10^{-6}$ & $7.40\times 10^{1}$ & $%
9.70\times 10^{-1}$ \\ \hline
\multicolumn{1}{l}{$\ \ \ \ 0.05\ \ \ \ $} & $\ \ \ \ 10^{9}\ \ \ \ $ & $%
1.74\times 10^{-5}\pm 2.57\times 10^{-6}$ & $1.54\times 10^{2}$ & $%
9.76\times 10^{-1}$ \\ \hline
\end{tabular}%
%TCIMACRO{\TeXButton{E}{\end{table}}}%
%BeginExpansion
\end{table}%
%EndExpansion

Let us note that in \cite{Gupta} a number of volatility models including one
and two factors proportional volatility models are\ examined. The
performance of the models is evaluated based on the accuracy of their
out-of-sample price prediction and their ability to hedge caps and floors.
This study reveals that in out-of-sample pricing accuracy the one- and two-
factor proportional volatility models outperform the other competing one-
and two- factor models, correspondingly. The one-factor BGM model
outperforms the proportional volatility model only in pricing tests, which
were not strictly out-of-sample. In terms of hedging performance, the
two-factor models provides significantly better results than the one-factor
models.

In our experiments we consider two factors, i.e., $d=2$. We use the same
parameters for (\ref{7.2.1}) as those found in \cite{Gupta} by calibrating
the model to the market prices of caps and floors across different
maturities and strike rates: 
\begin{equation}
\sigma _{1}=0.1043,\text{ }\sigma _{2}=0.1719,\ \kappa _{1}=0.052,\text{ }%
\kappa _{2}=0.035.  \label{7.2.2}
\end{equation}%
As the initial forward curve, we take the one used in numerical examples in 
\cite{GlassermanHJM}: 
\begin{equation}
f_{0}(T)=\log (150+48T)/100.  \label{7.2.3}
\end{equation}

Since the closed-form formula for caplet price is not available for the HJM
model (\ref{3})-(\ref{3b}) with the volatility (\ref{7.2.1}), we found the
reference caplet price by evaluating the price using Algorithm~5.3 with $%
h=0.00625,$ $\Delta =\alpha \sqrt[4]{h}$ with $\alpha $ from (\ref{alpha3}),
and taking the number of Monte Carlo runs $L=10^{9}$. This reference value
has the Monte Carlo error $2.56\times 10^{-6},$ which gives half of the size
of the confidence interval for the corresponding estimator with probability $%
0.95$.

Tables~\ref{tab5}, \ref{tab6}, and \ref{tab7} report the results of our
experiments for Algorithm~5.1 with $\Delta =h$, Algorithm~5.2 with $\Delta =%
\sqrt{h},$ and Algorithm~5.3 with $\Delta =\alpha \sqrt[4]{h},$ $\alpha $ is
from (\ref{alpha3}). As in the previous tables, the error column values
before \textquotedblleft $\pm $\textquotedblright\ are estimates of the bias
computed using the reference price value and the values after
\textquotedblleft $\pm $\textquotedblright\ reflect the Monte Carlo error
with probability $0.95$. As in the Vasicek model example, the Monte Carlo
error was made relatively small in order to be able to analyze the bias. One
can observe that the results demonstrate first order of convergence which is
in agreement with our theoretical results. The experiments also clearly
illustrate the computational superiority of Algorithm~5.3 whereas
Algorithm~5.1 is the slowest out of the three algorithms presented. The
computational times are consistent with the theoretical complexity of the
algorithms described in Remark~\ref{algcomplexity}.

\section*{\textbf{Acknowledgments}}

This research used the ALICE High Performance Computing Facility at the
University of Leicester. Part of this work was done while MVT was on study
leave granted by the University of Leicester.

\end{document}